\documentclass[
twocolumn,
aps,
reprint,
superscriptaddress,
citeautoscript
]{revtex4-1}
\usepackage[T1]{fontenc}
\usepackage[utf8]{inputenc}
\usepackage{lmodern}
\usepackage[dvipsnames]{xcolor}
\usepackage{graphicx}

\usepackage{amsmath}
\usepackage{booktabs}
\usepackage{xcolor}
\usepackage{epstopdf}
\usepackage[normalem]{ulem}
\usepackage[colorlinks=true, urlcolor=blue, citecolor=blue, linkcolor=blue]{hyperref}

\usepackage[utf8]{inputenc}
\usepackage{xspace}
\usepackage{graphicx}
\usepackage{epstopdf}
\usepackage{color,xcolor}
\usepackage{amsmath,amsfonts,amssymb}
\usepackage{bm}
\usepackage{upgreek}
\usepackage{physics}
\usepackage{algorithm}
\usepackage{algorithmic}
\usepackage{siunitx}
\usepackage[colorlinks=true]{hyperref}
\newcommand{\CrO}{Cr$_{2}$O$_{3}$\xspace}
\newcommand{\etal}{\textit{et~al.}\xspace}

\newcommand{\ii}{\mathrm{i}}
\newcommand{\Cpi}{\uppi}     
\newcommand{\Ce}{\mathrm{e}} 
\newcommand{\mper}{\,.} 		   
\newcommand{\mcom}{\,,}  		   

\newcommand{\xr}{\xi_{\text{R}}}
\newcommand{\xir}{\xi_{\text{IR}}}

\renewcommand{\vec}[1]{{{\bm #1}}}

\newcommand{\Fref}[2][]{Fig.~\ref{#2}\textcolor{black}{#1}} 

\newcommand{\cref}[1]{chapter~\ref{#1}\xspace}
\newcommand{\Cref}[1]{Chapter~\ref{#1}\xspace} 

\begin{document}
	
\title{A quantum heat engine based on dynamical materials design}

\author{G. Tulzer}
\affiliation{Institute for Theoretical Physics, Johannes Kepler University, Altenberger Strasse 69, 4040 Linz, Austria}

\author{M. Hoffmann}
\affiliation{Institute for Theoretical Physics, Johannes Kepler University, Altenberger Strasse 69, 4040 Linz, Austria}

\author{R. E. Zillich}
\affiliation{Institute for Theoretical Physics, Johannes Kepler University, Altenberger Strasse 69, 4040 Linz, Austria}

\begin{abstract}
We propose a novel type of quantum heat engine based on the ultrafast dynamical control of the magnetic properties of a nano-scale working body. The working principle relies on nonlinear phononics, an example for dynamical materials design. We describe the general recipe for identifying candidate materials, and also propose \CrO as a promising working body for a quantum Otto cycle. Using a spin Hamiltonian as a model for \CrO, we investigate the performance in terms of efficiency, output power, and quantum friction. To assess the assumptions underlying our effective spin Hamiltonian we also consider a working substance composed of several unit cells.
We show that even without an implementation of transitionless driving, the quantum friction is very low compared to the total produced work and the energy cost of counterdiabatic driving is negligible. This is an
advantage of the working substance, as experimentally
hard-to-implement shortcuts to adiabaticity are not needed. Moreover,
we discuss some remarkable thermodynamic features due to
the quantumness of the proposed system such as a non-monotonic dependence of the efficiency on the temperature of the hot-bath. Finally, we explore the dependence of the performance on the system parameters for a generic model of this type of quantum heat engine and identify properties of the energy spectrum required for a well-performing quantum heat engine.
\end{abstract}

\date{\today}

\maketitle
\section{Introduction}
\label{sec:intro}
Nowadays, experimentally feasible high-intensity terahertz (\si{\THz}) laser
pulses can reach  peak fields of \SI{300}{\kilo\volt\per\cm} and an energy
of \SI{5}{\nano\joule} \cite{Seifert} and can be used to excite long-wavelength
phonon modes in bulk materials and thin films. When the amplitudes 
of the excited lattice vibrations exceed several percents of the interatomic
distance, they can modify the structural material properties via
nonlinear effects \cite{Mankowsky2016_RepProgPhys_79_064503,
Juraschek2017_PhysRevLett.118.054101}. 
As a result, such nonlinear phononics allow an ultrafast (on
subpicosecond time scales) dynamical control of various sample material
properties and constitutes an example of what is called {\it dynamical
  materials design}. 

Nonlinear phononics was experimentally verified in 2011 
\cite{Forst2011_nphys2055,Forst2011_PhysRevB.84.241104,
Forst2013_j.ssc.2013.06.024}. Since then it has been studied in
many different materials, mainly transition metal oxides: 
It can favor superconductivity at ultra-high temperatures 
\cite{Cavalleri2018_00107514.2017.1406623},
induce multiferroicity \cite{Juraschek2017_PhysRevMaterials.1.014401}
and phase transitions \cite{Klein2020_PhysRevResearch.2.013336},
reduce promptly and highly nonlinearly charge order 
\cite{Esposito2017_PhysRevLett.118.247601},
cause an anisotropic effective electronic temperature
\cite{Schutt2018_PhysRevB.97.035135}, or 
interact in different ways with magnetism
\cite{Nova2017_nphys3925, 
Radaelli2018_PhysRevB.97.085145,
Fechner}.
This influence of phonons on magnetic properties is used in this work, where
the nonlinear effects are caused by an infrared (IR)
mode excited by terahertz-frequency optical pulses, which 
couples anharmonically to a Raman~(R) mode. The coupling of the 
resulting lattice displacements of the R and IR modes
($\xr$ and $\xir$, respectively)
is determined by the energy surface of nonlinear lattice vibrations.
It can be either fitted through first-principles calculations to an
anharmonic potential energy 
\cite{Subedi2014_PhysRevB.89.220301, Mankowsky2016_RepProgPhys_79_064503,Juraschek2018thesis},
or even probed by experiments \cite{vonHoegen2018_nature25484}.
The nonlinear dynamics governed by the anharmonic potential $V\big(\xr, \xir\big)$
was calculated for various different materials, e.g., 
the magnetoresistive manganites PrMnO$_{3}$ 
\cite{Subedi2014_PhysRevB.89.220301} or the 
magnetoelectric material Cr$_{2}$O$_{3}$ \cite{Fechner}.

It seems to be promising to extend this emerging field 
of dynamical materials design to applications in quantum
thermodynamics. The latter is 
a rapidly developing research field at the crossover
of quantum mechanics and statistical physics, where the central issue is whether and how the laws of thermodynamics and non-equilibrium statistical physics can be generalized to systems far away from the thermodynamic limit \cite{Gemmer2009,Vinjanampathy2016,Alicki2019,Jarzynski1997,Campisi,Campisi2011erratum,Talkner2007,Sokolov2014,Bochkov2013,Deffner2010,Quan2009}. Among the most interesting applications in this field are \emph{quantum} thermal machines, and many different aspects of them have been theoretically investigated recently: working principles \cite{Linden2010,Hubner2014,Chotorlishvili2016_10.1103/PhysRevE.94.032116,Fusco,Altintas,Hardal2017,Huang2018}, performance \cite{Esposito2010,Abah2012,Zagoskin2012,Abah2014,Rossnagel2014,DelCampo2014,Mukherjee2016}, improvements and variants of working quantum heat engines \cite{Stefanatos2018,Vepsalainen2017,Nguyen,Stefanatos,Xia}, but also studies aiming at a fundamental theoretical understanding of the features of such machines \cite{Quan2007,Lucia,Fadaie2018,Roman2019,Silveri,delCampo2018}. A general review on open-system modeling for quantum thermodynamics is given in \cite{Kosloff2019}, a review about employing the ubiquitous example of a quantum harmonic oscillator in a quantum Otto cycle is given in \cite{Rezek}, and an experimental characterization of a simple quantum heat engine has been published recently \cite{Peterson2019}.

In the present work, we propose a
new type of quantum heat engine based on a nano-scale working body
employing nonlinear phononics:
Intense laser pulses serve to dynamically control the
magnetic coupling and the associated energy spectrum of a spin system
through the anharmonic coupling of R and IR modes specific to the system, which can be described by an effective spin Hamiltonian. 
The critical issue for any quantum
thermodynamic engine is the energy spectrum of the working body,
because the 
work is produced by a parametric change of the energy levels, 
determined here by magnetic exchange interaction parameters controlled through
the dynamical materials design protocol -- the time-dependent 
microscopic lattice displacements from the phonon modes.

We demonstrate the concept of a laser induced magnetically driven 
quantum heat engine with a \CrO working body, containing 4~Cr atoms
in its unit cell. 
\CrO is in particular well suited because it was shown earlier that 
in \CrO the anharmonic coupling of an excited IR-active with a
symmetry-conserving Raman mode induces a time-dependent variation
of the magnetic coupling between the Cr ions \cite{Fechner}.
Moreover, several studies show that it is enough to consider 
the interactions inside one unit cell in order to provide
accurate magnetic properties of \CrO 
\cite{Shi2009_PhysRevB.79.104404,
Mostovoy2010_PhysRevLett.105.087202,Fechner}.
Compared to other models studied in quantum
thermodynamics, the proposed quantum heat engine has a nontrivial
spectrum (including, e.g., level crossings). Another merit of our
system is that the quantum friction turns out to be reasonably small even without an implementation of adiabatic shortcuts. 

To better understand the behavior of the heat engine, we generalize the \CrO~model Hamiltonian by considering different variations of the magnetic exchange interaction parameters. With these generic models, we systematically study the efficiency of the resulting heat engines and how they depend on the parametrically driven energy spectra.

The work is organized as follows: In section \textbf{\ref{sec:model}},
we briefly review the general steps to be taken in order to describe 
the material properties of the working body 
needed for our thermodynamic considerations
by a first-principles method. In particular, the
nonlinear lattice excitations and the magnetic coupling parameters are 
necessary and recalled for \CrO.
In section \textbf{\ref{sec:otto}}, we describe the used quantum Otto cycle as well as the concepts of quantum work, transitionless driving, and the thermalization of the working body. The results are analyzed in section \textbf{\ref{sec:results}}.
The generalization of the Hamiltonian to multiple unit cells is discussed in section~\textbf{\ref{sec:multi_cell}}.
We present a generic model Hamiltonian to describe systems which might be exploited as a quantum heat engine using a similar approach as for \CrO in section~\textbf{\ref{sec:alternative}}.

\section{Dynamical magnetic coupling}
\label{sec:model}

Below we propose a quantum heat engine driven by a time-dependent 
variation of the magnetic exchange interaction parameters $J_{ij}(t)$ in
a quantum spin Heisenberg Hamiltonian $\hat{H}$, 
\begin{align}
    \hat{H} & = \sum_{ij}^N J_{ij}(t) \hat{\vec{S}}_{i}\hat{\vec{S}}_{j}
        + \sum_{i=1}^{N}\bigg[ \mu B_z \hat{S}^z_i + D \big( \hat{S}^z_i \big)^2 \bigg] \,,
    \label{eq:Hamiltonian_general}
\end{align}
where $\hat{\vec{S}}_i$ are the spin operators. We include a Zeeman term
and a contribution for the magnetocrystalline anisotropy. This Hamiltonian
will determine the energy spectrum of our quantum system and, hence, 
the thermodynamic properties of the quantum heat engine.

While the particular origin of the time variation in $J_{ij}(t)$ 
will play only  a minor role for the quantum heat engine itself,
the form of the time dependence will have a strong impact on the 
thermodynamic properties. 
In connection with nonlinear phononics, the $J_{ij}(t)$
will follow the laser-induced lattice vibrations.
Those can be described within a semi-classical oscillator model for the
normal mode amplitude $\xi_\alpha$ 
of a phonon mode $\alpha$
\cite{Mankowsky2016_RepProgPhys_79_064503,Juraschek2018thesis}
\begin{align}
    \ddot{\xi}_\alpha+\kappa_\alpha \dot{\xi}_\alpha
      + \partial_\alpha V(\{ \xi \}) = f_\alpha (t)\mper
    \label{eq:nonlinear_phononics:oscillator_model_Jaruschek2018}
\end{align}
Here, $V$ is the potential energy of the lattice, $\kappa_\alpha$ is the 
damping constant (inverse lifetime) of the phonon mode $\alpha$, and $f_\alpha$ is the driving 
force of the laser pulse acting on mode $\alpha$.
$\kappa_\alpha$ can be either extracted from 
experimental data or calculated ab-initio using phonon methods that include 
third- or higher-order derivatives in the force constants.

The potential $V(\{ \xi \})$ has to be considered beyond the usual harmonic phonon potential 
with a general expansion in terms of all phonon modes and can be written
as \cite{Juraschek2018thesis}
\begin{align}
    V(\{ \xi \}) &= \sum_\alpha \frac{\omega^2_\alpha}{2} \xi_\alpha^2
        + \sum_{\alpha,\beta,\gamma} c_{\alpha\beta\gamma} 
            \xi_\alpha\xi_\beta\xi_\gamma\nonumber\\
       &+ \sum_{\alpha,\beta,\gamma,\delta} 
         d_{\alpha\beta\gamma\delta}
                    \xi_\alpha\xi_\beta\xi_\gamma\xi_\delta 
        + ...
   \label{eq:nonlinear_phononics:potential_expansion_Juraschek2018}
\end{align}
The first term represents the well-known linear approximation,
while the exact form and the nonzero terms in Eq.~\eqref{eq:nonlinear_phononics:potential_expansion_Juraschek2018} depend 
on the symmetries of the respective material 
\cite{Subedi2014_PhysRevB.89.220301}. 
In particular, $\xir$ and $\xr$ represent the displacements of the
IR- and Raman-active mode, respectively, for which 
the lattice symmetry of, e.g., La$_2$CuO$_4$ allows
a term $\propto \xr^2\xir^2$
\cite{Subedi2014_PhysRevB.89.220301},
while Juraschek \etal \cite{Juraschek2017_PhysRevLett.118.054101}
considered trilinear terms $\propto  \xi_{\text{IR}_1} \xi_{\text{IR}_2} \xr$
between two different IR modes.
For many other systems and most particular \CrO, a coupling term 
$\propto \xr \xir^2$ appears to be most relevant
due to its centrosymmetric symmetry, where the 
IR-active phonon modes are of odd parity while
Raman-active modes are even
\cite{Mankowsky2016_RepProgPhys_79_064503, Juraschek2018thesis,Fechner}.
The resulting potential is then
\cite{Subedi2014_PhysRevB.89.220301, Fechner}
\begin{align}
  V_\mathrm{NL}(\{ \xi \}) &=
    \frac{1}{2} \omega_\mathrm{R}^2  \xr^2 +
    \frac{1}{2} \omega_\mathrm{IR}^2  \xir^2 +
    c_{\mathrm{R,2IR}} \xr \xir^2 \nonumber\\ 
    & +
    d_\mathrm{R}   \xr^4 +
    d_\mathrm{IR}  \xir^4 \mper
    \label{eq:nonlinear_phononics:Cr2O3}
\end{align}
Note the missing factor $1/4$ with respect to the form in the work of Fechner 
\etal \cite{Fechner} in order to be consistent with 
Eq.~\eqref{eq:nonlinear_phononics:potential_expansion_Juraschek2018}.

The missing parameters in Eq.~\eqref{eq:nonlinear_phononics:Cr2O3} or in 
general in Eq.~\eqref{eq:nonlinear_phononics:potential_expansion_Juraschek2018}
can be obtained as mentioned in the introduction by fitting them to the probed
anharmonic potential surface. This was successfully done for several different
materials by theoretical calculations 
\cite{Subedi2014_PhysRevB.89.220301, 
Mankowsky2016_RepProgPhys_79_064503,Juraschek2018thesis},
or experiments \cite{vonHoegen2018_nature25484} and most importantly 
for \CrO \cite{Fechner}.

Finally, the potential in Eq.~\eqref{eq:nonlinear_phononics:Cr2O3}
is put into Eq.~\eqref{eq:nonlinear_phononics:oscillator_model_Jaruschek2018}
which results in a system of equations for the relevant modes, where only the 
IR-active mode is driven externally by the force
\begin{align}
    f_\mathrm{IR}(t) = F(t)\sin(\varOmega t)  \mcom
    \label{eq:nonlinear_phononics:driving}
\end{align}
where $F(t)$ could be a Gaussian envelope in general 
\cite{Mankowsky2016_RepProgPhys_79_064503}, but is assumed below to be
the amplitude of the driving $F(t)=E_\mathrm{drive}$ following \cite{Fechner}.
Such a system of equations provides then the normal mode amplitudes
$\xr(t)$ and  $\xir(t)$ explicitly or allows to obtain
the real atomic motions of atom $n$ in phonon mode $\alpha$
along the spatial direction $i$ via \cite{Juraschek2018thesis}
\begin{align}
    U_{n,\alpha,i}(t) = \xi_\alpha(t) \frac{q_{n,\alpha,i}}{\sqrt{M_n}}\mcom
\end{align}
with $q_{n,\alpha,i}$ being the eigenvector of atom $n$ along phonon mode 
$\alpha$, and $M_n$ is the mass of atom $n$.

\begin{figure}
	\begin{center}
		\includegraphics[width=\columnwidth]{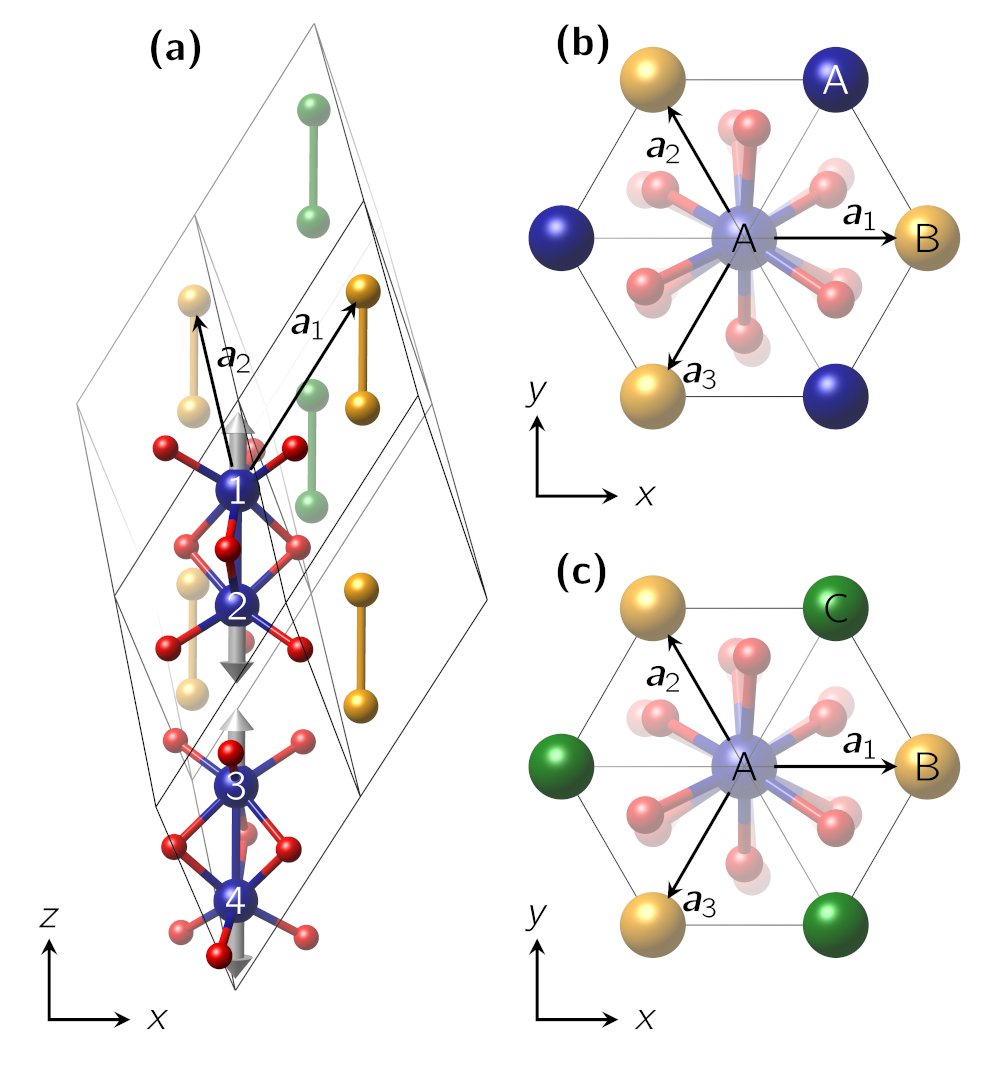}
	\end{center}
	\caption{Rhombohedral lattice structure of Cr$_2$O$_3$: (a) Unit cell with Cr
	(blue) and oxygen (red) sites numbered from 1 to 4 (from top to bottom). Yellow
	and green balls indicate Cr sites in the neighboring unit cells, considered in
	our multiple cell model below. The gray arrows show the magnetic moments at the
	Cr sites. (b) Top view of the structure with two unit cells as basis (named A
	and B). Note that only the uppermost Cr site can be shown. New lattice vectors
	can be expressed via the original single-unit-cell lattice vectors (shown as
	$\vec{a}_1$, $\vec{a}_2$, and $\vec{a}_3$): $\vec{a}_1+\vec{a}_2$\,,
	$\vec{a}_2+\vec{a}_3$\,, and $\vec{a}_1+\vec{a}_3$\,. (c) Top view of the
	structure with three unit cells as basis (named A, B, and C). They relate to
	the unit cells shown in (a). New lattice vectors for three unit cells are:
	$\vec{a}_1-\vec{a}_2$\,, $\vec{a}_1-\vec{a}_3$\,, and
	$\vec{a}_1+\vec{a}_2+\vec{a}_3$\,. The structure pictures are created with
	VESTA \cite{Momma2011jac}. }
	\label{fig:structure and spin}
\end{figure}

The displacement $\xr(t)$ 
of the Raman-active mode is the one which determines the 
interaction parameters $J_{ij}(t)$ in case of \CrO \cite{Fechner}.
Here, a THz-pulse excites an IR-active mode with amplitude $\xir$, where the
Cr and O atoms move in opposite directions,
see Fig.~\ref{fig:structure and spin} for a depiction of the lattice structure
of \CrO. The relative distance between Cr atoms hardly changes in the IR mode, leaving
the interaction parameters $J_{ij}$ for the magnetic coupling between Cr atoms essentially unchanged. For a
sufficiently strong pulse and a correspondingly large normal mode amplitude $\xir$,
the IR mode excites the optically inactive R mode via nonlinear coupling, described by the third term in Eq.~(\ref{eq:nonlinear_phononics:Cr2O3}). The R mode with large normal mode amplitude $\xr$,
changes the distances between Cr atoms appreciably, which in turn changes
the interaction parameters among Cr atoms. We follow the notation of \cite{Fechner}
and write $J_{ij}(t)=J_{ij}\big(\xr(t)\big)$ which shows that the time dependence of
$J_{ij}(t)$ comes from their dependence on $\xr(t)$. More details and calculations
for the case of \CrO can be found in \cite{Fechner}.

All these steps discussed above can be in general followed for any interesting 
material:
(a) The relevant phonon modes 
in Eq.~\eqref{eq:nonlinear_phononics:potential_expansion_Juraschek2018} have to 
be identified.
(b) Their coupling parameters have to be obtained.
(c) Eq.~\eqref{eq:nonlinear_phononics:oscillator_model_Jaruschek2018} has to be 
solved for all involved modes including a potential external driving.
(d) The most important interaction parameters for 
Eq.~\eqref{eq:Hamiltonian_general} are needed together with 
(e) the relation $J_{ij}\big(\xr(t)\big)$  for each $J_{ij}$.
Those steps involve tremendous numerical efforts for any new material, which 
explains also the limited number of different materials studied so far. 
Hence, we refrain in this work from studying new materials
but demonstrate how those \textit{ab initio} results could be 
used to allow a quantum heat engine with a realistic working 
body.

It turns out that the prototypical magnetoelectric material \CrO is a 
perfect candidate for quantum heat engine working body. 
Previous theoretical studies using the Vienna \textit{ab initio} simulation
package (VASP) showed that DFT calculations
can reproduce experimentally observable properties \cite{Shi2009_PhysRevB.79.104404, Mostovoy2010_PhysRevLett.105.087202},
whereas Fechner \etal 
\cite{Fechner} obtained in their study all necessary ingredients discussed 
above. 
\CrO crystallizes in the rhombohedrical
lattice structure with space group 167 (R$\overline{3}$c)
\cite{Sawada1994,Fechner}, while the lattice constant or 
angle play for our study only a minor role
(structure and lattices vectors $\vec{a}_1$ and
$\vec{a}_2$ are depicted in Fig.~\ref{fig:structure and spin}a).
Its magnetic ground state consists of antiferromagnetically oriented Cr 
ions (see Fig.~\ref{fig:structure and spin}a)). 

\begin{figure}[htp]
	\includegraphics[width=\columnwidth]{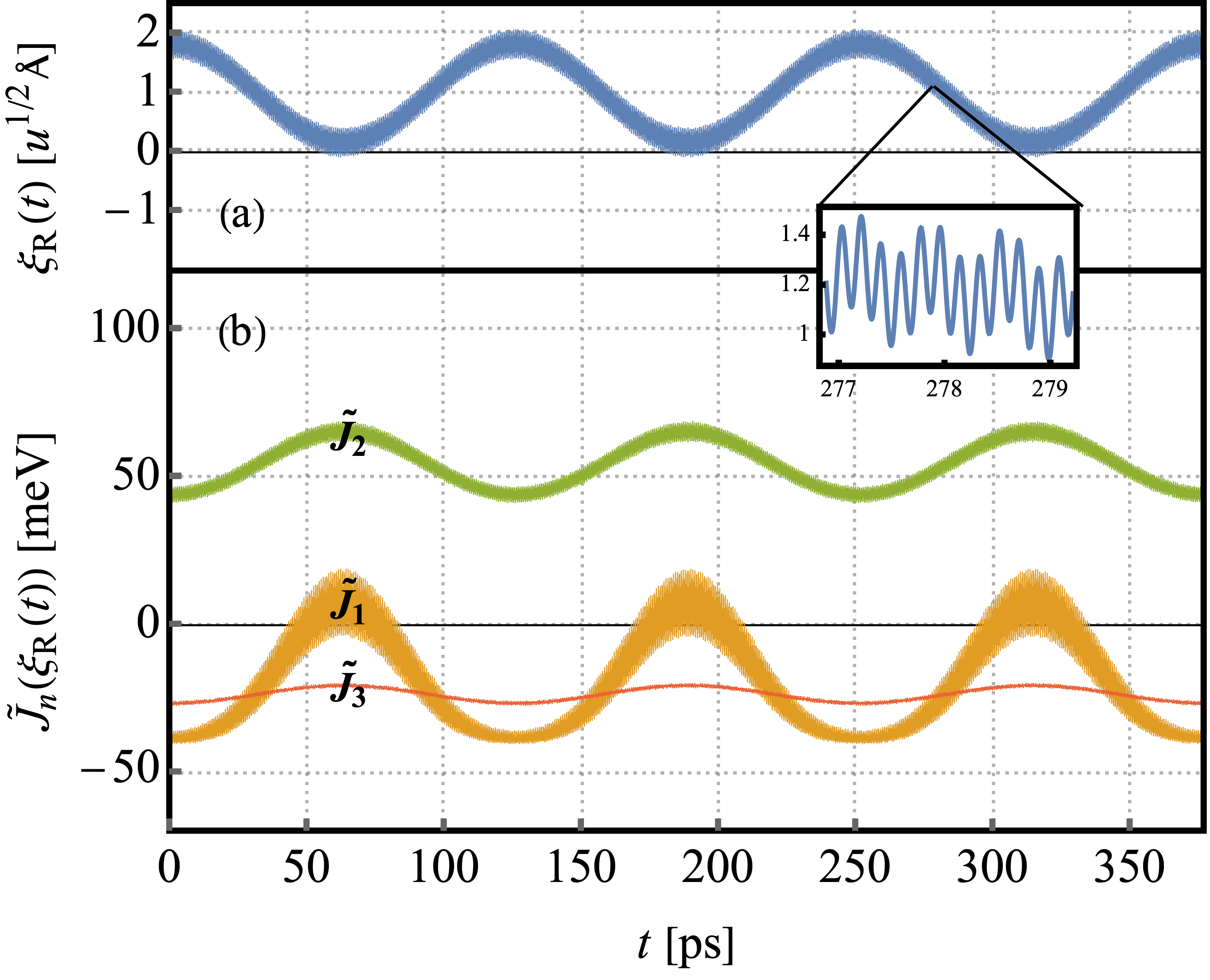}
	\caption{Time-dependence of (a) $\xr(t)$ and (b) $\tilde{J}_n(\xr(t))$ over three periods. Apart from the low-frequency
	    oscillations, there are two additional high-frequency terms yielding extremely fast oscillations of lower
	    amplitude (see inset).}
	\label{fgr:xr_Jn_dependence}
\end{figure}

The lattice displacements $\xr(t)$ due 
to the nonlinear phonon coupling are obtained from solving the 
system of equations described by 
Eq.~\eqref{eq:nonlinear_phononics:oscillator_model_Jaruschek2018}
with the correct potential \eqref{eq:nonlinear_phononics:Cr2O3}.
In this work, we slightly simplify the full lattice distortion
derived in \cite{Fechner} (see appendix) and restrict ourselves to a
dependence on the three most important frequencies
\begin{align}
\xr(t) =  \xi_{\text{R}0} + C_\text{R} &\cos(\tilde{\omega}_\text{R}t)
+ C_\text{IR} \cos(2 \tilde{\omega}_\text{IR}t)\nonumber\\
&+ C_{\varOmega_-} \cos[(\varOmega-\tilde{\omega}_\text{IR}) t]\,,
\label{eq:xr_definition}
\end{align}
where $\tilde{\omega}_\text{IR}$ and $\tilde{\omega}_\text{R}$ are the
renormalized infrared and Raman frequencies, respectively
\cite{Fechner}, and $\varOmega$ is the frequency of the THz pulse
in Eq.~\eqref{eq:nonlinear_phononics:driving}.
It results in a slow sinusoidal variation overlayed with a 
quick oscillation (see Fig.~\ref{fgr:xr_Jn_dependence}a
for our choice of parameters).

On the other hand, the interaction parameters
in \CrO can be calculated by comparing total energies of 
different magnetic configurations and fitting them 
to the Heisenberg Hamiltonian in Eq.~\eqref{eq:Hamiltonian_general}.
The resulting Hamiltonian of the single-unit-cell magnetoelectric chromium oxide Cr$_{2}$O$_{3}$ derived in \cite{Fechner} has the form
\begin{align}
    \hat{H}&=\tilde J_{1}\big(\xr(t)\big)\big(\hat{\vec{S}}_{1}\hat{\vec{S}}_{2}+\hat{\vec{S}}_{3}\hat{\vec{S}}_{4}\big)\nonumber\\
    &\qquad +\tilde J_{2}\big(\xr(t)\big)\big(\hat{\vec{S}}_{1}\hat{\vec{S}}_{4}+\hat{\vec{S}}_{2}\hat{\vec{S}}_{3}\big)\nonumber\\
    &\qquad +\tilde J_{3}\big(\xr(t)\big)\big(\hat{\vec{S}}_{1}\hat{\vec{S}}_{3}+\hat{\vec{S}}_{2}\hat{\vec{S}}_{4}\big)\nonumber\\
    &\qquad+\sum_{i=1}^{4}\bigg[ \mu B_z \hat{S}^z_i + D \big( \hat{S}^z_i \big)^2 \bigg] \,.
    \label{eq:Hamiltonian}
\end{align}
Here, the magnetocrystalline anisotropy constant is equal to
$D=\SI{-27}{\micro\eV}$ and $\hat{\vec{S}}_i$ are the spin operators of the four
chromium atoms with ${S=3/2}$. The $\tilde J_{n}\big(\xr(t)\big)$ are
effective interaction parameters between the spins of Cr
atoms, and depend on the R mode distortion $\xr(t)$, excited by
\si{\THz} pulses.  The three $\tilde J_{n}\big(\xr(t)\big)$ are
obtained from the five bare interaction parameters
$J_{n}\big(\xr(t)\big)$ for the nearest-neighbor to fifth-nearest-neighbor Cr atoms in the
Cr$_{2}$O$_{3}$ lattice by identifying the Cr atoms of other unit
cells with Cr atoms of the reference unit cell. The resulting mapping
is \cite{Fechner}
\begin{align}
\tilde{J}_1(\xr(t)) &= J_1(\xr(t)) + 3J_3(\xr(t)), \nonumber\\
\tilde{J}_2(\xr(t)) &= 3J_2(\xr(t)) + J_5(\xr(t)),\nonumber\\
\tilde{J}_3(\xr(t)) &= 6J_4(\xr(t)).
\end{align}

We assume that the interaction parameters are quadratic functions of $\xr(t)$ 
\cite{Fechner},
\begin{equation}
    J_n(\xr(t)) = J_n(0)+\xr\left.\pdv{J_n}{\xr}\right|_{\xr=0} + \frac{1}{2} \xr^2\left.\pdv[2]{J_n}{\xr}\right|_{\xr=0}\,.
    \label{eq:Jn_definition}
\end{equation}
For strong, experimentally feasible THz pulses with appropriate frequencies, 
the temporal variation of R modes $\xr(t)$ can lead to significant temporal 
variation of the interaction parameters -- even inverting the sign of
$\tilde J_{1}\big(\xr(t)\big)$ (see Fig.~\ref{fgr:xr_Jn_dependence}b).
The $\tilde{J}_n(\xr(t))$ exhibit slow
oscillations if $\varOmega\approx\tilde{\omega}_\text{IR}$ due to the
last term in Eq.~\eqref{eq:xr_definition}. One such oscillation of
period $126\,$ps constitutes the quantum Otto cycle described in
section~\textbf{\ref{sec:otto}}.  The other two terms in
Eq.~(\ref{eq:xr_definition}) lead to very fast oscillations with
smaller amplitudes than the low-frequency oscillation (see inset in
Fig.~\ref{fgr:xr_Jn_dependence}). For
more details concerning the interaction parameters $J_n\big(\xr(t)\big)$,
we refer to \cite{Fechner}, while details concerning the
simplifications as well as numerical values for the parameters in
Eq.~\eqref{eq:Jn_definition} and Eq.~\eqref{eq:xr_definition} are
given in the appendix.

The dependence of $J(\xr(t))$ as in Eq.~\eqref{eq:Jn_definition} is a valid 
approximation and the mechanism of steering the interaction parameters through a selectively activated Raman mode $\xr(t)$ might be a universal protocol of 
dynamical materials design, which could also be applied to other materials as 
well. Therefore, we assume this parametrical dependence of the interaction parameters in the generic model Hamiltonians in 
section~\textbf{\ref{sec:alternative}}.

\section{A quantum Otto cycle}\label{sec:otto}

We are suggesting a quantum heat engine based on dynamically driven
interaction parameters $J(\xr(t))$ in Eq.~\eqref{eq:Hamiltonian}. Here, we provide the 
definitions and theoretical background for the necessary 
thermodynamic properties in order to characterize our engine with the \CrO working body.

\subsection{Definition of the cycle}
We consider a quantum Otto cycle consisting of two quantum isochoric and two quantum adiabatic strokes. Quantum isochoric means that heat is exchanged between the working substance and heat baths, which reshuffles the level populations of the system, while the Hamiltonian is time independent during this period. Quantum adiabatic means that the level populations remain constant throughout the whole driving process, in which the system performs work because of a change of the interaction parameters $J_n$, while being detached from the heat baths. As a consequence the system does not remain in thermal equilibrium during the driving, as is usually assumed for an adiabatic stroke in classical thermodynamics.

The four strokes of our quantum cycle are thus defined as follows:
\begin{itemize}
	\setlength\itemsep{-.3em}
	\item isochoric heating: A heat bath at temperature $T_{\text{H}}$ heats the working substance while the $J_n$ are fixed.
	\item adiabatic driving: The system is driven by variation of the interaction parameters $J_n(t)$.
	\item isochoric cooling: The working substance is cooled by a heat bath at temperature $T_{\text{L}}$ while the $J_n$ are fixed.
	\item adiabatic driving: The system is driven back to the initial values of the interaction parameters.
\end{itemize}

\begin{figure}[bp]
    \centering
    \includegraphics[width=1\columnwidth]{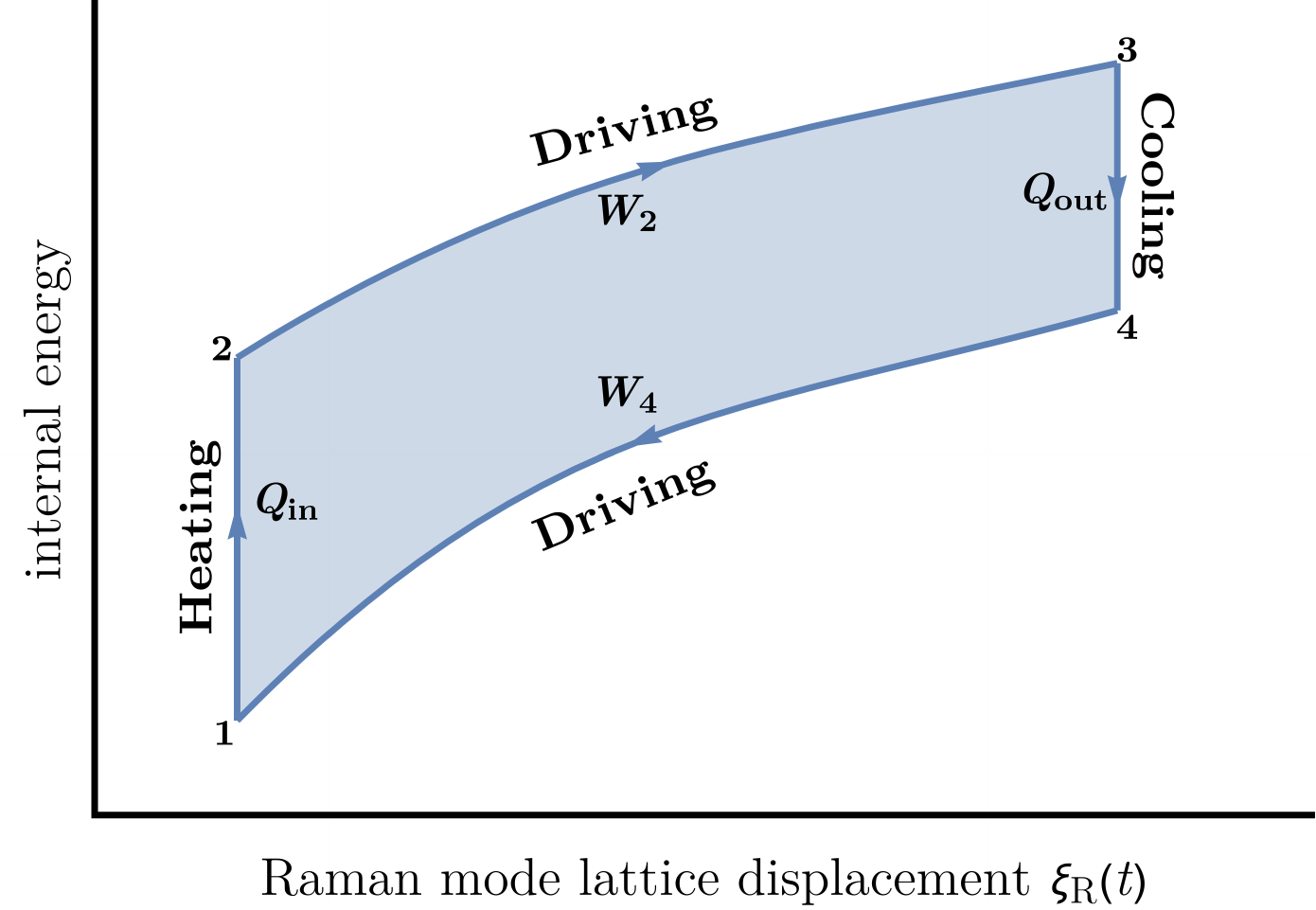}
    \caption{Schematics of the Otto cycle. Heating (1$\rightarrow$2) and cooling (3$\rightarrow$4) are isochoric processes, the work strokes $W_2$ (2$\rightarrow$3) and $W_4$ (4$\rightarrow$1) are quantum adiabatic.}
    \label{fgr:thermo_cycle}
\end{figure}

As for the \CrO~working body, we assume phonon-like baths, i.e., the baths consist of harmonic oscillators, which are not affected by the heat exchange with the system. The driving in terms of a variation of the $\tilde{J}_n\big(\xr(t)\big)$ is achieved via applied \si{\THz} pulses that change the Raman mode distortion between $\xr^{\text{(H)}}$ and $\xr^{\text{(L)}}$. The schematics of such a cycle is shown in Fig.~\ref{fgr:thermo_cycle}. We note that the separation of cycles is based on the assumption that the $\tilde{J}_n\big(\xr(t)\big)$, i.e., the energy spectrum of the system, is solely driven by IR and R modes and that the thermal bath has no impact on $\tilde{J}_n\big(\xr(t)\big)$. This is based on the following reasoning: When the frequencies of the anharmonic IR and R modes are different from the thermal phonon excitations, the mismatch between frequencies only allows a slow energy exchange between IR and R modes and thermal phonons.

At nonzero temperature the internal energy of the system can be evaluated as $U=\text{Tr}\big(\hat{\varrho}\hat{H}\big)$ where $\hat{\varrho}$ is the density matrix. For the change in the internal energy we deduce \cite{Quan2007}:
$\dd{U}=\sum_{n=1}^{N}\big(E_{n}\underbar{d}{\varrho}_{nn}+\varrho_{nn}\underbar{d}{E}_{n}\big)$, where $\underbar{d}$ denotes an inexact differential.
The first term ${\underbar{d} Q= \sum_{n=1}^N E_{n}\underbar{d}\varrho_{nn}}$  quantifies the heat exchange in terms of reshuffled level populations $\underbar{d}{\varrho}_{nn}\big(T)$ due to an infinitesimal change of the temperature $T$ for fixed eigenenergies $E_{n}$\,. The second term ${\underbar{d} W=\sum_{n=1}^N\varrho_{nn}\underbar{d}{E}_{n}}$ corresponds to the produced work due to the change of the energy spectrum $\underbar{d}{E}_{n}$ of the system.

\subsection{Quantum work}
\label{sec:quantum_work}
While the notion of heat and heat exchange between the system and the heat bath is precisely defined, the concept of quantum work is nontrivial as quantum work is not an observable. Rather, the exponential average of the quantum work is given  by time-ordered correlation functions of the exponentiated Hamiltonian and cannot be expressed through the expectation values of an operator representing the work (see Talkner et al.~\cite{Talkner2007}).

The work produced in a time interval $[0,t]$ of an arbitrary driving process can be quantified as \cite{DelCampo2014}
\begin{align}
\label{Works}
\langle W(t)\rangle\!=\!\sum\limits_{n,m}\big[E_m(t)-E_n(0)\big]P_{n}(0)P_{mn}(t),
\end{align}
where $E_n(t)$ denotes the energy levels at time $t$ and $P_n(0)$ indicates the level population at the beginning of the stroke, which, for a system prepared in thermal equilibrium at temperature $T = 1/\beta$, reads 
${P_{n}^{\beta}(0) =
	\Ce^{-\beta E_n(0)}/\sum_{m}\Ce^{-\beta E_m(0)}}$.
Next,
\begin{align}
P_{mn}(t) =
\frac{1}{\lambda_{m}\lambda_{n}}
\sum_{q=1}^{\lambda_{n}}\sum_{k=1}^{\lambda_{m}}
\vert\mel{\Phi_{m}^{k}(t)}{\hat{U}(t,0)}{\Phi_{n}^{q}(0)}\vert^2
\end{align}
is the transition probability between the eigenstates of the Hamiltonian $\hat{H}(t)$. Here, $\Phi_m^k(t)$ are the eigenfunctions of $\hat{H}(t)$, where we assume that the $m^{\text{th}}$ level is $\lambda_m$-fold degenerate. $\hat{U}(t,0)$ is  the time-evolution operator associated to $\hat{H}(t)$.
For quantum adiabatic work strokes we have no transitions, i.e., ${P_{mn}(t)=\delta_{mn}}$ in Eq.~\eqref{Works}, which means that the populations of all states do not change throughout the whole driving process. 

For a process that is performed in finite time rather than quantum adiabatically (in the sense of infinitely slow) there will be a finite probability for inter-level transitions, because the Hamiltonian does not commute with itself for different $t$. Since the driving requires more energy due to these transitions, this behavior is referred to as quantum friction \cite{Allahverdyan2005,Plastina2014,Alecce2015,Cakmak2016}. However, it is possible to engineer a finite-time process that -- while being nonadiabatic itself -- drives the system along the exact adiabatic states of the original Hamiltonian, i.e., as if the system had been driven perfectly adiabatically under the original Hamiltonian. This approach is thus termed a \emph{shortcut to adiabaticity} \cite{Jarzynski2013,DelCampo2012,DelCampo2013}.
In this work we will compare the evolution of the system under $\hat{H}(t)$ to the evolution performed by counter-diabatic (CD) driving
\cite{Demirplak2003,Berry2009,DelCampo2014}, using the general formulation valid also for degenerate spectra \cite{XueKeSong}. This results in a modified, CD Hamiltonian $\hat{H}_{\text{CD}}(t)$ which reads
\begin{eqnarray}
\hat{H}_{\text{CD}}(t)=\hat{H}(t)+\hat{H}_{1}(t)\,,
\end{eqnarray}
where is $\hat{H}(t)$ the original system Hamiltonian, {Eq.~\eqref{eq:Hamiltonian}}, and the
second term
\begin{align}
\label{Shortcut}
\hat{H}_{1}(t) \!=
\ii \!\displaystyle
\sum_{m\neq n}\sum_{q=1}^{\lambda_{n}} \sum_{k=1}^{\lambda_{m}}
\frac{
	\ket{\varPhi_{m}^{k}}\!\!
	\mel{\varPhi_{m}^{k}}{\partial_{t}\hat{H}(t)}{\varPhi_{n}^{q}}\!
	\bra{\varPhi_{n}^{q}}
}{E_n - E_m}
\end{align}
compensates the nonadiabatic effects of quantum friction.

In order to make CD driving work the Hamiltonians $\hat{H}(t)$ and $\hat{H}_\text{CD}(t)$ need to coincide both at the beginning and at the end of the stroke, i.e., ${\hat{H}_1(0)=\hat{H}_1(\tau)=0}$\,. In our model for \CrO this is achieved by slightly changing the temporal dependence of the atomic displacement (see appendix).
Note that for our system the experimental realization of the CD term $ \hat{H}_{1}(t)$ appears to be extremely difficult. However, our numerical results below show that even without the CD term the quantum friction is rather small.

The adiabaticity of the cycle can be assessed by investigating the non-equilibrium work fluctuations $\delta W(t)$,
\begin{equation}
\delta W(t)=\langle W(t)\rangle-\langle W_{\text{ad}}(t)\rangle,
\label{eq:non_equilibrium_work_for_calculation}
\end{equation}
where $W_{\text{ad}}$ is the work done by a quantum-adiabatic process. While for a perfectly quantum adiabatic process we have $\delta W(t)=0$ for all $t$, the use of shortcuts to adiabaticity yields $\delta W(\tau)=0$,i.e., only for $t=\tau$, which means that the final system state is as if it had been driven perfectly quantum-adiabatically.  We can therefore employ $\delta W(\tau)$ to quantify the quantum friction for a given stroke, as the latter is usually defined as \cite{Cakmak2016}
\begin{equation}
\ev*{W_\text{fric}} = \ev*{W}-\ev*{W_{\tau\rightarrow\infty}},
\end{equation} 
where $\ev*{W_{\tau\rightarrow\infty}}$ denotes the work performed in an infinitely slow process. In our case this is just the adiabatically performed work $\ev*{W_\text{ad}(\tau)}$.

\subsection{Efficency and output power}

There are two quantities to be investigated in order to assess the performance of a quantum heat engine: its efficiency and its output power.

The efficiency of a cycle is -- similarly to the classical case -- defined as
\begin{equation}
\eta=\frac{Q_{\text{in}}-Q_{\text{out}}}{Q_{\text{in}}} = -\frac{\langle W_{2}\rangle+\langle W_{4}\rangle}{Q_{\text{in}}}.
\label{eq:efficiency}
\end{equation}
Here, $\langle W_{2}\rangle$ and $\langle W_{4}\rangle$ denote the work produced during the respective work strokes.

As the implementation of CD driving also requires some energetic costs, these also need to be taken into account in the efficiency calculation \cite{Abah2017,Abah2018,Abah2019,Cakmak2019}, leading to the corrected definition
\begin{equation}
\eta_{\text{CD}} = -\frac{\langle W_{2}\rangle+\langle W_{4}\rangle}{Q_{\text{in}}+\langle \hat{H}_{1}^{(2)}\rangle_{\tau}+\langle \hat{H}_{1}^{(4)}\rangle_{\tau}}.
\label{eq:correctedefficiency}
\end{equation}
Here 
\begin{align}
\ev*{\hat{H}_{1}}_{\tau} &=
\frac{1}{\tau}\!\int\limits_0^\tau\!\!\!\ev*{\hat{H}_{\text{CD}}(t)}\!-\!\ev*{\hat{H}(t)}\text{d}t = \frac{1}{\tau}\int\limits_0^\tau\!\!\delta W(t)\text{d}t 
\label{eq:total_STA_cost}
\end{align}
is the time average of the CD driving term, which can also be written in terms of the non-equilibrium work fluctuations. The superscript in Eq.~\eqref{eq:correctedefficiency} indicates the work strokes, i.e., $i=2,4$.

The output power of the quantum Otto cycle can be computed as
\begin{eqnarray}
\mathcal{P}=
-\frac{\langle W_{2}\rangle+\langle W_{4}\rangle}{\tau_1+\tau_2+\tau_3+\tau_4}\,.
\label{outputpower}
\end{eqnarray}
Here $\tau_1$, $\tau_3$ are the relaxation times of the working substance in contact with the hot and cold phonon baths, respectively, and $\tau_2=\tau_4$ are the durations of the adiabatic strokes.

As for the efficiency the output power of the engine is corrected due to additional energetic costs of CD driving, yielding
\begin{eqnarray}
\mathcal{P}_\text{CD}=
-\frac{\langle W_{2}\rangle+\langle W_{4}\rangle-\langle \hat{H}_{1}^{(2)}\rangle_{\tau}-\langle \hat{H}_{1}^{(4)}\rangle_{\tau}}{\tau_1+\tau_2+\tau_3+\tau_4}\,.
\label{eq:correctedoutputpower}
\end{eqnarray}

\subsection{Thermalization of the working body}
\label{sec:thermalization}

For heating and cooling of the spin system, we need a heat bath which we assume is provided by the thermal phonons of the crystal.
The thermal coupling between the phonons and the spins is approximated.
We describe the energy exchange between the bath and the spin system through the coupling of phonons with the spin components $\hat S_j^{x}$\,. The total Hamiltonian during this stroke comprises the Hamiltonian of the spin system, the Hamiltonian of the phonon bath $\hat{H}_{\text{bath}}$\,, and the system-bath interaction $\hat{H}_{\text{int}}$:
\begin{align}
\label{interactions}
\hat{H}_{\text{tot}}  & =
\hat{H}+\hat{H}_{\text{int}}+\hat{H}_{\text{bath}} \,,\nonumber\\
\hat{H}_{\text{bath}} & =\int \dd{k} \omega_k\,
\hat{b}^{\dag}_{k}\hat{b}_{k} \,,\nonumber\\
\hat{H}_{\text{int}}  & = \displaystyle\sum_{j=1}^4 \hat{S}^{x}_{j}
\int \dd{k} g_k(\hat{b}^{\dag}_{k}+\hat{b}_{k}) \,.
\end{align}
Here $\hat{b}^{\dag}_{k}$ and $\hat{b}_{k}$ are the phonon creation and annihilation operators, respectively, and $g_k$ is the coupling constant between spins and phonons. A straightforward derivation leads to a master equation for the density matrix \cite{Breuer2002}

\begin{align}
\label{master equation}
\frac{\dd \rho_{\text{S}}(t)}{\dd t}
& = -\ii\comm{\hat{H}}{\rho_\text{S}(t)} + \sum_{\omega,j} \gamma(\omega)\\
&\times
\left(\hat{S}_j^x(\omega)\rho_S(t)\hat{S}_j^x(\omega)-\hat{S}_j^x(\omega)\hat{S}_j^x(\omega)\rho_S(t)\right) + \text{h.c.}, \nonumber
\end{align}
where
\begin{eqnarray}
\label{bathcorrelation}
&&\gamma(\omega)=g_k\begin{cases} \frac{1}{\exp[-\beta\omega]-1}\,, &  \omega< 0 \,, \\
\frac{1}{\exp[\beta\omega]-1}+1\,, & \omega>0 \,.  \end{cases}
\end{eqnarray}
and 
\begin{equation}
\hat{S}^{x}_{j}(\omega) = \!\!
\sum_{\omega=E_m-E_k}\!\!
\dyad{\Phi_k}{\Phi_k}\hat{S}^{x}_{j}\dyad{\Phi_m}{\Phi_m} \,.
\end{equation}

Since $\gamma(\omega)$ is positive and hermitian, the master equation \eqref{master equation} could also be brought into Lindblad form.

\section{Performance of the Chromium Oxide Heat Engine}
\label{sec:results}

We now investigate the properties of a quantum heat engine based on \CrO, using the already available data from \cite{Fechner} as a basis for the time-dependent displacements $\xr$ and for the resulting values for the $J_n(\xr(t))$. The exact numerical values are presented in the appendix for reference. In our simulations, we use a dimensionless coupling constant between phonons and spins $g_k=0.1$ and apply a magnetic field in the $z$ direction. The strength of the magnetic field is set to \SI{0.45}{\tesla}.

\subsection{Spectrum of the Hamiltonian}

\begin{figure}[htp]
	\includegraphics[width=\columnwidth]{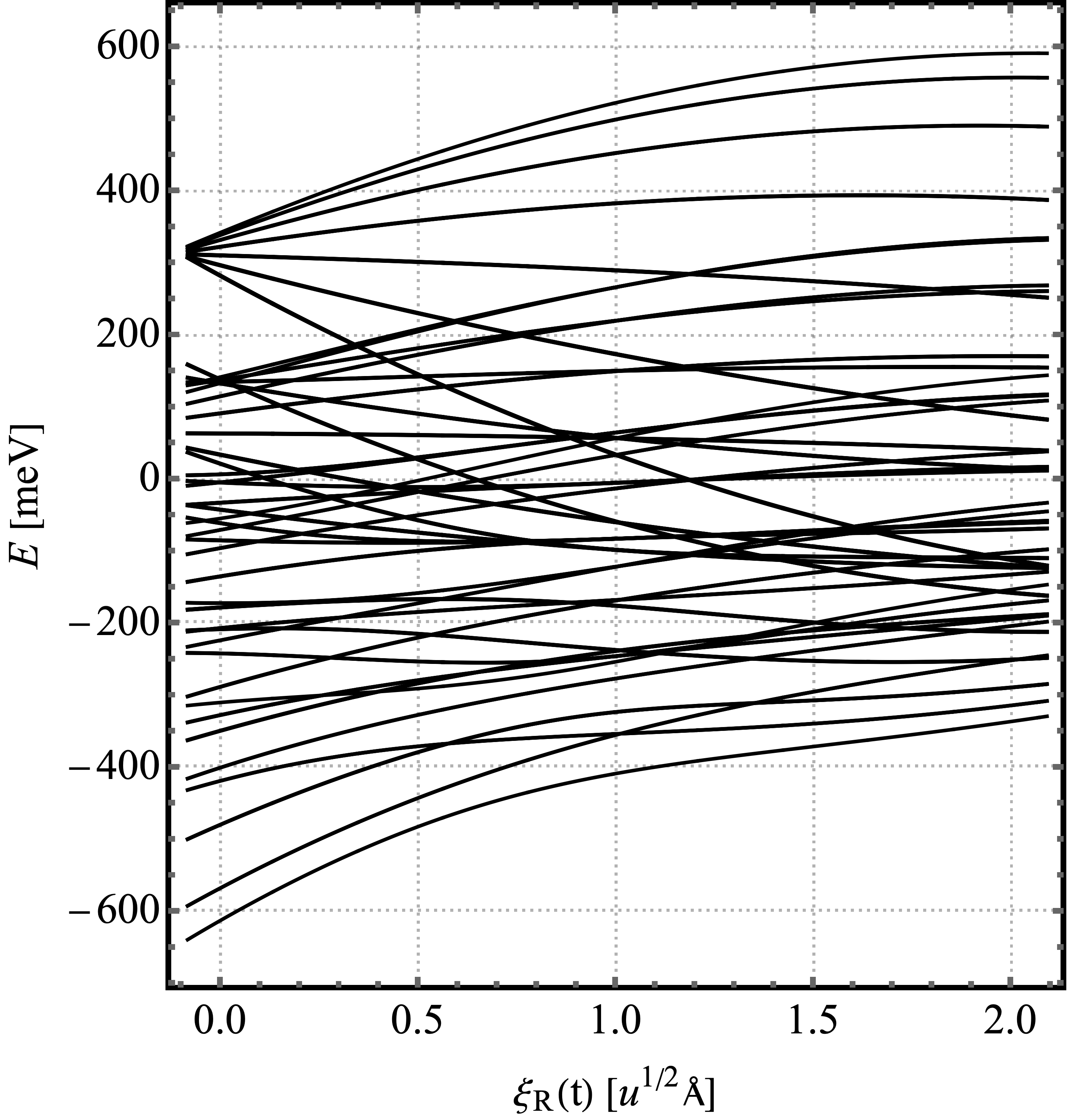}
	\caption{Spectrum of the Hamiltonian depending on the lattice displacement $\xr$. Some of the levels are very close together such that not each level can be distinguished in this figure. The heating stroke happens at  $\xr=0.4269$, followed by a work stroke from $\xr(\Cpi/0.05)=0.4269$ to $\xr(2\Cpi/0.05)=2.0891$. At the latter value, the cooling is applied, after which the second work stroke takes place, bringing the lattice displacement back to $\xr = 0.4269$. Note that $\xr = 0.4269$ is not the minimal $\xr$-value attained throughout the stroke.
	}
	\label{fgr:spectrum}
\end{figure}

To gain some insight into our working substance, we first explore the instantaneous energy spectrum of the spin Hamiltonian, Eq.~\eqref{eq:Hamiltonian}, as a function of the lattice displacement $\xr$\,. For each value of $\xr$, we perform an exact diagonalization and use the fidelity 
\begin{equation}
F\big(\Phi_k(\xr+\Delta\xi,\Phi_l(\xr)\big) = \big\vert\braket{\Phi_k(\xr+\Delta\xi)}{\Phi_l(\xr)}\big\vert^2
\label{eq:fidelity}
\end{equation} 
to track the energy levels through (avoided) level crossings. This spectrum is shown in Fig.~\ref{fgr:spectrum}. We note again that the value of $\xr(t)$ is given by a superposition of different frequencies, leading to a rather complicated behavior of the spectrum over time.

Obviously, the spectrum in Fig.~\ref{fgr:spectrum} is a lot more complicated than the spectrum of other models usually employed in the literature for quantum heat engines so far \cite{Abah2012,DelCampo2013,DelCampo2014,Rezek}. In particular, most of previous studies use self-similar spectra, which means that the spectrum is only scaled throughout the driving process like in the case of a harmonic oscillator, which in turn ensures that the system is always in a thermal equilibrium state. This is not the case in our system, as we note that while the ground state stays the same for the whole driving there are level crossings for most of the excited states. Moreover, there are many avoided crossings, with splittings too small to be visible in Fig.~\ref{fgr:spectrum}. One needs thus to be careful when considering the adiabatic driving since all (avoided) crossings must be tracked carefully. As a consequence of the stated spectrum properties, the system will not arrive in a thermal equilibrium state at the end of a work stroke.

In order to understand which energy levels play a role in the description of the system, we also consider the level populations in Gibbs equilibrium for the Hamiltonian (Eq.~\eqref{eq:Hamiltonian}) at $\xr = 2.0891$, which is the value during the cooling stroke of the system. It turns out that at \SI{50}{\kelvin} basically only the ground state is occupied, while at \SI{300}{\kelvin} 12 levels and at \SI{1000}{\kelvin} 65 levels have a diagonal element in the density matrix larger than $10^{-3}$. The latter result indicates that the behavior of the system as a heat engine can only be explained when taking into account a considerable part of the spectrum, which is in turn responsible for the rich features due to non-self-similarity of the spectrum and (avoided) level crossings.

\subsection{Non-equilibrium work fluctuations}

We employ the quantum adiabatic, the nonadiabatic, and the counter-diabatic driving scheme for both work strokes after complete thermalization to temperatures between 50$\,$K and 1500$\,$K and use the obtained data to compute the non-equilibrium work fluctuations $\delta W(\tau)$ (Eq.~\eqref{eq:non_equilibrium_work_for_calculation}) at the end of the strokes. For the nonadiabatic driving scheme we use a Runge-Kutta method to solve the von-Neumann equation (i.e.~$\dot{\rho}(t) = -\frac{\ii}{\hbar}\comm*{\hat{H}(t)}{\rho(t)}$), while we construct the density matrices for adiabatic and CD driving from the eigenvectors obtained from exact diagonalization.

\begin{figure}[!htb]
	\includegraphics[width=\columnwidth]{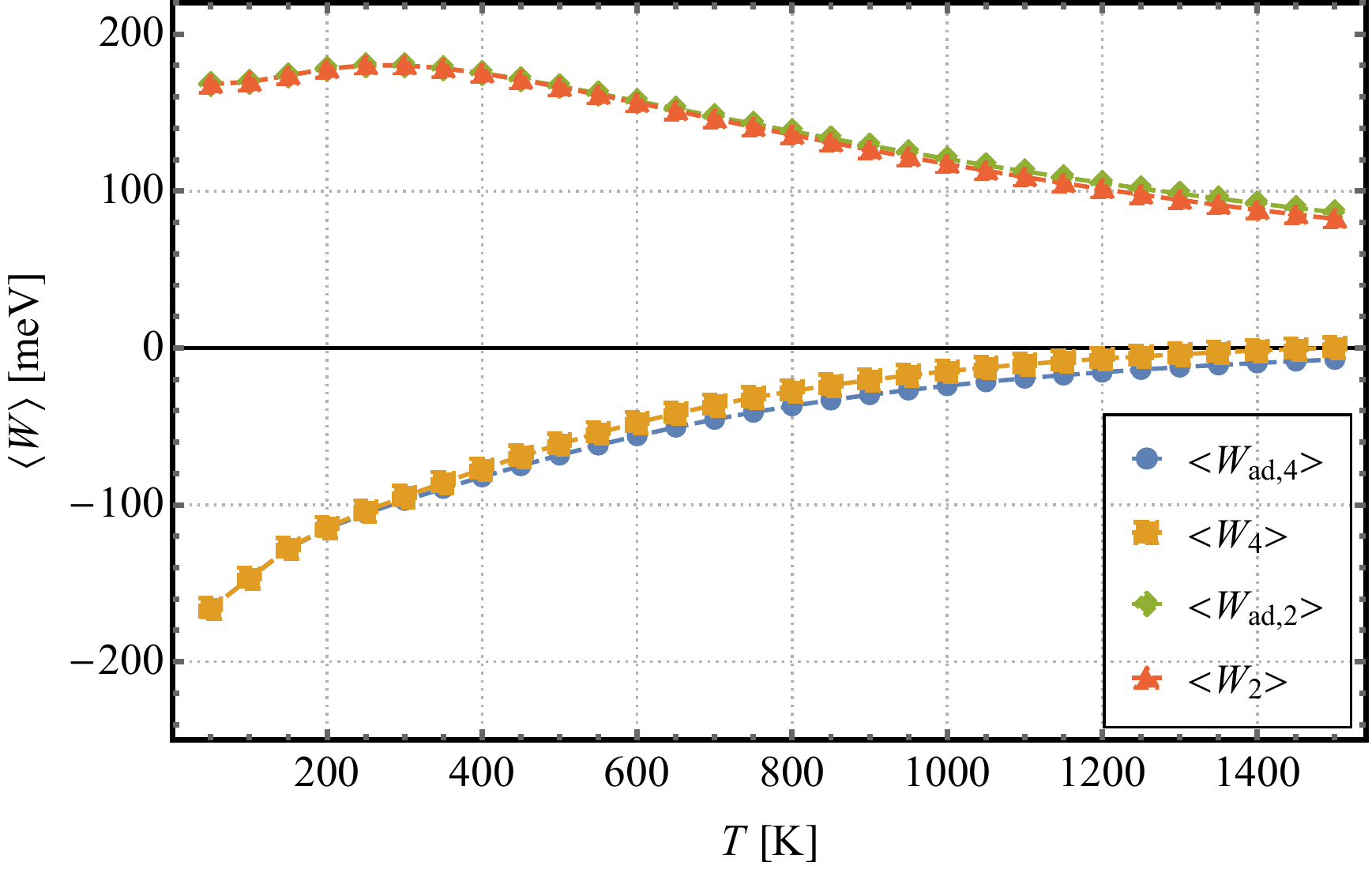}
	\caption{Total performed work (Eq.~\eqref{Works}) for nonadiabatic and CD-driving schemes as a function of the temperature at the beginning of the stroke. The difference between the two schemes is always very small compared to the absolute value.
	}
	\label{fgr:total_work}
\end{figure}

\begin{figure}[!htb]
	\includegraphics[width=\columnwidth]{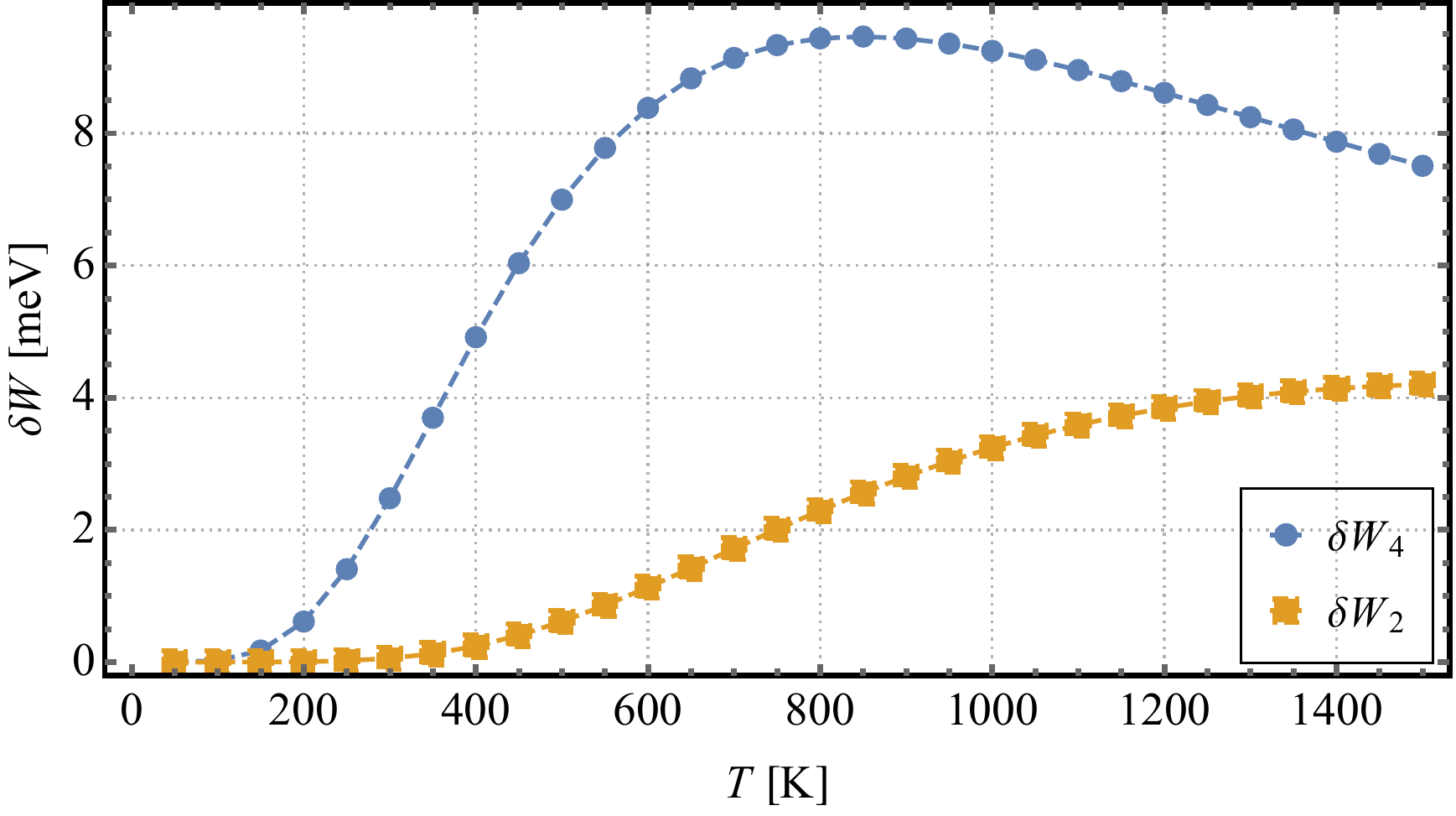}
	\caption{Non-equilibrium work fluctuations (Eq.~\eqref{eq:non_equilibrium_work_for_calculation}) as a function of the temperature at the beginning of the stroke.
	}
	\label{fgr:deltaW_at_end}
\end{figure}

The total performed work (Eq.~\eqref{Works}) using the nonadiabatic and the CD driving schemes for both work strokes and different temperatures is shown in Fig.~\ref{fgr:total_work}. It is worth noting that the difference between the two schemes is always very small compared to the absolute values. Since the work done during the CD-driven strokes actually coincides with $\ev*{W_\text{ad}}$, this difference just constitutes $\delta W(\tau)$ (see Eq.~\eqref{eq:non_equilibrium_work_for_calculation}). We therefore consider this quantity in detail in Fig.~\ref{fgr:deltaW_at_end}. We observe that the values are extremely small ($< \!1\,$meV) after cooling to a temperature below 200$\,$K, which is going to be the relevant temperature range (see next subsection). On the other hand, $\delta W(\tau)$ increases with temperature for strokes after heating, but the largest values are in the range of 4$\,$meV, which is -- as mentioned -- far below the total performed work.

\begin{figure}[!htb]
	\includegraphics[width=\columnwidth]{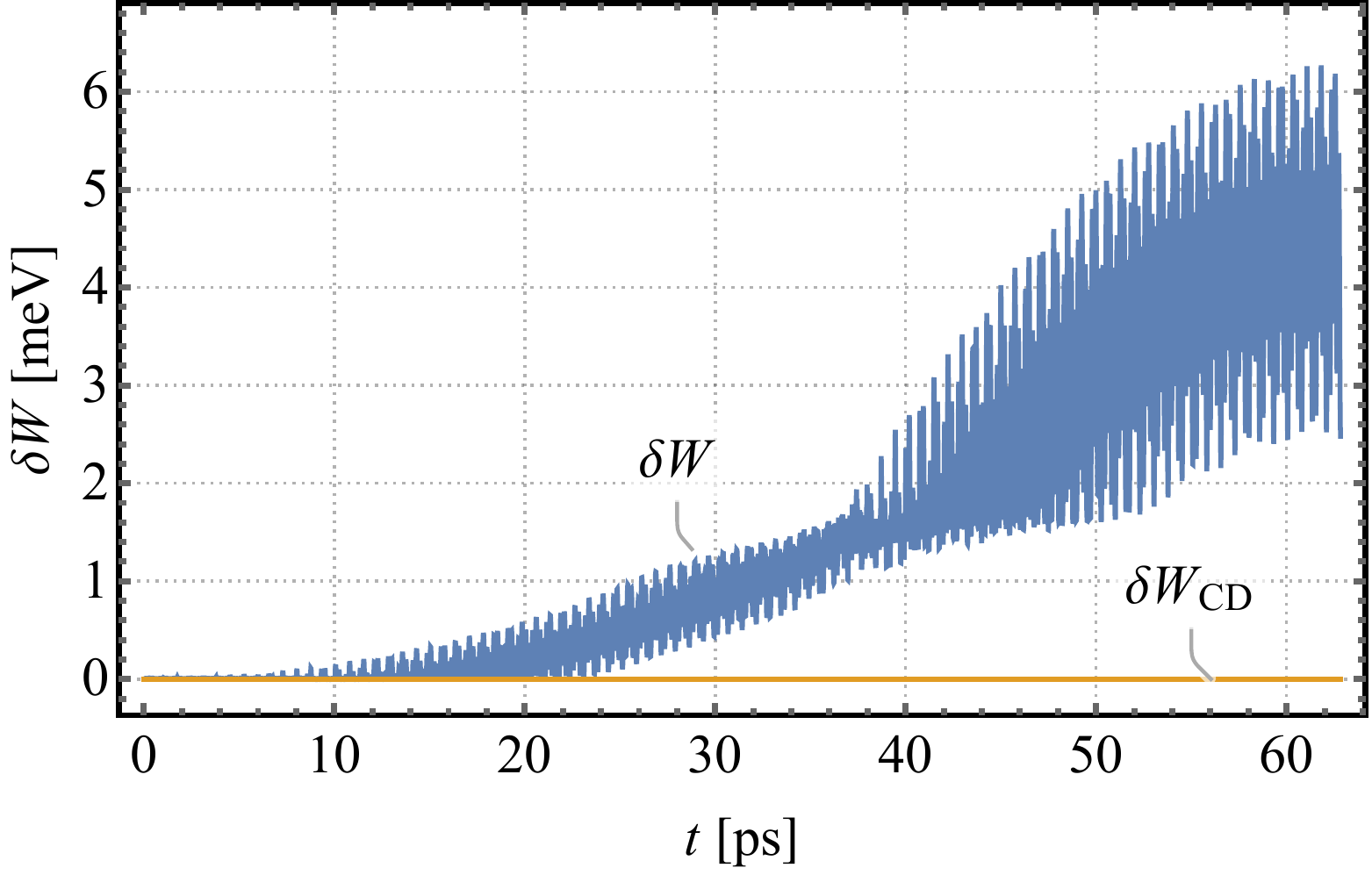}
	\caption{Non-equilibrium work fluctuations for the stroke after cooling to $T=300\,$K in the presence and absence of the CD driving term calculated for the single unit cell. CD driving yields a vanishing $\delta W_{\text{CD}}$ throughout the stroke.}
	\label{fgr:non-equilibrium work fluctuation}
\end{figure}

To shed more light onto the dynamics of a single stroke, we consider the time dependence of $\delta W(t)$ during the stroke after cooling to 300$\,$K for both driving schemes, which is shown in Fig.~\ref{fgr:non-equilibrium work fluctuation}. The non-equilibrium work fluctuations for nonadiabatic driving arise only after \SIrange{20}{25}{\pico\second} and continue to grow until the end of the stroke, where the fast oscillations due to the high-frequency terms in $\xr(t)$ (see Fig.~\ref{fgr:xr_Jn_dependence}) are clearly visible. On the other hand, there are no fluctuations at all throughout the CD driving scheme despite the high frequencies in the driving. This is a remarkable result since we would expect nonzero $\delta W(t)$ and only $\delta W(\tau)=0$ in general. To better understand the situation we will further discuss this finding later in section~\textbf{\ref{sec:alternative}} on the basis of the generic model.

Summarizing, while the chromium oxide heat engine is friction-free for CD-driven strokes by construction, we would like to emphasize again that even without CD driving the non-equilibrium work fluctuations are much smaller than the total work performed during the stroke (3-4 meV compared to $\approx$100 meV), despite the fast oscillations of the interaction parameters on top of the slow oscillation of the Otto cycle.  This is convenient since the experimental realization of CD driving is, if feasible at all, a complicated task.

\subsection{Efficiency}

Next, we turn our attention to the performance of our heat engine. To this end, we study one whole cycle, where we assume that the baths are coupled to the system until complete Gibbs equilibration. As the previous section has shown that nonadiabatic driving only leads to small non-equilibrium fluctuations, we consider both the nonadiabatic and the CD-driving scheme. Note that we do not need to simulate the thermalization processes for the Gibbs state. We calculate the efficiency $\eta$ as given in Eq.~\eqref{eq:efficiency} for both cases, as the small correction terms due to CD driving can be safely neglected.

\begin{figure}[!htb]
	\includegraphics[width=\columnwidth]{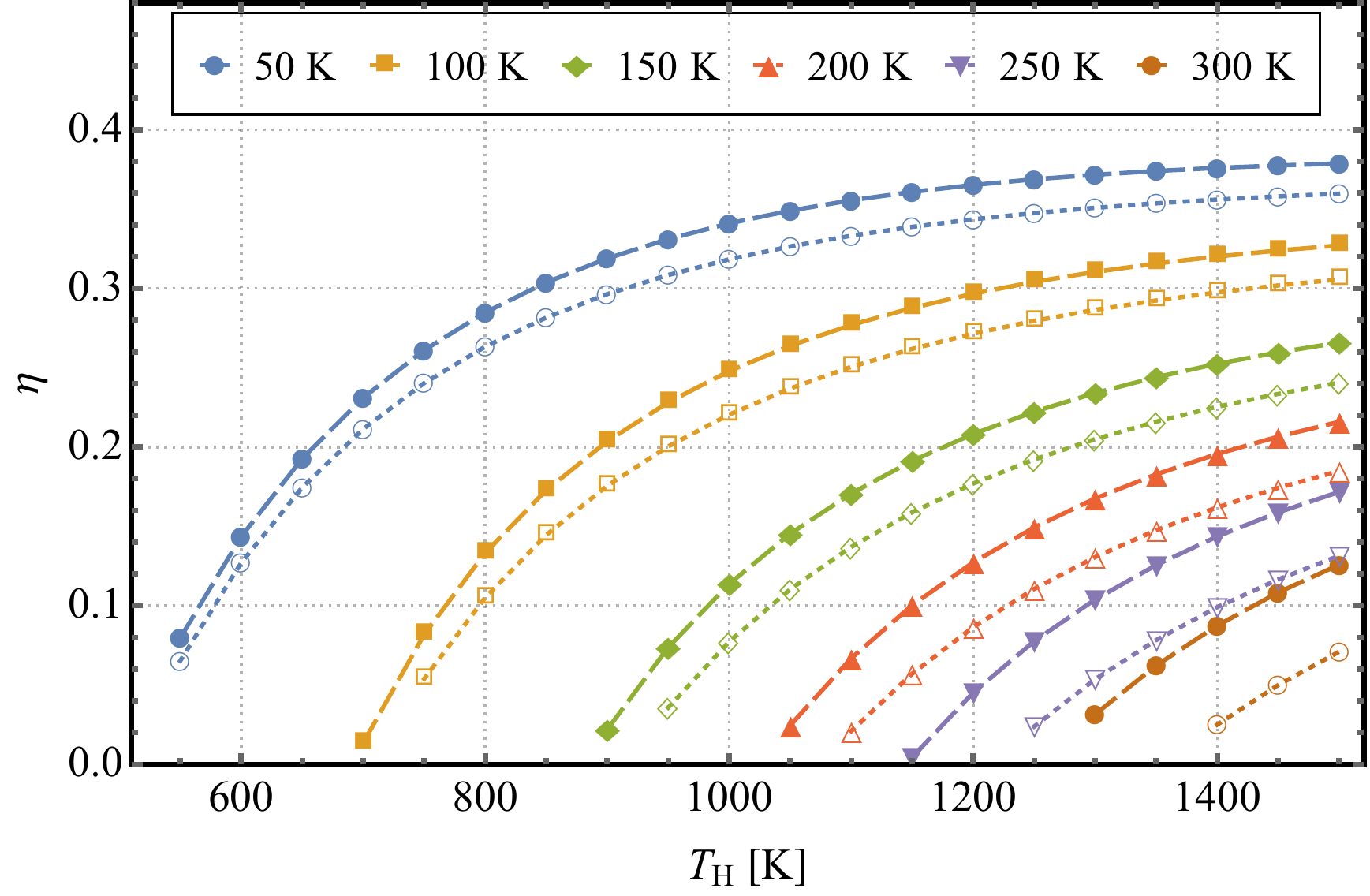}
	\caption{Efficiency vs.\ hot bath temperatures for different cold bath temperatures using the CD-driving (closed markers) and the nonadiabatic (open markers) driving scheme. The different curves correspond to different cooling-bath temperatures (see labels). In these calculations, the system is thermalized until thermal equilibrium is reached. The efficiency values are well below the Carnot limit.}
	\label{fgr:efficiency}
\end{figure}

We show the results for different hot and cold bath temperatures and both driving schemes in Fig.~\ref{fgr:efficiency}. As expected, for a fixed temperature of the cold bath  $T_{\text{L}}$ the efficiency increases with temperature of the hot bath $T_{\text{H}}$ within reasonable temperature ranges. It turns out that there can be a shallow maximum (see Fig.~\ref{fgr:compare_spin_1/2_3/2} in section~\textbf{\ref{sec:multi_cell}}) for low-temperature cooling baths for heating temperatures above \SI{1900}{\kelvin}. However such a high temperature seems beyond the limit of applicability of our \CrO-based quantum heat engine. Nevertheless, there is a certain saturation value of efficiency for fixed $T_\text{L}$\,.  Moreover, we observe that all efficiency values are well below the Carnot limit $\eta_{\text{Carnot}}=1-T_{\text{L}}/T_{\text{H}}$\,. The emergence of shallow maxima as well as a saturation value below the Carnot limit are features of the quantum mechanical behavior of the system. They are in particular based on the discreteness of the spectrum, which will be further discussed in section~\textbf{\ref{sec:multi_cell}}.

It is worth noting that the efficiency of the heat engine is only slightly lower if we refrain from performing CD driving, which is highly relevant since its experimental realization appears to be very difficult if at all doable.

We note that not all combinations of hot and cold baths lead to a positive efficiency value. In particular, if the hot bath temperature is too low, the system is cooled by both the hot and the cold bath, which does not constitute a proper Otto cycle. Note moreover that the theoretical output power (see following subsection) is zero here, since we assume an infinitely long thermalization time (corresponding to infinitely long isochoric strokes) to obtain a perfect Gibbs ensemble.

\subsection{Output power}

\begin{figure}[!htb]
	\includegraphics[width=\columnwidth]{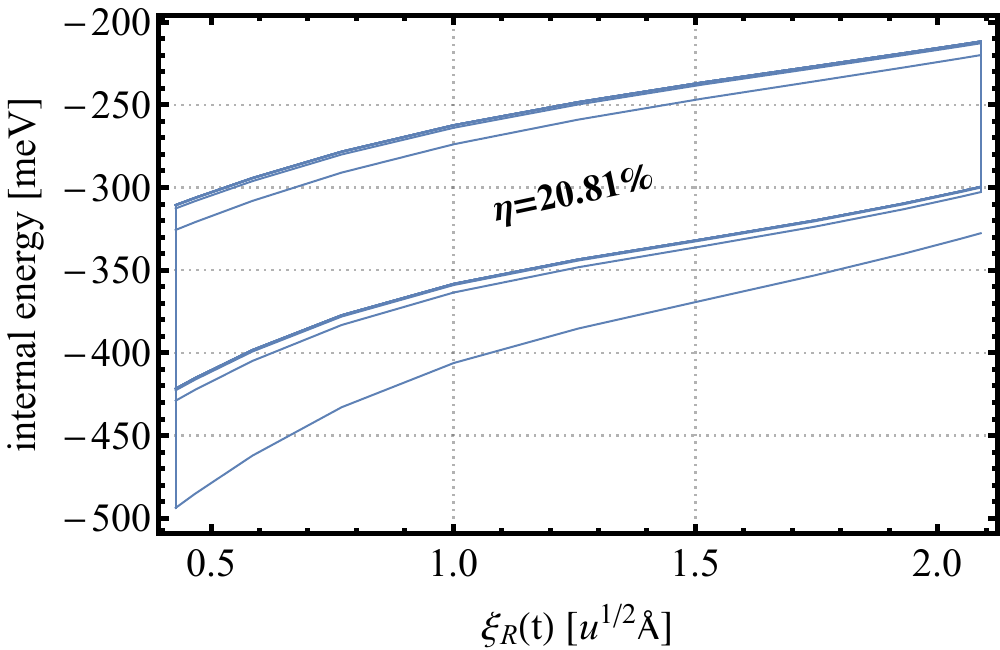}
	\caption{Trajectory and limit cycle for short-time thermalization. The parameters for these cycles (see text)
		are $\tau_\text{L}=\SI{8}{\pico\second}$, $T_{\text{L}}=\SI{50}{\K}$,
		$\tau_\text{H}=\SI{4.6}{\pico\second}$, ${T_{\text{H}}=\SI{1500}{\K}}$\,.
		Clearly, convergence to a limit cycle is obtained after just a few cycles.}
	\label{fgr:limit_cycle}
\end{figure}

We now fix the bath temperatures to $T_{\text{H}}=\SI{1500}{\kelvin}$ and $T_{\text{L}}=\SI{50}{\kelvin}$ and consider different finite-time durations of the isochoric thermalization strokes for the evaluation of the output power, where we use Eq.~\eqref{outputpower}, neglecting the small CD driving costs again. In order to calculate the state after thermalization, we solve the Lindblad equation (Eq.~\eqref{master equation}) using again a Runge-Kutta scheme. We use durations between $\SI{0.66}{\pico\second}$ to \SI{66}{\pico\second} for heating, and durations between $\SI{1.3}{\pico\second}$ and
$\SI{330}{\pico\second}$ for cooling. To obtain meaningful results, i.e., to make sure that the engine describes a closed-cycle loop, we let the engine perform several cycles to converge to a limit cycle. Fortunately, the necessary number of cycles is not too large as illustrated in Fig.~\ref{fgr:limit_cycle} for one particular example: using
a thermalization time $\tau_\text{L}=\SI{8}{\pico\second}$ for cooling with the cold bath and a thermalization time $\tau_\text{H}\approx\SI{4.6}{\pico\second}$ for heating with the hot bath, we see that convergence to the limit cycle happens within a few cycles.
Once the limit cycle is attained, we calculate both output power and efficiency (the latter
is $\eta=\SI{20.81}{\percent}$ for the cycle in Fig.~\ref{fgr:limit_cycle}).

\begin{figure}[!htb]
	\includegraphics[width=\columnwidth]{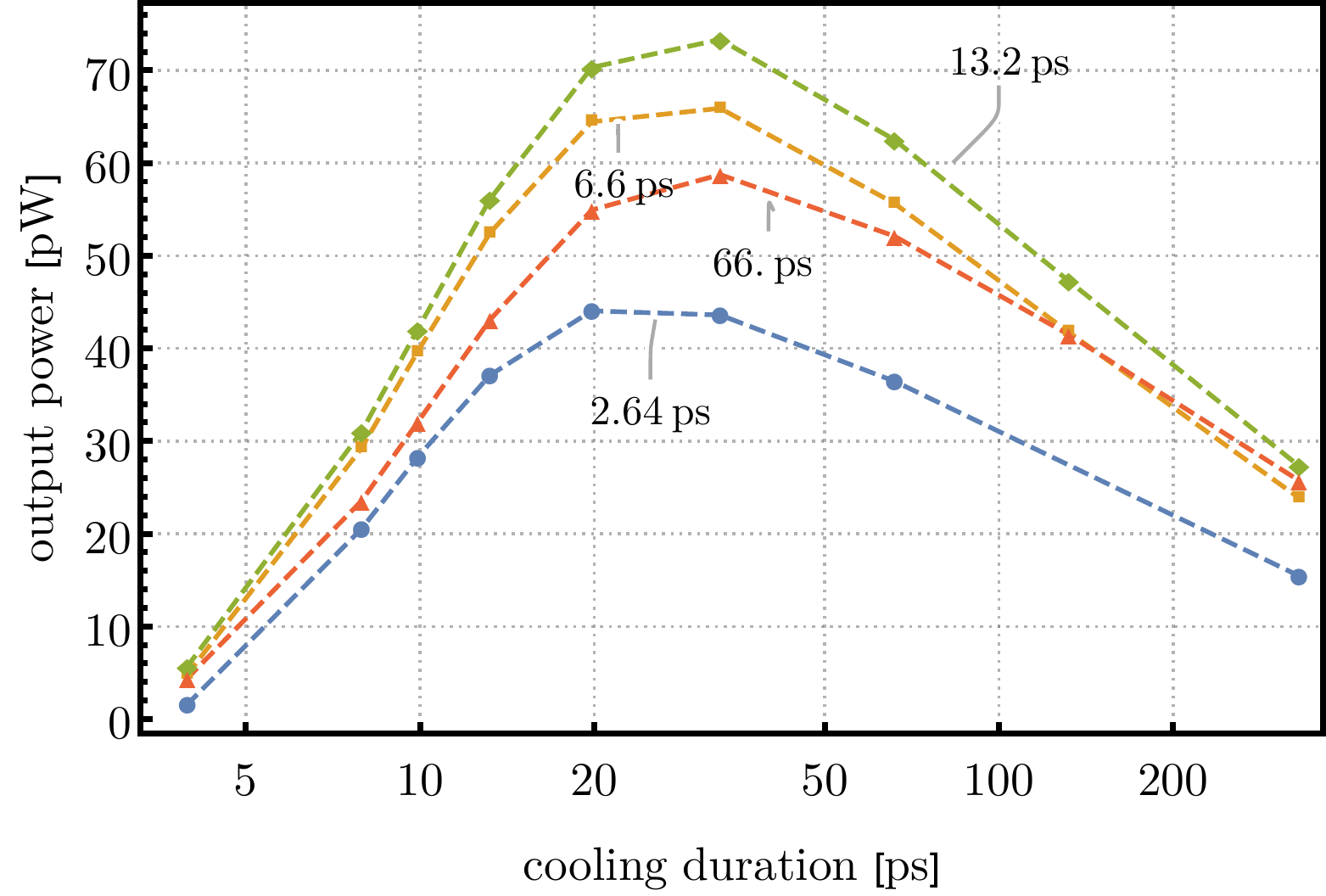}
	\caption{Output power calculated for a single \CrO unit cell. The lines connect data with the same heating duration (value indicated in the figure). The output power is maximal around \SI{33}{\pico\second} of cooling, regardless of the heating duration. Efficiency values of the cycles at the maximal output power are \SI{32.9}{\percent}, \SI{36.6}{\percent}, \SI{37.3}{\percent}, and \SI{37.6}{\percent}, respectively, for increasing heat duration.}
	\label{fgr:output_power}
\end{figure}

The results for the output power for different durations of heating and cooling strokes are shown in Fig.~\ref{fgr:output_power}. We observe distinct maxima for fixed durations of either heating or cooling strokes (only fixed heating curves indicated). Interestingly, for most given heating durations, the peaks lie around \SI{33}{\pico\second} of cooling, while for most given cooling durations, the maxima are around \SI{13}{\pico\second} of heating (not shown). Hence, for the chosen system parameters the optimal quantum isochoric stroke durations are even a little shorter than the work stroke durations $\tau_{2,4}=63\,$ps.

For the indicated curves in Fig.~\ref{fgr:output_power}, the efficiency values at the maximum output power are in the range of \SIrange{32.9}{37.6}{\percent} (\SI{37.3}{\percent} at the maximal output power), which is very close to the efficiency when thermalizing until thermal equilibrium is obtained (see Fig.~\ref{fgr:efficiency}, efficiency for {$T_{\text{H}}=\SI{1500}{\K}$} and $T_{\text{L}}=\SI{50}{\K}$ is \SI{37.9}{\percent}).

The large efficiency values estimated for the optimal output powers indicate that it is not necessary to apply the isochoric strokes until the system is in perfect thermal equilibrium (Gibbs state). This fact is an advantage as the decrease of the cycle durations enhances the output power.

\section{The working substance constructed from several unit cells}
\label{sec:multi_cell}

The four-spin unit-cell model has been shown to successfully describe experimentally observable material properties of the crystal such as the N\'eel temperature, the band gap or the spectral density \cite{Shi2009_PhysRevB.79.104404, Mostovoy2010_PhysRevLett.105.087202}. However, it is unclear whether work statistics are also well-captured by the corresponding effective Hamiltonian. In a working body consisting of several unit cells, the number of the energy levels and all types of interlevel transitions between them are drastically increased.  This may influence the produced work and other thermodynamic characteristics of the engine. Therefore, we consider working bodies consisting of several unit cells and try to infer the scaling with size. We will look at a similar crystal, but use spin-1/2 particles instead of the spin-3/2 of the Cr atoms for computational capacity reasons. We will show that the qualitative behavior is quite similar.

Most of the interactions taken into account are between atoms of neighboring unit cells. As explained in section~\textbf{\ref{sec:model}}, in the single-unit-cell model we use an effective interaction approach by projecting the interactions with atoms of neighboring unit cells onto the respective atoms of the reference unit cell. In this case, the periodic continuation is straightforward. We now generalize the effective Hamiltonian to the settings of two and three unit cells (see Figs.~\ref{fig:structure and spin}b and \ref{fig:structure and spin}c), respectively, by considering a periodically extendable arrangement of unit cells. For example, when building the Hamiltonian for two unit cells, we keep the interactions between atoms of the original first and second unit cell as they are and project everything else on the corresponding atoms.

The Hamiltonian for the two-cell case then reads
\begin{align}
\label{eq:hamilton_two_cells}
\hat{H}_2(t) &:= \nonumber\\
&J_1(\xr(t))\big(\hat{\vec{S}}_1 \hat{\vec{S}}_2+\hat{\vec{S}}_3 \hat{\vec{S}}_4+\hat{\vec{S}}_5 \hat{\vec{S}}_6+\hat{\vec{S}}_7 \hat{\vec{S}}_8\big)\nonumber\\
&+ 3J_2(\xr(t))\big(\hat{\vec{S}}_1 \hat{\vec{S}}_4+\hat{\vec{S}}_5 \hat{\vec{S}}_8+\hat{\vec{S}}_2 \hat{\vec{S}}_7+\hat{\vec{S}}_3 \hat{\vec{S}}_6\big)\nonumber\\
&+
3J_3(\xr(t))\big(\hat{\vec{S}}_1 \hat{\vec{S}}_6+\hat{\vec{S}}_2 \hat{\vec{S}}_5+\hat{\vec{S}}_3 \hat{\vec{S}}_8+\hat{\vec{S}}_4 \hat{\vec{S}}_7\big) \nonumber\\
&+
3J_4(\xr(t))\big(\hat{\vec{S}}_1 \hat{\vec{S}}_3+\hat{\vec{S}}_1 \hat{\vec{S}}_7+\hat{\vec{S}}_2 \hat{\vec{S}}_4 + \hat{\vec{S}}_2 \hat{\vec{S}}_8  \nonumber\\
&\qquad\qquad\qquad+ \hat{\vec{S}}_3 \hat{\vec{S}}_5+\hat{\vec{S}}_4 \hat{\vec{S}}_6+\hat{\vec{S}}_5 \hat{\vec{S}}_7+\hat{\vec{S}}_6 \hat{\vec{S}}_8\big)\nonumber\\
&+
J_5(\xr(t))\big(\hat{\vec{S}}_1 \hat{\vec{S}}_8+\hat{\vec{S}}_2 \hat{\vec{S}}_3+\hat{\vec{S}}_4 \hat{\vec{S}}_5+\hat{\vec{S}}_6 \hat{\vec{S}}_7\big)\nonumber\\
&+\sum_{i=1}^{8}\bigg[ \mu B_z \hat{S}^z_i + D \big( \hat{S}^z_i \big)^2 \bigg] \,.
\end{align}
Here, site numbers 1 to 4 represent the blue cell
(A, counting from top to bottom),
while site numbers 5 to 8 represent the yellow cell (B)
in Fig.~\ref{fig:structure and spin}b.
Note that here and in the following spin Hamiltonian $J_n(\xr(t))$ are the {\em bare}
interaction parameters.

We do the same for the three-unit-cell case, where the resulting Hamiltonian reads
\begin{widetext}
	\begin{align}
	\label{eq:hamilton_three_cells}
	\hat{H}_3(t) &:=
	J_1(\xr(t))\big(\hat{\vec{S}}_1 \hat{\vec{S}}_2+\hat{\vec{S}}_3 \hat{\vec{S}}_4+\hat{\vec{S}}_5 \hat{\vec{S}}_6+\hat{\vec{S}}_7 \hat{\vec{S}}_8+\hat{\vec{S}}_9 \hat{\vec{S}}_{10}+\hat{\vec{S}}_{11} \hat{\vec{S}}_{12}\big)\nonumber\\
	&+ 3J_2(\xr(t))\big(\hat{\vec{S}}_1 \hat{\vec{S}}_{12}+\hat{\vec{S}}_2 \hat{\vec{S}}_7+\hat{\vec{S}}_3 \hat{\vec{S}}_{10}+\hat{\vec{S}}_4 \hat{\vec{S}}_5+\hat{\vec{S}}_6 \hat{\vec{S}}_{11}+\hat{\vec{S}}_{8} \hat{\vec{S}}_{9}\big)\nonumber\\
	&+
	3J_3(\xr(t))\big(\hat{\vec{S}}_1 \hat{\vec{S}}_6+\hat{\vec{S}}_2 \hat{\vec{S}}_9+\hat{\vec{S}}_3 \hat{\vec{S}}_8+\hat{\vec{S}}_4 \hat{\vec{S}}_{11}+\hat{\vec{S}}_5 \hat{\vec{S}}_{10}+\hat{\vec{S}}_{7} \hat{\vec{S}}_{12}\big) \nonumber\\
	&+
	3J_4(\xr(t))\big(\hat{\vec{S}}_1 \hat{\vec{S}}_7+\hat{\vec{S}}_1 \hat{\vec{S}}_{11}+\hat{\vec{S}}_2 \hat{\vec{S}}_8+\hat{\vec{S}}_2 \hat{\vec{S}}_{12} + \hat{\vec{S}}_3 \hat{\vec{S}}_5+\hat{\vec{S}}_3 \hat{\vec{S}}_9+\nonumber\\
	&\qquad\qquad\qquad\qquad+ \hat{\vec{S}}_4 \hat{\vec{S}}_6+\hat{\vec{S}}_4 \hat{\vec{S}}_{10} +\hat{\vec{S}}_5 \hat{\vec{S}}_{11}+\hat{\vec{S}}_6 \hat{\vec{S}}_{12}+\hat{\vec{S}}_7 \hat{\vec{S}}_9+\hat{\vec{S}}_8 \hat{\vec{S}}_{10}\big)\nonumber\\
	&+
	J_5(\xr(t))\big(\hat{\vec{S}}_1 \hat{\vec{S}}_4+\hat{\vec{S}}_2 \hat{\vec{S}}_3+\hat{\vec{S}}_5 \hat{\vec{S}}_8+\hat{\vec{S}}_6 \hat{\vec{S}}_7+\hat{\vec{S}}_9 \hat{\vec{S}}_{12}+\hat{\vec{S}}_{10} \hat{\vec{S}}_{11}\big)\nonumber\\
	&+\sum_{i=1}^{12}\bigg[ \mu B_z \hat{S}^z_i + D \big( \hat{S}^z_i \big)^2 \bigg] \,.
	\end{align}
\end{widetext}
In addition to sites 1 to 8, we added sites 9 to 12 in the green unit cell (C)
in Fig.~\ref{fig:structure and spin}c.
All non-nearest-neighbor interaction parameters between the spins of different cells in Eq.~\eqref{eq:hamilton_two_cells} and Eq.~\eqref{eq:hamilton_three_cells}
mimic the inter-cell coupling.

\subsection{Multi-cell efficiencies}

\begin{figure}[!htb]
	\includegraphics[width=\columnwidth]{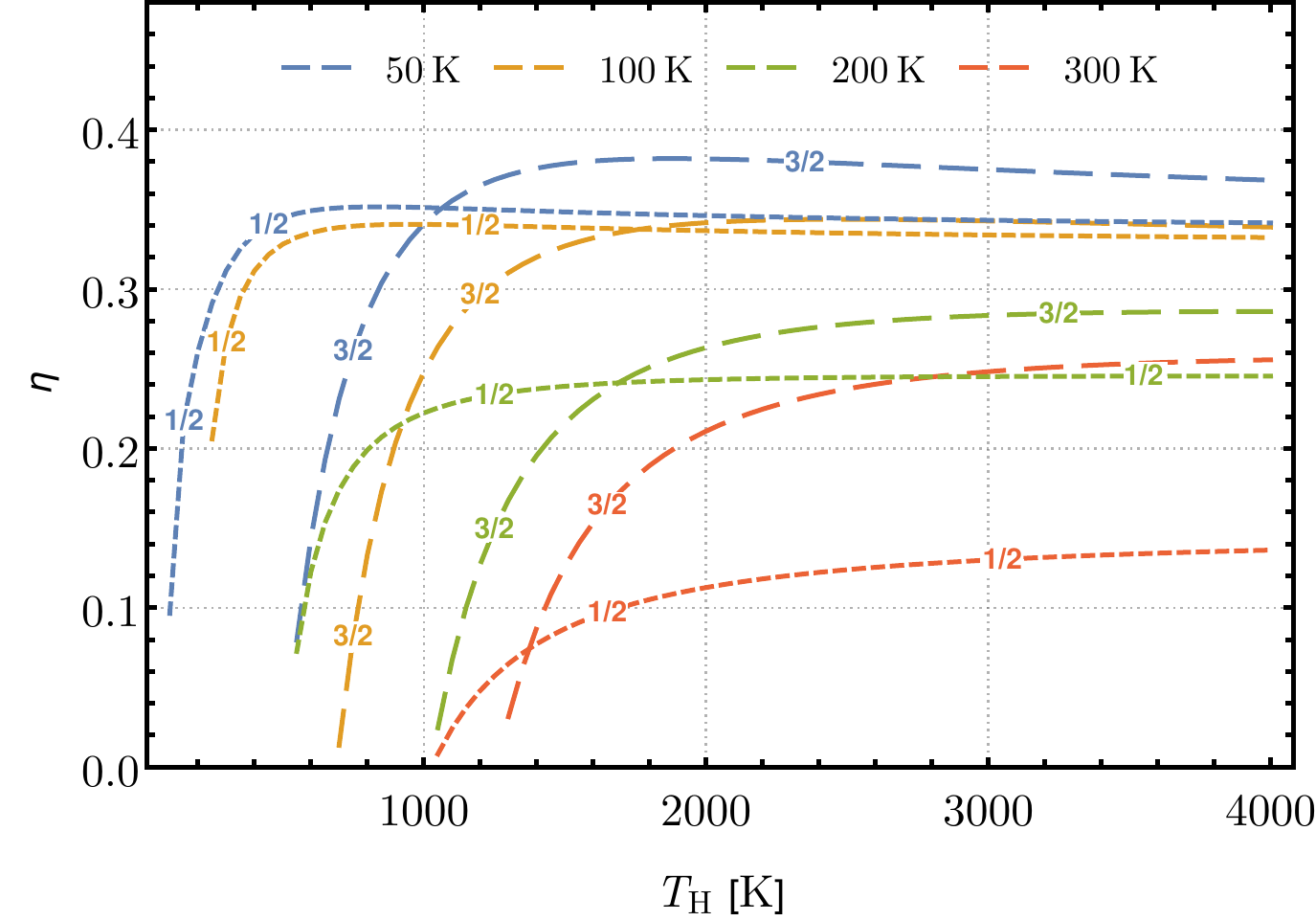}
	\caption{Comparison of the efficiencies of the spin-1/2 and the spin-3/2 case (indicated by markers). Each curve corresponds to a specific cooling temperature (see legend). Despite the different scaling the results are qualitatively similar. 
	}
	\label{fgr:compare_spin_1/2_3/2}
\end{figure}

First, we compare the efficiencies  of the spin-1/2 and the spin-3/2 systems in Fig.~\ref{fgr:compare_spin_1/2_3/2}, where we use a single unit cell for both cases. We observe that the qualitative behavior (fast monotonic increase for low heating temperatures, build-up of a shallow maximum for low-temperature cooling, saturation for high-temperature heating), is rather similar, even if the scales of temperatures are different. Note that temperatures above \SI{1500}{\K} serve only as a comparison of the two models and will not yield physically reasonable results since our model will most likely not be valid  due to changes in the material properties for such high temperatures.

\begin{figure}[!htb]
	\includegraphics[width=\columnwidth]{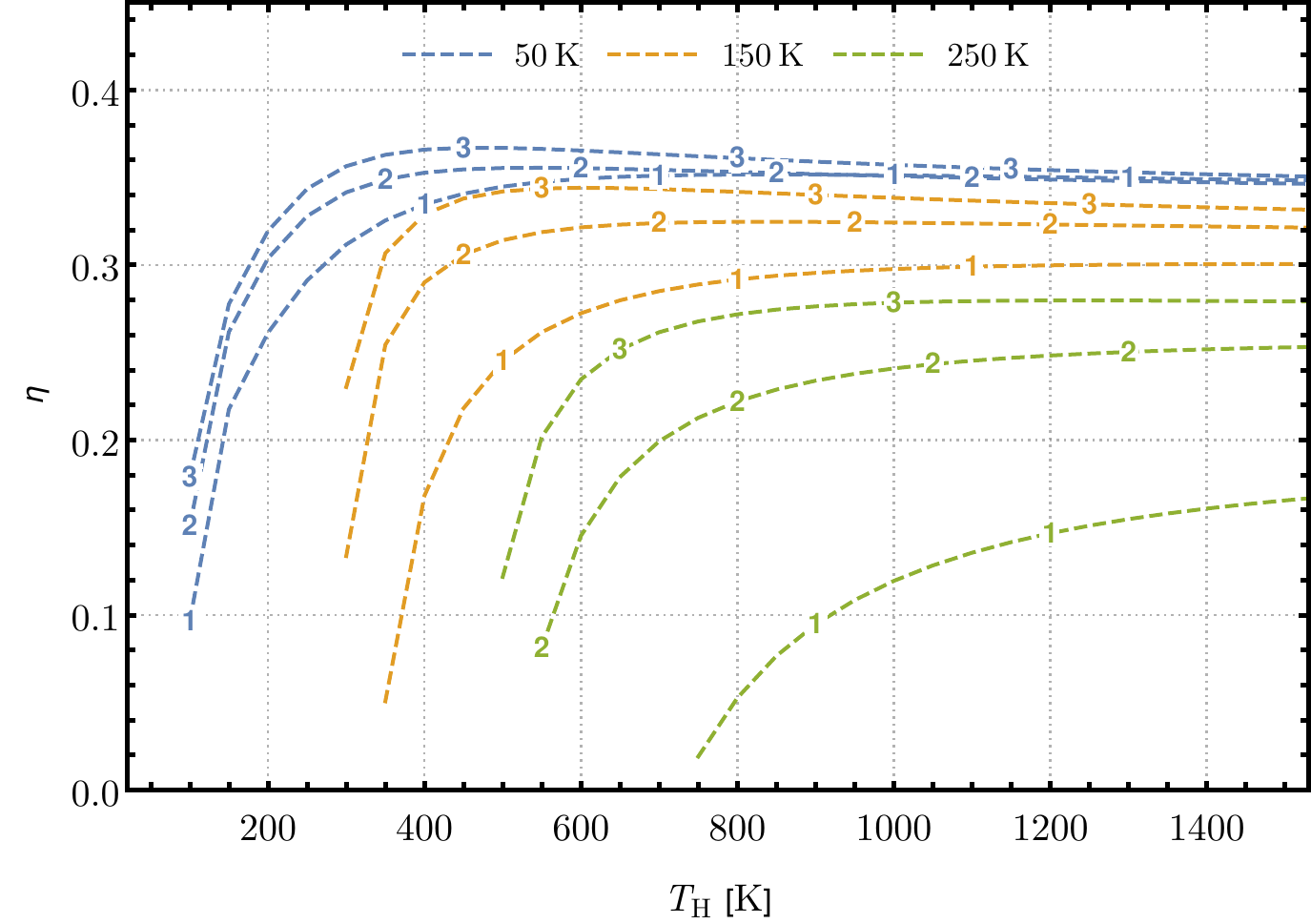}
	\caption{Efficiencies for one, two and three unit cells (indicated by markers) of the spin-1/2 system. Each curve corresponds to a specific cooling temperature (see legend). The shallow maximum for low-temperature cooling is remarkable.}
	\label{fgr:multi_cell_efficiencies}
\end{figure}

We now address the resulting efficiencies for two and for three unit cells, for which the Hamiltonians have been given in Eq.~\eqref{eq:hamilton_two_cells} and \eqref{eq:hamilton_three_cells}, where we again assume perfect CD driving and show the results for different bath temperatures in Fig.~\ref{fgr:multi_cell_efficiencies}. The range of the efficiencies is very similar for different numbers of cells, although the heat engine seems to be more efficient when using more unit cells. We note that, especially when cooling to very low temperatures, e.g., $T_\text{L}=50\,$K, there is only a very small difference between the efficiencies obtained for different numbers of cells and that the saturation values are also extremely close. The single-unit-cell model seems thus to be a very good approximation for the whole system.

As mentioned already above, we observe that the efficiency is not increasing monotonously with increasing temperature $T_\text{H}$ of the hot bath. As an example, a shallow maximum appears for a cooling temperature of \SI{50}{\K} and a heating temperature of \SI{450}{\K} for the three-unit-cell case. This is a quantum effect related to the properties of the energy spectrum and can be explained by the following observation. Looking at the spectrum of the Hamiltonian (Fig.~\ref{fgr:spectrum}), most of the energy levels (especially the lowest ones) show a monotonic energy increase with increasing $\xr$\,. Nevertheless, some of the levels show exactly the opposite behavior. It is therefore clear that if such a level is occupied during the work stroke, it is going to lower the mean work of the whole system, such that the engine would be more efficient if this level was not occupied.

\section{Generic model Hamiltonian}
\label{sec:alternative}

The quantum heat engine proposed in this work is not particularly tailored for
\CrO alone, because the same mechanism could also be applied to other materials 
as discussed above. 

\begin{figure}[htp]
	\centering
	\includegraphics[width=\columnwidth]{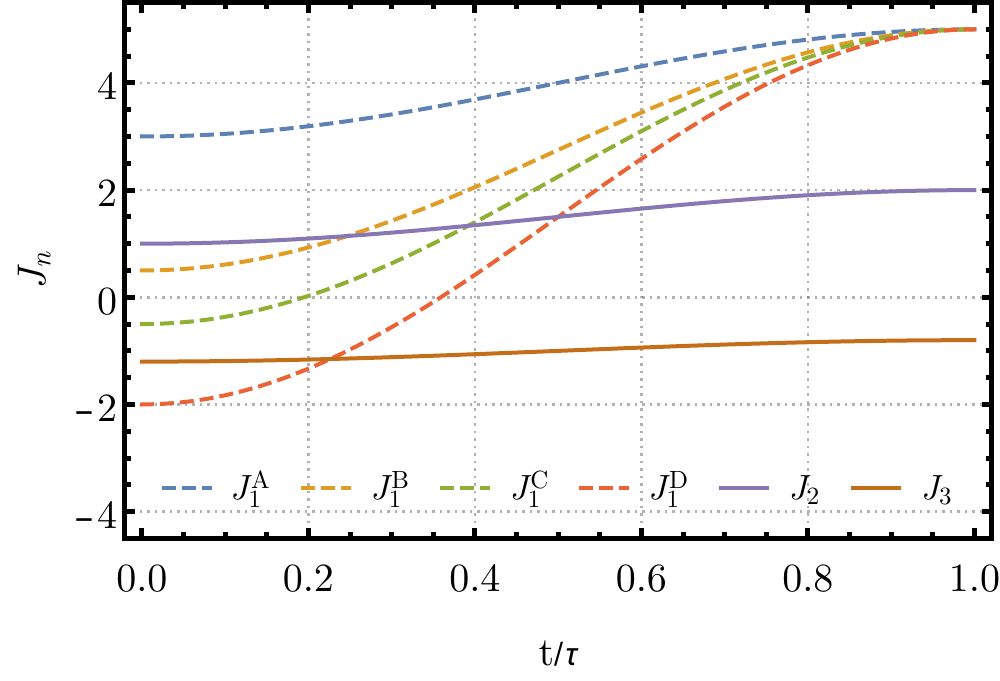}
	\caption{Exchange interactions for the four model systems. The dashed lines represent the different choices for $J_1$.}
	\label{fgr:model_exchange_interactions}
\end{figure}

Inspired by the spin Hamiltonian Eq.~\eqref{eq:Hamiltonian} for \CrO, we 
introduce a generic Heisenberg unit-cell Hamiltonian for a system of four 
spin-1/2 particles, where nearest neighbor $J_1$, next-nearest neighbor $J_2$, 
and next-next-nearest neighbor $J_3$ interactions together with a 
magneto-crystalline anisotropy term $D$ and a Zeeman term are taken into 
account,
\begin{align}
\hat{H}^{\text{A,B,C,D}}(t) &= J_1^{\text{A,B,C,D}}(t)\bigg(\hat{\vec{S}}_1\hat{\vec{S}}_2+\hat{\vec{S}}_2\hat{\vec{S}}_3+\hat{\vec{S}}_3 \hat{\vec{S}}_4\bigg) \nonumber\\
&+ J_2(t)\bigg(\hat{\vec{S}}_1\hat{\vec{S}}_3+\hat{\vec{S}}_2\hat{\vec{S}}_4\bigg)\nonumber\\
&+
J_3(t)\hat{\vec{S}}_1\hat{\vec{S}}_4\nonumber\\
&+D\sum\limits_{j=1}^4 (\hat{S}_j^{(z)})^2 + \mu B_z\sum\limits_{j=1}^4 \hat{S}_j^{(z)}.
\label{eq:model_Hamiltonians}
\end{align}
In particular, the model systems A,B,C,D differ in the change of the nearest-neighbor interaction, where the lowest values for $J_1$ range from $J_1^\text{A}(0)=3$ to $J_1^\text{D}(0)=-2$ (see Fig.~\ref{fgr:model_exchange_interactions}).

\begin{figure}[htp]
	\centering
	\includegraphics[width=\columnwidth]{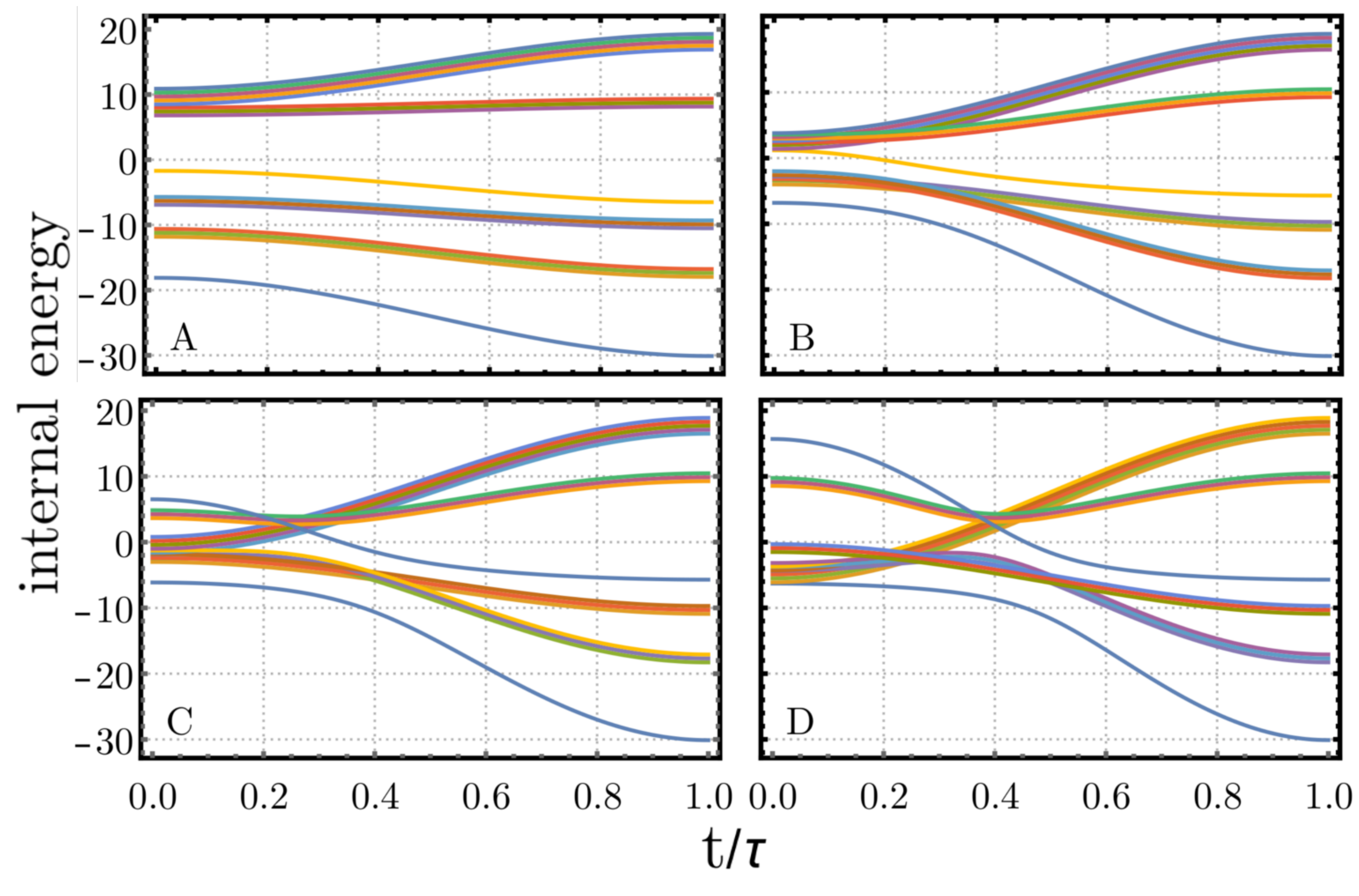}
	\caption{Instantaneous spectra for the systems A,B,C,D during the variation of the interaction parameters according to the scheme in Fig.~\ref{fgr:model_exchange_interactions}.}
	\label{fgr:model_spectra}
\end{figure}

The characteristic differences between these Hamiltonians lie in the type of their energy spectra (see Fig.~\ref{fgr:model_spectra}): The energy spectrum of $\hat{H}^{\text{A}}$ contains no level crossing at all, while the spectra of $\hat{H}^{\text{B}}$ and $\hat{H}^{\text{C}}$ contain a few level crossings. We observe that some levels of $\hat{H}^{\text{D}}$ cross almost all of the other levels, rendering them almost the lowest ones at $t=0$ and the highest ones at $t=\tau$. Altogether, we note that the energy levels are close together at $t=0$ while they are spread further apart at $t=\tau$ for all models. Finally, it is worth noting that the ground state always stays the same throughout the driving in all four systems. 

Comparing these spectra to the one of the \CrO~system (see Fig.~\ref{fgr:spectrum}) we observe that it is most similar to $\hat{H}^{\text{B}}$ and $\hat{H}^{\text{C}}$ in the sense of sharing the following features:
\begin{itemize}
	\setlength\itemsep{-.3em}
	\item There are no quantum phase transitions during the strokes.
	\item Some of the energy levels increase during the stroke, and some of the levels decrease, leading to several level crossings. 
	\item The level crossings do not lead to an inversion-like behavior as in $\hat{H}^{\text{D}}$. 
\end{itemize}

\subsection{Efficiencies of the generic models}

\begin{figure}[htp]
	\centering
	\includegraphics[width=\columnwidth]{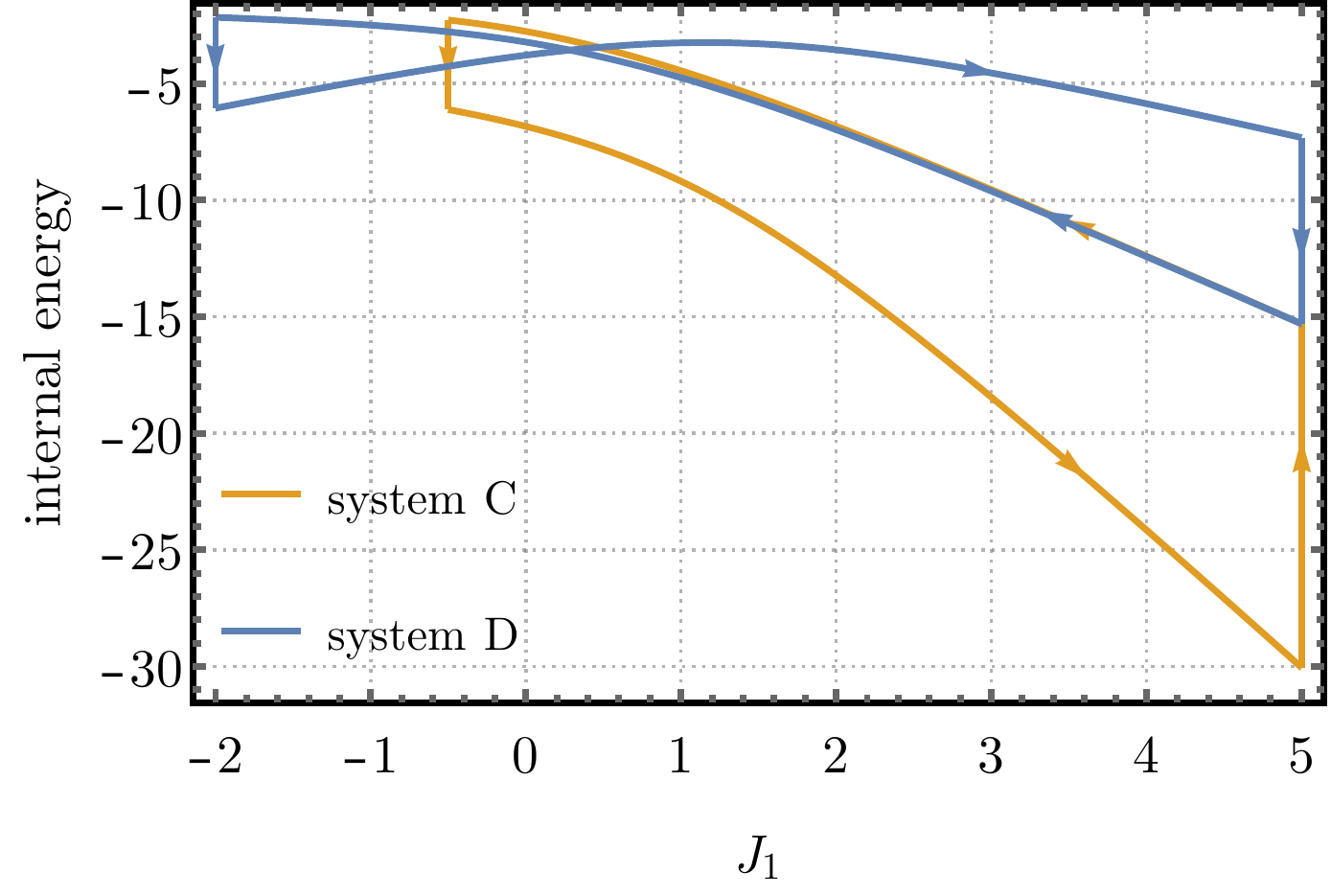}
	\caption{Typical cycles for models C and D. While model C yields a proper Otto cycle, model D does not.}
	\label{fgr:model_cycles}
\end{figure}

We now investigate the efficiencies of Otto cycles as described in section~\textbf{\ref{sec:otto}} for the model systems A,B,C,D. Two typical cycles for model systems C and D are shown in Fig.~\ref{fgr:model_cycles}. 
For system C, we observe that the total energy describes a closed loop representing a proper Otto cycle. On the contrary, the loop described by system D contains a crossing such that both of the baths are actually cooling the system. Thus system D does not yield a proper Otto cycle and will therefore not be included for further investigation.

\begin{figure}[htp]
	\includegraphics[width=\columnwidth]{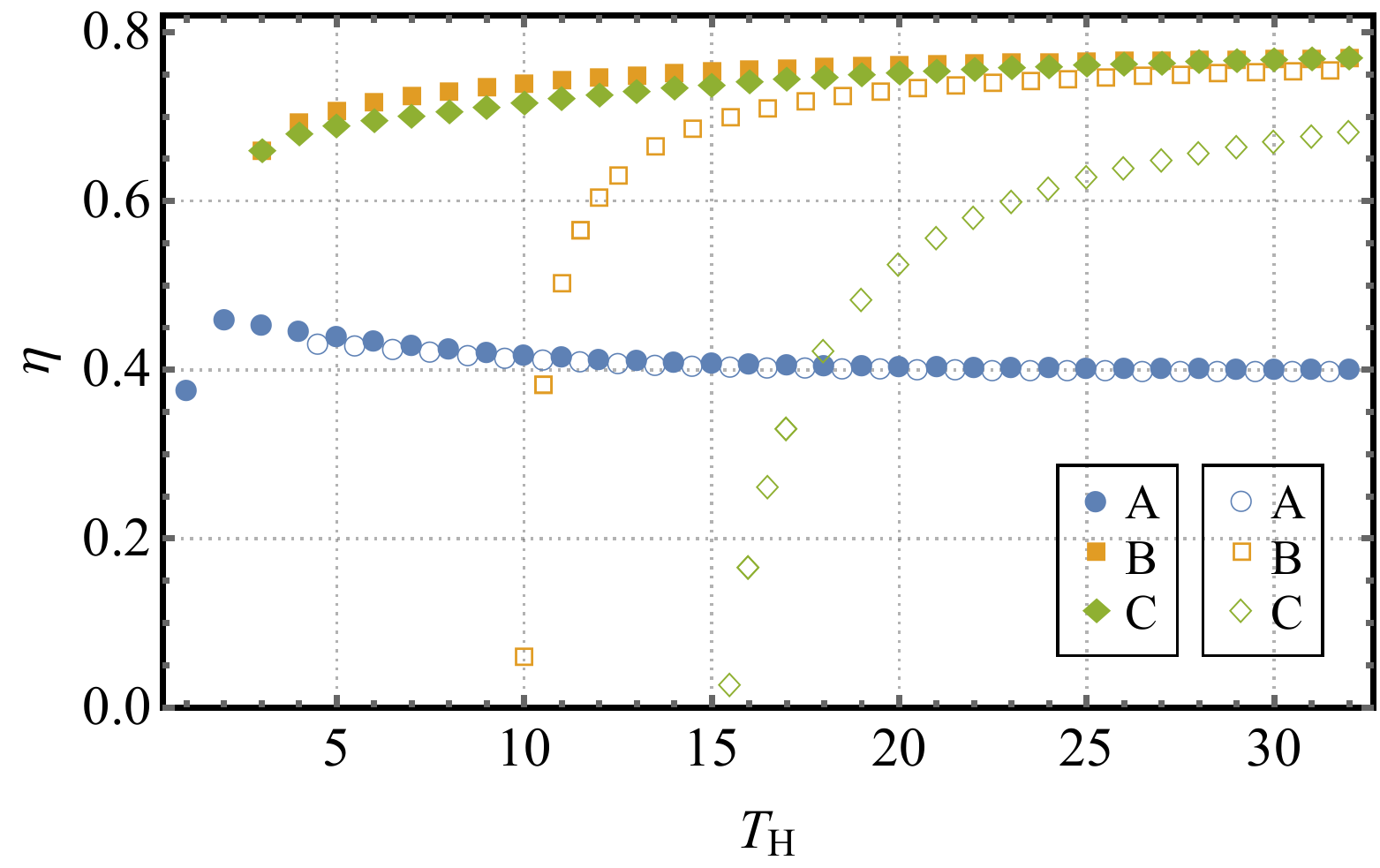}
	\caption{Efficiencies for systems A,B,C for cooling temperatures $T_\text{L}=0.5$ (closed markers) and $T_\text{L}=2.0$ (open markers) as functions of the heating temperature.}
	\label{fgr:model_efficiencies}
\end{figure}

We now look at the efficiencies of the systems A, B, and C for two different cooling temperatures $T_\text{L}=0.5$ and $T_\text{L}=2.0$ in Fig.~\ref{fgr:model_efficiencies}, where we use the definition in Eq.~\eqref{eq:efficiency}. Apparently, systems B and C show a much better performance than system A. The reason for this appears to be the occupation of levels that strongly spread out during driving in systems B and C, which is less pronounced for system A. It can be therefore suspected that heat engines of this type are especially effective if they show a spectrum that resembles the ones of systems B or C concerning the features described in section~\textbf{\ref{sec:model}}.

\subsection{Energy costs of counterdiabatic driving}
\label{costs}

\begin{figure}[htp]
	\centering
	\includegraphics[width=\columnwidth]{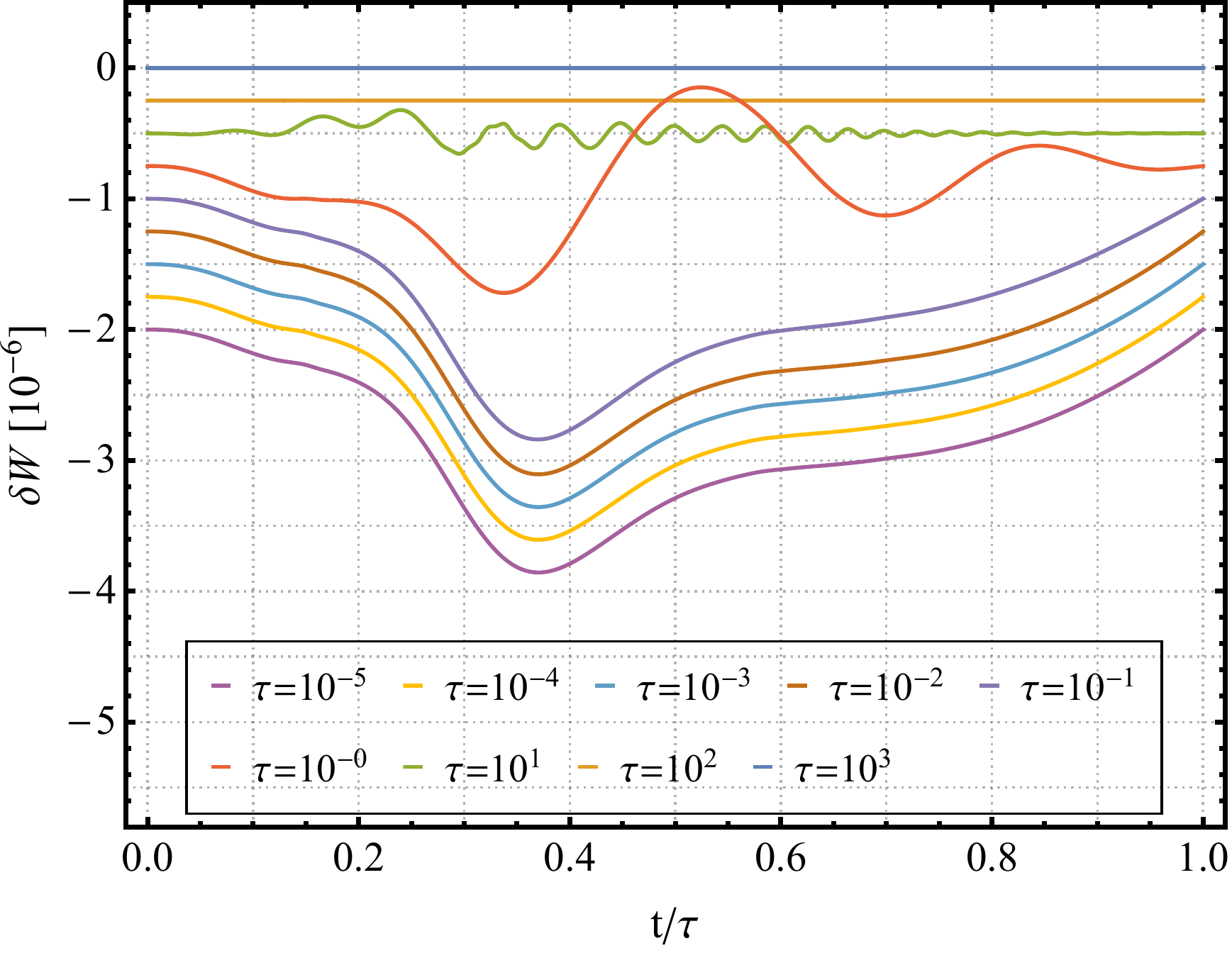}
	\caption{$\delta W(t)$ for system B for different durations $\tau$ of work stroke 4 as a function of the scaled time $t/\tau$. Each of the curves is offset by -0.25 from the previous one in order to increase visibility, they all actually start and end at $\delta W=0$. The cost function is similar for stroke durations below $10^{-2}$, and is zero for very slow driving.}
	\label{fgr:deltaW_cold}
\end{figure}

\begin{figure}[htp]
	\centering
	\includegraphics[width=\columnwidth]{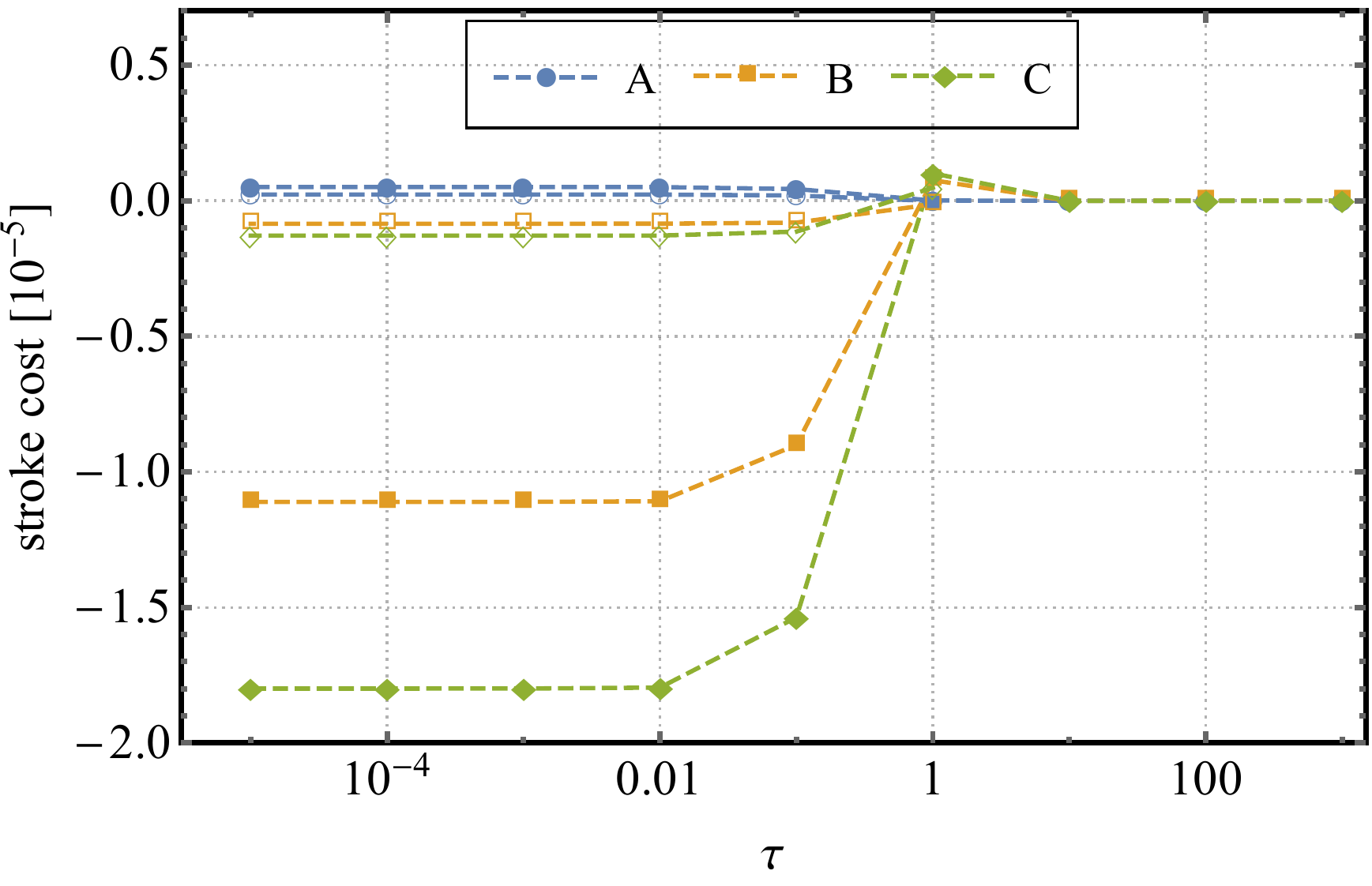}
	\caption{Total costs for the different stroke durations $\tau$ for stroke 2 (open markers) and stroke 4 (closed markers) for systems A,B,C. For stroke durations below $10^{-2}$, the cost is constant and very small compared to the total energy put into the system during heating.}
	\label{fgr:total_cost}
\end{figure}

We now turn to the additional driving costs due to CD driving. As an example, we calculate $\delta W(t)$ as defined in Eq.~\eqref{eq:non_equilibrium_work_for_calculation} for the work stroke after cooling of model B, which shows a high efficiency and therefore seems more suitable for application. While the bath temperatures are fixed to $T_\text{L}=0.5$ and $T_{\text{H}}=15$, we consider different stroke durations between $\tau = 10^{-5}$ and $\tau=10^3$ in Fig.~\ref{fgr:deltaW_cold}. 
We observe small values, where in particular $\delta W(t)$ is zero throughout the whole stroke for $\tau\geq 10^2$. Comparing these stroke durations to the \CrO~model with its different scaling of the spectrum, we conclude that the duration of the strokes in \CrO corresponds to about $\tau=10^3$ for the generic model B. The fact that $\delta W(t)$ is essentially zero throughout the whole stroke in \CrO is thus completely consistent with the results obtained for the generic models. It also turns out that $\delta W(t)$ as a function of the scaled time $t/\tau$ is always the same for stroke durations shorter than $\tau=10^{-2}$. As a result, the total CD driving cost $\ev*{\hat{H}_1^{(i)}}_\tau$ (Eq.~\eqref{eq:total_STA_cost}) is independent of $\tau$, which is also observed in Fig.~\ref{fgr:total_cost}, where the total CD driving costs for models A, B, C for different stroke durations are depicted. The magnitude of the CD driving costs is always below $2\cdot 10^{-5}$. For reference, the energies supplied to the system throughout heating strokes are 14.98 for model A, 14.69 for B, and 14.73 for C, which is several orders of magnitude larger.

This result implies that $\ev*{\hat{H}_1^{(i)}}_\tau$ can be safely neglected in the calculation of the efficiency (Eq.~\eqref{eq:correctedefficiency}) and the output power (Eq.~\eqref{eq:correctedoutputpower}), which is in stark contrast to harmonic-oscillator systems, for which the driving cost usually increases at faster driving \cite{Abah2019}.

\section{Conclusions}
\label{sec:conclusions}

We proposed a new type of quantum heat engine using a nano-scale working body described by an effective spin Hamiltonian, where the working principle is based on the fast tunability of the interaction parameters between the spins. The latter is achieved by employing nonlinear phononics, where intense laser pulses dynamically control the magnetic coupling through anharmonic coupling of infrared and Raman modes that are specific to the system.

While we describe the general recipe to identify possible materials for 
obtaining such a quantum heat engine, we already have a promising candidate 
material at hand: As it has been shown \cite{Fechner} that \CrO exactly yields 
the necessary behavior upon exciting an infrared-active phonon mode. Here, \CrO 
is described by a unit cell consisting of four spin-3/2 particles (the Cr 
atoms), where an effective Heisenberg Hamiltonian can be employed.

We defined a quantum Otto cycle for the four spin-3/2 system, where we also 
employ methods for transitionless driving for the work strokes. Based on the 
results in \cite{Fechner} we then investigated the performance of the engine 
using a \CrO working body.

The rich energy spectrum with many (avoided) level crossings  is the cause for 
a lot of remarkable features. Our main finding is that apart from obtaining a 
functioning heat engine, it turns out that the quantum friction - calculated 
via 
the non-equilibrium work fluctuations - is very small compared to the total 
work performed during a cycle. That means that one can avoid the 
presumably infeasible experimental construction of a counterdiabatic 
Hamiltonian without too much performance penalty.

Investigating the heat engine efficiency when thermalizing until equilibrium, 
we 
found that not all combinations of hot and cold bath temperatures can 
be used. In particular, this does not mean that the resulting engine could 
work as a refrigerator in those cases, but rather that there is no proper Otto 
cycle at all. This fact together with the observation of maxima of the 
efficiency with varying 
hot bath temperature are features that we can attribute to the discreteness of the 
spectrum and its dependence on the Raman mode displacement. Finally, we 
can conclude that the 
efficiency saturates for high temperature baths and lies always well below the 
Carnot limit.

We also calculated the output power, which depends on the 
duration of a cycle. Very long thermalization, i.e.~a very long cycle, leads to 
a vanishing output power. In our case, the output power shows maxima for much 
shorter heating and cooling durations than needed for full thermalization of 
the system. Having these maxima together with high efficiency values is the
basis to actually construct a experimentally suitable working cycle with 
finite-time cycle durations.

In order to explore the approximation of considering an effective single-unit-cell 
Hamiltonian for the \CrO system, we also considered a model with multiple unit 
cells employing spin-1/2 particles. Increasing the number of cells enhances the 
cycle efficiency, but at least for low cold bath temperatures, the efficiency depends only little on the number of unit cells used for the calculation.

To get a feeling for the dependence of the engine performance on its 
parameters, we defined several generic models, where different variations of 
the nearest-neighbor interaction were considered. Investigating the performance 
of these generic models, we observe high efficiencies when the spectrum shows 
a few level crossings and when the levels spread out during the parametrical 
driving, whereas there is no proper Otto cycle if the spectrum shows an 
inversion-like behavior, i.e.~when too many levels change their order by 
energy. Comparing the realistic \CrO~system to the generic models, it appears 
to belong to the class of highly-efficient quantum heat engines based on a 
parametric variation of the interaction parameters.

Moreover, it turns out that the additional cost for CD driving is negligible in the generic models compared to the amount of heat put into the system during heating for all stroke durations and vanishes for slow enough driving. Comparing these results to the \CrO~case again it turns out that obtaining zero non-equilibrium work fluctuations is not a coincidence for our investigated setup. Rather, the driving is slow enough to make the fluctuations dissappear.

Our results indicate that \CrO~is a suitable material for realizing a quantum 
heat engine based on dynamical material design, which calls for further 
investigations -- theoretical as well as experimental -- of this system.

\section*{Acknowledgment}

The authors would like to thank Jamal Berakdar, Adolfo del Campo, Levan Chotorlishvili, and Arthur Ernst for many useful suggestions and fruitful discussions.


\section*{Appendix}

\subsection{Lattice displacements and interaction parameters in \CrO}
\setcounter{equation}{0}
\renewcommand{\theequation}{A\arabic{equation}}

The time dependence of the interaction parameters $J_n(\xr(t))$ follows from the time dependence of the lattice displacements $\xr$\,. The latter are derived from solving a differential equation system governing a potential energy surface as given in Eq.~\eqref{eq:nonlinear_phononics:Cr2O3} for \CrO
and the external driving field as given in Eq.~\eqref{eq:nonlinear_phononics:driving}. Hence, the lattice displacements themselves are time dependent:
\begin{align}
\xr(t) &=  \xi_{\text{R}0} + C_\text{R} \cos(\tilde{\omega}_\text{R}t)
+ C_\text{IR} \cos(2 \tilde{\omega}_\text{IR}t) \nonumber\\
& +  C_{\varOmega} \cos(2\varOmega t)
+ C_{\varOmega_+} \sin[(\varOmega+\tilde{\omega}_\text{IR}) t] \nonumber\\
& + C_{\varOmega_-} \sin[(\varOmega-\tilde{\omega}_\text{IR}) t]\,.
\label{eq:original_displacement}
\end{align}

Therein, the five amplitudes $C_\text{R}$\,, $C_\text{IR}$\,,
$C_\varOmega$\,, and $C_{\varOmega_{\pm}}$ depend on the renormalized
frequencies $\tilde{\omega}_\text{R}\approx\SI{9}{\THz}$,
$\tilde{\omega}_\text{IR}\approx\SI{17}{\THz}$ (see \cite{Fechner} for
the complete expressions) and on the driving frequency~$\varOmega$ (see
Tab.~\ref{tbl:xi_amplitudes}). Those lattice displacements can then be
connected with the magnetic coupling constants via a continuous
quadratic function $J_n(\xr)$ \cite{Fechner} (see Eq.~\eqref{eq:Jn_definition}), where the
parameters are tabulated in Tab.~\ref{tbl:spin_phonon_coupling_constants}.
Here, we consider only the variation of $J_n$ with the R mode because
the variation due to the IR mode has hardly any effect on the coupling constants.

\begin{table}[htp]
	\renewcommand{\arraystretch}{1.4}
	\setlength{\tabcolsep}{9pt}
	\caption{Amplitudes of $\xr(t)$ and $\xr$ in Eq.~\eqref{eq:original_displacement}
		for ${\varOmega=\SI{16.95}{\THz}}$ (taken from \cite{Fechner}).
		The units are $\sqrt{u}\si{\angstrom}$\,.}
	\label{tbl:xi_amplitudes}
	\begin{tabular*}{\linewidth}{c@{\extracolsep{\fill}} ccccc}
		\hline\hline
		$\xi_{\text{R}0}$ & $C_\text{R}$ & $C_\text{IR}$ & $C_\varOmega$
		& $C_{\varOmega_+}$ & $C_{\varOmega_-}$\\
		\hline
		1  & 0.0754  & 0.1826 & 0.0246 & 0.0705 & 0.8311 \\
		\hline\hline
	\end{tabular*}
\end{table}

\begin{table}[bp]
	\renewcommand{\arraystretch}{1.4}
	\setlength{\tabcolsep}{9pt}
	\caption{The ground state magnetic interaction constants ($J_n(\xr=0)$) and their first and second derivatives for the $A_{1\text{g}}(9)$ mode in Cr$_{2}$O$_{3}$\,. The values are obtained from a quadratic fit to the $J_n(\xr)$ variation depicted in Fig.~4b in \cite{Fechner}. The units are meV, meV/($\sqrt{u}$\si{\angstrom}), and meV/($\sqrt{u}$\si{\angstrom})$^2$, respectively.}
	\label{tbl:spin_phonon_coupling_constants}
	\begin{tabular*}{\linewidth}{l@{\extracolsep{\fill}} ccccc}
		\hline\hline
		& $J_1$& $J_2$ & $J_3$ & $J_4$ & $J_5$\\
		\hline
		$J_{n}(0)$  & 26.54 & 21.2 & -3.9 & -3.3 & 4.2\\
		$\pdv{J_n(\xi)}{\xi}$ & -48.38 & -4.4 & -0.1 & -0.6 & -1.0\\
		$\pdv[2]{J_n(\xi)}{\xi}$ & 20.74 & 0.4 & 0.2 & 0.0 & 0.2\\
		\hline\hline
	\end{tabular*}
\end{table}

We now explain the simplifications made to obtain Eq.~\eqref{eq:xr_definition}. The last term in Eq.~\eqref{eq:original_displacement} forms a long-wave solution, which dominates the variation of $\xr(t)$. The term including $\tilde{\omega}_{\text{R}}$ yields a wave with frequency $\approx\SI{9}{\THz}$\,. The other components with $2\tilde{\omega}_\text{IR}$\,, $2\varOmega$\,, and $(\varOmega+\tilde{\omega}_\text{IR})$ have very similar frequencies at $\approx\SI{34}{\THz}$, yielding rapid oscillations. We therefore decided to only keep the term with the largest contribution, $C_{\text{IR}}$\,, for simplicity, yielding
\begin{align}
\xr(t) \approx \xi_{\text{R}0} + C_\text{R} &\cos(\tilde{\omega}_\text{R}t)
+ C_\text{IR} \cos(2 \tilde{\omega}_\text{IR}t)\nonumber\\
&+ C_{\varOmega_-} \sin[(\varOmega-\tilde{\omega}_\text{IR}) t]\,.
\end{align}

As a next step, we obtain Eq.~\eqref{eq:xr_definition} by shifting by half a period of the lowest frequency, yielding $\cos\left[(\varOmega-\tilde{\omega}_{\text{IR}})t\right]$ instead of the corresponding $\sin\left[(\varOmega-\tilde{\omega}_{\text{IR}})t\right]$\,.
This is advantageous since it yields $\partial_t H(0) = \partial_t H(\tau) = 0$, which is in turn important for the CD Hamiltonian to coincide with the original one at the beginning and at the end of the strokes.

Finally, we plug $\xr(t)$ into the quadratic Taylor expansions of the $J_n$ (Eq.~\eqref{eq:Jn_definition}) and use these for the definition of the time-dependent spin Hamiltonian~(Eq.~\eqref{eq:Hamiltonian}) for $E_\text{drive} = \SI{0.6}{\mega \volt \per \cm}$ at a driving frequency of $\varOmega = \SI{16.95}{\THz}$\,. The latter choice follows Fig.~8 in \cite{Fechner}, showing a reasonable slow modulation of the $\tilde{J}_n$  -- almost five periods during \SI{500}{\ps}. The thermodynamic strokes follow then as described in the main text (\Fref{fgr:thermo_cycle}).

\bibliography{lib}

\begin{thebibliography}{75}%
\makeatletter
\providecommand \@ifxundefined [1]{%
 \@ifx{#1\undefined}
}%
\providecommand \@ifnum [1]{%
 \ifnum #1\expandafter \@firstoftwo
 \else \expandafter \@secondoftwo
 \fi
}%
\providecommand \@ifx [1]{%
 \ifx #1\expandafter \@firstoftwo
 \else \expandafter \@secondoftwo
 \fi
}%
\providecommand \natexlab [1]{#1}%
\providecommand \enquote  [1]{``#1''}%
\providecommand \bibnamefont  [1]{#1}%
\providecommand \bibfnamefont [1]{#1}%
\providecommand \citenamefont [1]{#1}%
\providecommand \href@noop [0]{\@secondoftwo}%
\providecommand \href [0]{\begingroup \@sanitize@url \@href}%
\providecommand \@href[1]{\@@startlink{#1}\@@href}%
\providecommand \@@href[1]{\endgroup#1\@@endlink}%
\providecommand \@sanitize@url [0]{\catcode `\\12\catcode `\$12\catcode
  `\&12\catcode `\#12\catcode `\^12\catcode `\_12\catcode `\%12\relax}%
\providecommand \@@startlink[1]{}%
\providecommand \@@endlink[0]{}%
\providecommand \url  [0]{\begingroup\@sanitize@url \@url }%
\providecommand \@url [1]{\endgroup\@href {#1}{\urlprefix }}%
\providecommand \urlprefix  [0]{URL }%
\providecommand \Eprint [0]{\href }%
\providecommand \doibase [0]{http://dx.doi.org/}%
\providecommand \selectlanguage [0]{\@gobble}%
\providecommand \bibinfo  [0]{\@secondoftwo}%
\providecommand \bibfield  [0]{\@secondoftwo}%
\providecommand \translation [1]{[#1]}%
\providecommand \BibitemOpen [0]{}%
\providecommand \bibitemStop [0]{}%
\providecommand \bibitemNoStop [0]{.\EOS\space}%
\providecommand \EOS [0]{\spacefactor3000\relax}%
\providecommand \BibitemShut  [1]{\csname bibitem#1\endcsname}%
\let\auto@bib@innerbib\@empty
\bibitem [{\citenamefont {Seifert}\ \emph {et~al.}(2017)\citenamefont
  {Seifert}, \citenamefont {Jaiswal}, \citenamefont {Sajadi}, \citenamefont
  {Jakob}, \citenamefont {Winnerl}, \citenamefont {Wolf}, \citenamefont
  {Kl{\"{a}}ui},\ and\ \citenamefont {Kampfrath}}]{Seifert}%
  \BibitemOpen
  \bibfield  {author} {\bibinfo {author} {\bibfnamefont {T.}~\bibnamefont
  {Seifert}}, \bibinfo {author} {\bibfnamefont {S.}~\bibnamefont {Jaiswal}},
  \bibinfo {author} {\bibfnamefont {M.}~\bibnamefont {Sajadi}}, \bibinfo
  {author} {\bibfnamefont {G.}~\bibnamefont {Jakob}}, \bibinfo {author}
  {\bibfnamefont {S.}~\bibnamefont {Winnerl}}, \bibinfo {author} {\bibfnamefont
  {M.}~\bibnamefont {Wolf}}, \bibinfo {author} {\bibfnamefont {M.}~\bibnamefont
  {Kl{\"{a}}ui}}, \ and\ \bibinfo {author} {\bibfnamefont {T.}~\bibnamefont
  {Kampfrath}},\ }\href {\doibase 10.1063/1.4986755} {\bibfield  {journal}
  {\bibinfo  {journal} {Appl. Phys. Lett.}\ }\textbf {\bibinfo {volume}
  {110}},\ \bibinfo {pages} {252402} (\bibinfo {year} {2017})}\BibitemShut
  {NoStop}%
\bibitem [{\citenamefont {Mankowsky}\ \emph {et~al.}(2016)\citenamefont
  {Mankowsky}, \citenamefont {F{\"{o}}rst},\ and\ \citenamefont
  {Cavalleri}}]{Mankowsky2016_RepProgPhys_79_064503}%
  \BibitemOpen
  \bibfield  {author} {\bibinfo {author} {\bibfnamefont {R.}~\bibnamefont
  {Mankowsky}}, \bibinfo {author} {\bibfnamefont {M.}~\bibnamefont
  {F{\"{o}}rst}}, \ and\ \bibinfo {author} {\bibfnamefont {A.}~\bibnamefont
  {Cavalleri}},\ }\href {\doibase 10.1088/0034-4885/79/6/064503} {\bibfield
  {journal} {\bibinfo  {journal} {Reports Prog. Phys.}\ }\textbf {\bibinfo
  {volume} {79}},\ \bibinfo {pages} {064503} (\bibinfo {year}
  {2016})}\BibitemShut {NoStop}%
\bibitem [{\citenamefont {Juraschek}\ \emph
  {et~al.}(2017{\natexlab{a}})\citenamefont {Juraschek}, \citenamefont
  {Fechner},\ and\ \citenamefont
  {Spaldin}}]{Juraschek2017_PhysRevLett.118.054101}%
  \BibitemOpen
  \bibfield  {author} {\bibinfo {author} {\bibfnamefont {D.~M.}\ \bibnamefont
  {Juraschek}}, \bibinfo {author} {\bibfnamefont {M.}~\bibnamefont {Fechner}},
  \ and\ \bibinfo {author} {\bibfnamefont {N.~A.}\ \bibnamefont {Spaldin}},\
  }\href {\doibase 10.1103/PhysRevLett.118.054101} {\bibfield  {journal}
  {\bibinfo  {journal} {Phys. Rev. Lett.}\ }\textbf {\bibinfo {volume} {118}},\
  \bibinfo {pages} {054101} (\bibinfo {year} {2017}{\natexlab{a}})}\BibitemShut
  {NoStop}%
\bibitem [{\citenamefont {F{\"{o}}rst}\ \emph
  {et~al.}(2011{\natexlab{a}})\citenamefont {F{\"{o}}rst}, \citenamefont
  {Manzoni}, \citenamefont {Kaiser}, \citenamefont {Tomioka}, \citenamefont
  {Tokura}, \citenamefont {Merlin},\ and\ \citenamefont
  {Cavalleri}}]{Forst2011_nphys2055}%
  \BibitemOpen
  \bibfield  {author} {\bibinfo {author} {\bibfnamefont {M.}~\bibnamefont
  {F{\"{o}}rst}}, \bibinfo {author} {\bibfnamefont {C.}~\bibnamefont
  {Manzoni}}, \bibinfo {author} {\bibfnamefont {S.}~\bibnamefont {Kaiser}},
  \bibinfo {author} {\bibfnamefont {Y.}~\bibnamefont {Tomioka}}, \bibinfo
  {author} {\bibfnamefont {Y.}~\bibnamefont {Tokura}}, \bibinfo {author}
  {\bibfnamefont {R.}~\bibnamefont {Merlin}}, \ and\ \bibinfo {author}
  {\bibfnamefont {A.}~\bibnamefont {Cavalleri}},\ }\href {\doibase
  10.1038/nphys2055} {\bibfield  {journal} {\bibinfo  {journal} {Nat. Phys.}\
  }\textbf {\bibinfo {volume} {7}},\ \bibinfo {pages} {854} (\bibinfo {year}
  {2011}{\natexlab{a}})}\BibitemShut {NoStop}%
\bibitem [{\citenamefont {F{\"{o}}rst}\ \emph
  {et~al.}(2011{\natexlab{b}})\citenamefont {F{\"{o}}rst}, \citenamefont
  {Tobey}, \citenamefont {Wall}, \citenamefont {Bromberger}, \citenamefont
  {Khanna}, \citenamefont {Cavalieri}, \citenamefont {Chuang}, \citenamefont
  {Lee}, \citenamefont {Moore}, \citenamefont {Schlotter}, \citenamefont
  {Turner}, \citenamefont {Krupin}, \citenamefont {Trigo}, \citenamefont
  {Zheng}, \citenamefont {Mitchell}, \citenamefont {Dhesi}, \citenamefont
  {Hill},\ and\ \citenamefont {Cavalleri}}]{Forst2011_PhysRevB.84.241104}%
  \BibitemOpen
  \bibfield  {author} {\bibinfo {author} {\bibfnamefont {M.}~\bibnamefont
  {F{\"{o}}rst}}, \bibinfo {author} {\bibfnamefont {R.~I.}\ \bibnamefont
  {Tobey}}, \bibinfo {author} {\bibfnamefont {S.}~\bibnamefont {Wall}},
  \bibinfo {author} {\bibfnamefont {H.}~\bibnamefont {Bromberger}}, \bibinfo
  {author} {\bibfnamefont {V.}~\bibnamefont {Khanna}}, \bibinfo {author}
  {\bibfnamefont {A.~L.}\ \bibnamefont {Cavalieri}}, \bibinfo {author}
  {\bibfnamefont {Y.-D.}\ \bibnamefont {Chuang}}, \bibinfo {author}
  {\bibfnamefont {W.~S.}\ \bibnamefont {Lee}}, \bibinfo {author} {\bibfnamefont
  {R.}~\bibnamefont {Moore}}, \bibinfo {author} {\bibfnamefont {W.~F.}\
  \bibnamefont {Schlotter}}, \bibinfo {author} {\bibfnamefont {J.~J.}\
  \bibnamefont {Turner}}, \bibinfo {author} {\bibfnamefont {O.}~\bibnamefont
  {Krupin}}, \bibinfo {author} {\bibfnamefont {M.}~\bibnamefont {Trigo}},
  \bibinfo {author} {\bibfnamefont {H.}~\bibnamefont {Zheng}}, \bibinfo
  {author} {\bibfnamefont {J.~F.}\ \bibnamefont {Mitchell}}, \bibinfo {author}
  {\bibfnamefont {S.~S.}\ \bibnamefont {Dhesi}}, \bibinfo {author}
  {\bibfnamefont {J.~P.}\ \bibnamefont {Hill}}, \ and\ \bibinfo {author}
  {\bibfnamefont {A.}~\bibnamefont {Cavalleri}},\ }\href {\doibase
  10.1103/PhysRevB.84.241104} {\bibfield  {journal} {\bibinfo  {journal} {Phys.
  Rev. B}\ }\textbf {\bibinfo {volume} {84}},\ \bibinfo {pages} {241104}
  (\bibinfo {year} {2011}{\natexlab{b}})}\BibitemShut {NoStop}%
\bibitem [{\citenamefont {F{\"{o}}rst}\ \emph {et~al.}(2013)\citenamefont
  {F{\"{o}}rst}, \citenamefont {Mankowsky}, \citenamefont {Bromberger},
  \citenamefont {Fritz}, \citenamefont {Lemke}, \citenamefont {Zhu},
  \citenamefont {Chollet}, \citenamefont {Tomioka}, \citenamefont {Tokura},
  \citenamefont {Merlin}, \citenamefont {Hill}, \citenamefont {Johnson},\ and\
  \citenamefont {Cavalleri}}]{Forst2013_j.ssc.2013.06.024}%
  \BibitemOpen
  \bibfield  {author} {\bibinfo {author} {\bibfnamefont {M.}~\bibnamefont
  {F{\"{o}}rst}}, \bibinfo {author} {\bibfnamefont {R.}~\bibnamefont
  {Mankowsky}}, \bibinfo {author} {\bibfnamefont {H.}~\bibnamefont
  {Bromberger}}, \bibinfo {author} {\bibfnamefont {D.}~\bibnamefont {Fritz}},
  \bibinfo {author} {\bibfnamefont {H.}~\bibnamefont {Lemke}}, \bibinfo
  {author} {\bibfnamefont {D.}~\bibnamefont {Zhu}}, \bibinfo {author}
  {\bibfnamefont {M.}~\bibnamefont {Chollet}}, \bibinfo {author} {\bibfnamefont
  {Y.}~\bibnamefont {Tomioka}}, \bibinfo {author} {\bibfnamefont
  {Y.}~\bibnamefont {Tokura}}, \bibinfo {author} {\bibfnamefont
  {R.}~\bibnamefont {Merlin}}, \bibinfo {author} {\bibfnamefont
  {J.}~\bibnamefont {Hill}}, \bibinfo {author} {\bibfnamefont {S.}~\bibnamefont
  {Johnson}}, \ and\ \bibinfo {author} {\bibfnamefont {A.}~\bibnamefont
  {Cavalleri}},\ }\href {\doibase 10.1016/j.ssc.2013.06.024} {\bibfield
  {journal} {\bibinfo  {journal} {Solid State Commun.}\ }\textbf {\bibinfo
  {volume} {169}},\ \bibinfo {pages} {24} (\bibinfo {year} {2013})}\BibitemShut
  {NoStop}%
\bibitem [{\citenamefont
  {Cavalleri}(2018)}]{Cavalleri2018_00107514.2017.1406623}%
  \BibitemOpen
  \bibfield  {author} {\bibinfo {author} {\bibfnamefont {A.}~\bibnamefont
  {Cavalleri}},\ }\href {\doibase 10.1080/00107514.2017.1406623} {\bibfield
  {journal} {\bibinfo  {journal} {Contemp. Phys.}\ }\textbf {\bibinfo {volume}
  {59}},\ \bibinfo {pages} {31} (\bibinfo {year} {2018})}\BibitemShut {NoStop}%
\bibitem [{\citenamefont {Juraschek}\ \emph
  {et~al.}(2017{\natexlab{b}})\citenamefont {Juraschek}, \citenamefont
  {Fechner}, \citenamefont {Balatsky},\ and\ \citenamefont
  {Spaldin}}]{Juraschek2017_PhysRevMaterials.1.014401}%
  \BibitemOpen
  \bibfield  {author} {\bibinfo {author} {\bibfnamefont {D.~M.}\ \bibnamefont
  {Juraschek}}, \bibinfo {author} {\bibfnamefont {M.}~\bibnamefont {Fechner}},
  \bibinfo {author} {\bibfnamefont {A.~V.}\ \bibnamefont {Balatsky}}, \ and\
  \bibinfo {author} {\bibfnamefont {N.~A.}\ \bibnamefont {Spaldin}},\ }\href
  {\doibase 10.1103/PhysRevMaterials.1.014401} {\bibfield  {journal} {\bibinfo
  {journal} {Phys. Rev. Mater.}\ }\textbf {\bibinfo {volume} {1}} (\bibinfo
  {year} {2017}{\natexlab{b}}),\ 10.1103/PhysRevMaterials.1.014401}\BibitemShut
  {NoStop}%
\bibitem [{\citenamefont {Klein}\ \emph {et~al.}(2020)\citenamefont {Klein},
  \citenamefont {Christensen},\ and\ \citenamefont
  {Fernandes}}]{Klein2020_PhysRevResearch.2.013336}%
  \BibitemOpen
  \bibfield  {author} {\bibinfo {author} {\bibfnamefont {A.}~\bibnamefont
  {Klein}}, \bibinfo {author} {\bibfnamefont {M.~H.}\ \bibnamefont
  {Christensen}}, \ and\ \bibinfo {author} {\bibfnamefont {R.~M.}\ \bibnamefont
  {Fernandes}},\ }\href {\doibase 10.1103/PhysRevResearch.2.013336} {\bibfield
  {journal} {\bibinfo  {journal} {Phys. Rev. Res.}\ }\textbf {\bibinfo {volume}
  {2}},\ \bibinfo {pages} {013336} (\bibinfo {year} {2020})}\BibitemShut
  {NoStop}%
\bibitem [{\citenamefont {Esposito}\ \emph {et~al.}(2017)\citenamefont
  {Esposito}, \citenamefont {Fechner}, \citenamefont {Mankowsky}, \citenamefont
  {Lemke}, \citenamefont {Chollet}, \citenamefont {Glownia}, \citenamefont
  {Nakamura}, \citenamefont {Kawasaki}, \citenamefont {Tokura}, \citenamefont
  {Staub}, \citenamefont {Beaud},\ and\ \citenamefont
  {F{\"{o}}rst}}]{Esposito2017_PhysRevLett.118.247601}%
  \BibitemOpen
  \bibfield  {author} {\bibinfo {author} {\bibfnamefont {V.}~\bibnamefont
  {Esposito}}, \bibinfo {author} {\bibfnamefont {M.}~\bibnamefont {Fechner}},
  \bibinfo {author} {\bibfnamefont {R.}~\bibnamefont {Mankowsky}}, \bibinfo
  {author} {\bibfnamefont {H.}~\bibnamefont {Lemke}}, \bibinfo {author}
  {\bibfnamefont {M.}~\bibnamefont {Chollet}}, \bibinfo {author} {\bibfnamefont
  {J.~M.}\ \bibnamefont {Glownia}}, \bibinfo {author} {\bibfnamefont
  {M.}~\bibnamefont {Nakamura}}, \bibinfo {author} {\bibfnamefont
  {M.}~\bibnamefont {Kawasaki}}, \bibinfo {author} {\bibfnamefont
  {Y.}~\bibnamefont {Tokura}}, \bibinfo {author} {\bibfnamefont
  {U.}~\bibnamefont {Staub}}, \bibinfo {author} {\bibfnamefont
  {P.}~\bibnamefont {Beaud}}, \ and\ \bibinfo {author} {\bibfnamefont
  {M.}~\bibnamefont {F{\"{o}}rst}},\ }\href {\doibase
  10.1103/PhysRevLett.118.247601} {\bibfield  {journal} {\bibinfo  {journal}
  {Phys. Rev. Lett.}\ }\textbf {\bibinfo {volume} {118}},\ \bibinfo {pages}
  {247601} (\bibinfo {year} {2017})}\BibitemShut {NoStop}%
\bibitem [{\citenamefont {Sch{\"{u}}tt}\ \emph {et~al.}(2018)\citenamefont
  {Sch{\"{u}}tt}, \citenamefont {Orth}, \citenamefont {Levchenko},\ and\
  \citenamefont {Fernandes}}]{Schutt2018_PhysRevB.97.035135}%
  \BibitemOpen
  \bibfield  {author} {\bibinfo {author} {\bibfnamefont {M.}~\bibnamefont
  {Sch{\"{u}}tt}}, \bibinfo {author} {\bibfnamefont {P.~P.}\ \bibnamefont
  {Orth}}, \bibinfo {author} {\bibfnamefont {A.}~\bibnamefont {Levchenko}}, \
  and\ \bibinfo {author} {\bibfnamefont {R.~M.}\ \bibnamefont {Fernandes}},\
  }\href {\doibase 10.1103/PhysRevB.97.035135} {\bibfield  {journal} {\bibinfo
  {journal} {Phys. Rev. B}\ }\textbf {\bibinfo {volume} {97}},\ \bibinfo
  {pages} {035135} (\bibinfo {year} {2018})}\BibitemShut {NoStop}%
\bibitem [{\citenamefont {Nova}\ \emph {et~al.}(2017)\citenamefont {Nova},
  \citenamefont {Cartella}, \citenamefont {Cantaluppi}, \citenamefont
  {F{\"{o}}rst}, \citenamefont {Bossini}, \citenamefont {Mikhaylovskiy},
  \citenamefont {Kimel}, \citenamefont {Merlin},\ and\ \citenamefont
  {Cavalleri}}]{Nova2017_nphys3925}%
  \BibitemOpen
  \bibfield  {author} {\bibinfo {author} {\bibfnamefont {T.~F.}\ \bibnamefont
  {Nova}}, \bibinfo {author} {\bibfnamefont {A.}~\bibnamefont {Cartella}},
  \bibinfo {author} {\bibfnamefont {A.}~\bibnamefont {Cantaluppi}}, \bibinfo
  {author} {\bibfnamefont {M.}~\bibnamefont {F{\"{o}}rst}}, \bibinfo {author}
  {\bibfnamefont {D.}~\bibnamefont {Bossini}}, \bibinfo {author} {\bibfnamefont
  {R.~V.}\ \bibnamefont {Mikhaylovskiy}}, \bibinfo {author} {\bibfnamefont
  {A.~V.}\ \bibnamefont {Kimel}}, \bibinfo {author} {\bibfnamefont
  {R.}~\bibnamefont {Merlin}}, \ and\ \bibinfo {author} {\bibfnamefont
  {A.}~\bibnamefont {Cavalleri}},\ }\href {\doibase 10.1038/nphys3925}
  {\bibfield  {journal} {\bibinfo  {journal} {Nat. Phys.}\ }\textbf {\bibinfo
  {volume} {13}},\ \bibinfo {pages} {132} (\bibinfo {year} {2017})}\BibitemShut
  {NoStop}%
\bibitem [{\citenamefont {Radaelli}(2018)}]{Radaelli2018_PhysRevB.97.085145}%
  \BibitemOpen
  \bibfield  {author} {\bibinfo {author} {\bibfnamefont {P.~G.}\ \bibnamefont
  {Radaelli}},\ }\href {\doibase 10.1103/PhysRevB.97.085145} {\bibfield
  {journal} {\bibinfo  {journal} {Phys. Rev. B}\ }\textbf {\bibinfo {volume}
  {97}},\ \bibinfo {pages} {085145} (\bibinfo {year} {2018})}\BibitemShut
  {NoStop}%
\bibitem [{\citenamefont {Fechner}\ \emph {et~al.}(2018)\citenamefont
  {Fechner}, \citenamefont {Sukhov}, \citenamefont {Chotorlishvili},
  \citenamefont {Kenel}, \citenamefont {Berakdar},\ and\ \citenamefont
  {Spaldin}}]{Fechner}%
  \BibitemOpen
  \bibfield  {author} {\bibinfo {author} {\bibfnamefont {M.}~\bibnamefont
  {Fechner}}, \bibinfo {author} {\bibfnamefont {A.}~\bibnamefont {Sukhov}},
  \bibinfo {author} {\bibfnamefont {L.}~\bibnamefont {Chotorlishvili}},
  \bibinfo {author} {\bibfnamefont {C.}~\bibnamefont {Kenel}}, \bibinfo
  {author} {\bibfnamefont {J.}~\bibnamefont {Berakdar}}, \ and\ \bibinfo
  {author} {\bibfnamefont {N.~A.}\ \bibnamefont {Spaldin}},\ }\href {\doibase
  10.1103/PhysRevMaterials.2.064401} {\bibfield  {journal} {\bibinfo  {journal}
  {Phys. Rev. Mater.}\ }\textbf {\bibinfo {volume} {2}},\ \bibinfo {pages}
  {064401} (\bibinfo {year} {2018})}\BibitemShut {NoStop}%
\bibitem [{\citenamefont {Subedi}\ \emph {et~al.}(2014)\citenamefont {Subedi},
  \citenamefont {Cavalleri},\ and\ \citenamefont
  {Georges}}]{Subedi2014_PhysRevB.89.220301}%
  \BibitemOpen
  \bibfield  {author} {\bibinfo {author} {\bibfnamefont {A.}~\bibnamefont
  {Subedi}}, \bibinfo {author} {\bibfnamefont {A.}~\bibnamefont {Cavalleri}}, \
  and\ \bibinfo {author} {\bibfnamefont {A.}~\bibnamefont {Georges}},\ }\href
  {\doibase 10.1103/PhysRevB.89.220301} {\bibfield  {journal} {\bibinfo
  {journal} {Phys. Rev. B}\ }\textbf {\bibinfo {volume} {89}},\ \bibinfo
  {pages} {220301} (\bibinfo {year} {2014})}\BibitemShut {NoStop}%
\bibitem [{\citenamefont {Juraschek}(2018)}]{Juraschek2018thesis}%
  \BibitemOpen
  \bibfield  {author} {\bibinfo {author} {\bibfnamefont {D.~M.}\ \bibnamefont
  {Juraschek}},\ }\emph {\bibinfo {title} {{Coherent Optical Phononics}}},\
  \href {\doibase 10.3929/ethz-b-000315499} {\bibinfo {type} {Doctoral
  thesis}},\ \bibinfo  {school} {ETH Zurich} (\bibinfo {year}
  {2018})\BibitemShut {NoStop}%
\bibitem [{\citenamefont {von Hoegen}\ \emph {et~al.}(2018)\citenamefont {von
  Hoegen}, \citenamefont {Mankowsky}, \citenamefont {Fechner}, \citenamefont
  {F{\"{o}}rst},\ and\ \citenamefont {Cavalleri}}]{vonHoegen2018_nature25484}%
  \BibitemOpen
  \bibfield  {author} {\bibinfo {author} {\bibfnamefont {A.}~\bibnamefont {von
  Hoegen}}, \bibinfo {author} {\bibfnamefont {R.}~\bibnamefont {Mankowsky}},
  \bibinfo {author} {\bibfnamefont {M.}~\bibnamefont {Fechner}}, \bibinfo
  {author} {\bibfnamefont {M.}~\bibnamefont {F{\"{o}}rst}}, \ and\ \bibinfo
  {author} {\bibfnamefont {A.}~\bibnamefont {Cavalleri}},\ }\href {\doibase
  10.1038/nature25484} {\bibfield  {journal} {\bibinfo  {journal} {Nature}\
  }\textbf {\bibinfo {volume} {555}},\ \bibinfo {pages} {79} (\bibinfo {year}
  {2018})}\BibitemShut {NoStop}%
\bibitem [{\citenamefont {Gemmer}\ \emph {et~al.}(2009)\citenamefont {Gemmer},
  \citenamefont {Michel},\ and\ \citenamefont {Mahler}}]{Gemmer2009}%
  \BibitemOpen
  \bibfield  {author} {\bibinfo {author} {\bibfnamefont {J.}~\bibnamefont
  {Gemmer}}, \bibinfo {author} {\bibfnamefont {M.}~\bibnamefont {Michel}}, \
  and\ \bibinfo {author} {\bibfnamefont {G.}~\bibnamefont {Mahler}},\ }\href
  {\doibase 10.1007/978-3-540-70510-9} {\emph {\bibinfo {title} {Quantum
  thermodynamics: Emergence of thermodynamic behavior within composite quantum
  systems}}},\ Vol.\ \bibinfo {volume} {784}\ (\bibinfo  {publisher}
  {Springer},\ \bibinfo {year} {2009})\BibitemShut {NoStop}%
\bibitem [{\citenamefont {Vinjanampathy}\ and\ \citenamefont
  {Anders}(2016)}]{Vinjanampathy2016}%
  \BibitemOpen
  \bibfield  {author} {\bibinfo {author} {\bibfnamefont {S.}~\bibnamefont
  {Vinjanampathy}}\ and\ \bibinfo {author} {\bibfnamefont {J.}~\bibnamefont
  {Anders}},\ }\href {\doibase 10.1080/00107514.2016.1201896} {\bibfield
  {journal} {\bibinfo  {journal} {Contemporary Physics}\ }\textbf {\bibinfo
  {volume} {57}},\ \bibinfo {pages} {545} (\bibinfo {year} {2016})}\BibitemShut
  {NoStop}%
\bibitem [{\citenamefont {Alicki}\ and\ \citenamefont
  {Kosloff}(2018)}]{Alicki2019}%
  \BibitemOpen
  \bibfield  {author} {\bibinfo {author} {\bibfnamefont {R.}~\bibnamefont
  {Alicki}}\ and\ \bibinfo {author} {\bibfnamefont {R.}~\bibnamefont
  {Kosloff}},\ }in\ \href {\doibase 10.1007/978-3-319-99046-0_1} {\emph
  {\bibinfo {booktitle} {Thermodynamics in the Quantum Regime. Fundamental
  Theories of Physics}}},\ Vol.\ \bibinfo {volume} {195},\ \bibinfo {editor}
  {edited by\ \bibinfo {editor} {\bibfnamefont {F.}~\bibnamefont {Binder}},
  \bibinfo {editor} {\bibfnamefont {L.~A.}\ \bibnamefont {Correa}}, \bibinfo
  {editor} {\bibfnamefont {C.}~\bibnamefont {Gogolin}}, \bibinfo {editor}
  {\bibfnamefont {J.}~\bibnamefont {Anders}}, \ and\ \bibinfo {editor}
  {\bibfnamefont {G.}~\bibnamefont {Adesso}}}\ (\bibinfo  {publisher}
  {Springer},\ \bibinfo {address} {Cham},\ \bibinfo {year} {2018})\
  Chap.~\bibinfo {chapter} {1}, p.\ \bibinfo {pages} {127}\BibitemShut
  {NoStop}%
\bibitem [{\citenamefont {Jarzynski}(1997)}]{Jarzynski1997}%
  \BibitemOpen
  \bibfield  {author} {\bibinfo {author} {\bibfnamefont {C.}~\bibnamefont
  {Jarzynski}},\ }\href {\doibase 10.1103/PhysRevLett.78.2690} {\bibfield
  {journal} {\bibinfo  {journal} {Phys. Rev. Lett.}\ }\textbf {\bibinfo
  {volume} {78}},\ \bibinfo {pages} {2690} (\bibinfo {year}
  {1997})}\BibitemShut {NoStop}%
\bibitem [{\citenamefont {Campisi}\ \emph
  {et~al.}(2011{\natexlab{a}})\citenamefont {Campisi}, \citenamefont
  {H{\"{a}}nggi},\ and\ \citenamefont {Talkner}}]{Campisi}%
  \BibitemOpen
  \bibfield  {author} {\bibinfo {author} {\bibfnamefont {M.}~\bibnamefont
  {Campisi}}, \bibinfo {author} {\bibfnamefont {P.}~\bibnamefont
  {H{\"{a}}nggi}}, \ and\ \bibinfo {author} {\bibfnamefont {P.}~\bibnamefont
  {Talkner}},\ }\href {\doibase 10.1103/RevModPhys.83.771} {\bibfield
  {journal} {\bibinfo  {journal} {Rev. Mod. Phys.}\ }\textbf {\bibinfo {volume}
  {83}},\ \bibinfo {pages} {771} (\bibinfo {year}
  {2011}{\natexlab{a}})}\BibitemShut {NoStop}%
\bibitem [{\citenamefont {Campisi}\ \emph
  {et~al.}(2011{\natexlab{b}})\citenamefont {Campisi}, \citenamefont
  {H{\"{a}}nggi},\ and\ \citenamefont {Talkner}}]{Campisi2011erratum}%
  \BibitemOpen
  \bibfield  {author} {\bibinfo {author} {\bibfnamefont {M.}~\bibnamefont
  {Campisi}}, \bibinfo {author} {\bibfnamefont {P.}~\bibnamefont
  {H{\"{a}}nggi}}, \ and\ \bibinfo {author} {\bibfnamefont {P.}~\bibnamefont
  {Talkner}},\ }\href {\doibase 10.1103/RevModPhys.83.1653} {\bibfield
  {journal} {\bibinfo  {journal} {Rev. Mod. Phys.}\ }\textbf {\bibinfo {volume}
  {83}},\ \bibinfo {pages} {1653} (\bibinfo {year}
  {2011}{\natexlab{b}})}\BibitemShut {NoStop}%
\bibitem [{\citenamefont {Talkner}\ \emph {et~al.}(2007)\citenamefont
  {Talkner}, \citenamefont {Lutz},\ and\ \citenamefont
  {H{\"{a}}nggi}}]{Talkner2007}%
  \BibitemOpen
  \bibfield  {author} {\bibinfo {author} {\bibfnamefont {P.}~\bibnamefont
  {Talkner}}, \bibinfo {author} {\bibfnamefont {E.}~\bibnamefont {Lutz}}, \
  and\ \bibinfo {author} {\bibfnamefont {P.}~\bibnamefont {H{\"{a}}nggi}},\
  }\href {\doibase 10.1103/PhysRevE.75.050102} {\bibfield  {journal} {\bibinfo
  {journal} {Phys. Rev. E}\ }\textbf {\bibinfo {volume} {75}},\ \bibinfo
  {pages} {050102} (\bibinfo {year} {2007})}\BibitemShut {NoStop}%
\bibitem [{\citenamefont {Sokolov}(2014)}]{Sokolov2014}%
  \BibitemOpen
  \bibfield  {author} {\bibinfo {author} {\bibfnamefont {I.~M.}\ \bibnamefont
  {Sokolov}},\ }\href {\doibase 10.1038/nphys2831} {\bibfield  {journal}
  {\bibinfo  {journal} {Nat. Phys.}\ }\textbf {\bibinfo {volume} {10}},\
  \bibinfo {pages} {7} (\bibinfo {year} {2014})}\BibitemShut {NoStop}%
\bibitem [{\citenamefont {Bochkov}\ and\ \citenamefont
  {Kuzovlev}(2013)}]{Bochkov2013}%
  \BibitemOpen
  \bibfield  {author} {\bibinfo {author} {\bibfnamefont {G.~N.}\ \bibnamefont
  {Bochkov}}\ and\ \bibinfo {author} {\bibfnamefont {Y.~E.}\ \bibnamefont
  {Kuzovlev}},\ }\href {\doibase 10.3367/UFNe.0183.201306d.0617} {\bibfield
  {journal} {\bibinfo  {journal} {Physics-Uspekhi}\ }\textbf {\bibinfo {volume}
  {56}},\ \bibinfo {pages} {590} (\bibinfo {year} {2013})}\BibitemShut
  {NoStop}%
\bibitem [{\citenamefont {Deffner}\ and\ \citenamefont
  {Lutz}(2010)}]{Deffner2010}%
  \BibitemOpen
  \bibfield  {author} {\bibinfo {author} {\bibfnamefont {S.}~\bibnamefont
  {Deffner}}\ and\ \bibinfo {author} {\bibfnamefont {E.}~\bibnamefont {Lutz}},\
  }\href {\doibase 10.1103/PhysRevLett.105.170402} {\bibfield  {journal}
  {\bibinfo  {journal} {Phys. Rev. Lett.}\ }\textbf {\bibinfo {volume} {105}},\
  \bibinfo {pages} {170402} (\bibinfo {year} {2010})}\BibitemShut {NoStop}%
\bibitem [{\citenamefont {Quan}\ and\ \citenamefont
  {Cucchietti}(2009)}]{Quan2009}%
  \BibitemOpen
  \bibfield  {author} {\bibinfo {author} {\bibfnamefont {H.~T.}\ \bibnamefont
  {Quan}}\ and\ \bibinfo {author} {\bibfnamefont {F.~M.}\ \bibnamefont
  {Cucchietti}},\ }\href {\doibase 10.1103/PhysRevE.79.031101} {\bibfield
  {journal} {\bibinfo  {journal} {Phys. Rev. E}\ }\textbf {\bibinfo {volume}
  {79}},\ \bibinfo {pages} {031101} (\bibinfo {year} {2009})}\BibitemShut
  {NoStop}%
\bibitem [{\citenamefont {Linden}\ \emph {et~al.}(2010)\citenamefont {Linden},
  \citenamefont {Popescu},\ and\ \citenamefont {Skrzypczyk}}]{Linden2010}%
  \BibitemOpen
  \bibfield  {author} {\bibinfo {author} {\bibfnamefont {N.}~\bibnamefont
  {Linden}}, \bibinfo {author} {\bibfnamefont {S.}~\bibnamefont {Popescu}}, \
  and\ \bibinfo {author} {\bibfnamefont {P.}~\bibnamefont {Skrzypczyk}},\
  }\href {\doibase 10.1103/PhysRevLett.105.130401} {\bibfield  {journal}
  {\bibinfo  {journal} {Phys. Rev. Lett.}\ }\textbf {\bibinfo {volume} {105}},\
  \bibinfo {pages} {130401} (\bibinfo {year} {2010})}\BibitemShut {NoStop}%
\bibitem [{\citenamefont {H{\"{u}}bner}\ \emph {et~al.}(2014)\citenamefont
  {H{\"{u}}bner}, \citenamefont {Lefkidis}, \citenamefont {Dong}, \citenamefont
  {Chaudhuri}, \citenamefont {Chotorlishvili},\ and\ \citenamefont
  {Berakdar}}]{Hubner2014}%
  \BibitemOpen
  \bibfield  {author} {\bibinfo {author} {\bibfnamefont {W.}~\bibnamefont
  {H{\"{u}}bner}}, \bibinfo {author} {\bibfnamefont {G.}~\bibnamefont
  {Lefkidis}}, \bibinfo {author} {\bibfnamefont {C.~D.}\ \bibnamefont {Dong}},
  \bibinfo {author} {\bibfnamefont {D.}~\bibnamefont {Chaudhuri}}, \bibinfo
  {author} {\bibfnamefont {L.}~\bibnamefont {Chotorlishvili}}, \ and\ \bibinfo
  {author} {\bibfnamefont {J.}~\bibnamefont {Berakdar}},\ }\href {\doibase
  10.1103/PhysRevB.90.024401} {\bibfield  {journal} {\bibinfo  {journal} {Phys.
  Rev. B}\ }\textbf {\bibinfo {volume} {90}},\ \bibinfo {pages} {024401}
  (\bibinfo {year} {2014})}\BibitemShut {NoStop}%
\bibitem [{\citenamefont {Chotorlishvili}\ \emph {et~al.}(2016)\citenamefont
  {Chotorlishvili}, \citenamefont {Azimi}, \citenamefont {Stagraczy{\'{n}}ski},
  \citenamefont {Toklikishvili}, \citenamefont {Sch{\"{u}}ler},\ and\
  \citenamefont {Berakdar}}]{Chotorlishvili2016_10.1103/PhysRevE.94.032116}%
  \BibitemOpen
  \bibfield  {author} {\bibinfo {author} {\bibfnamefont {L.}~\bibnamefont
  {Chotorlishvili}}, \bibinfo {author} {\bibfnamefont {M.}~\bibnamefont
  {Azimi}}, \bibinfo {author} {\bibfnamefont {S.}~\bibnamefont
  {Stagraczy{\'{n}}ski}}, \bibinfo {author} {\bibfnamefont {Z.}~\bibnamefont
  {Toklikishvili}}, \bibinfo {author} {\bibfnamefont {M.}~\bibnamefont
  {Sch{\"{u}}ler}}, \ and\ \bibinfo {author} {\bibfnamefont {J.}~\bibnamefont
  {Berakdar}},\ }\href {\doibase 10.1103/PhysRevE.94.032116} {\bibfield
  {journal} {\bibinfo  {journal} {Phys. Rev. E}\ }\textbf {\bibinfo {volume}
  {94}},\ \bibinfo {pages} {032116} (\bibinfo {year} {2016})}\BibitemShut
  {NoStop}%
\bibitem [{\citenamefont {Fusco}\ \emph {et~al.}(2016)\citenamefont {Fusco},
  \citenamefont {Paternostro},\ and\ \citenamefont {{De Chiara}}}]{Fusco}%
  \BibitemOpen
  \bibfield  {author} {\bibinfo {author} {\bibfnamefont {L.}~\bibnamefont
  {Fusco}}, \bibinfo {author} {\bibfnamefont {M.}~\bibnamefont {Paternostro}},
  \ and\ \bibinfo {author} {\bibfnamefont {G.}~\bibnamefont {{De Chiara}}},\
  }\href {\doibase 10.1103/PhysRevE.94.052122} {\bibfield  {journal} {\bibinfo
  {journal} {Phys. Rev. E}\ }\textbf {\bibinfo {volume} {94}},\ \bibinfo
  {pages} {052122} (\bibinfo {year} {2016})}\BibitemShut {NoStop}%
\bibitem [{\citenamefont {T{\"{u}}rkpen{\c{c}}e}\ \emph
  {et~al.}(2017)\citenamefont {T{\"{u}}rkpen{\c{c}}e}, \citenamefont
  {Altintas}, \citenamefont {Paternostro},\ and\ \citenamefont
  {M{\"{u}}stecaplioğlu}}]{Altintas}%
  \BibitemOpen
  \bibfield  {author} {\bibinfo {author} {\bibfnamefont {D.}~\bibnamefont
  {T{\"{u}}rkpen{\c{c}}e}}, \bibinfo {author} {\bibfnamefont {F.}~\bibnamefont
  {Altintas}}, \bibinfo {author} {\bibfnamefont {M.}~\bibnamefont
  {Paternostro}}, \ and\ \bibinfo {author} {\bibfnamefont {{\"{O}}.~E.}\
  \bibnamefont {M{\"{u}}stecaplioğlu}},\ }\href {\doibase
  10.1209/0295-5075/117/50002} {\bibfield  {journal} {\bibinfo  {journal}
  {Europhysics Lett.}\ }\textbf {\bibinfo {volume} {117}},\ \bibinfo {pages}
  {50002} (\bibinfo {year} {2017})}\BibitemShut {NoStop}%
\bibitem [{\citenamefont {Hardal}\ \emph {et~al.}(2017)\citenamefont {Hardal},
  \citenamefont {Aslan}, \citenamefont {Wilson},\ and\ \citenamefont
  {M{\"u}stecapl{\i}o{\u{g}}lu}}]{Hardal2017}%
  \BibitemOpen
  \bibfield  {author} {\bibinfo {author} {\bibfnamefont {A.~{\"U}.}\
  \bibnamefont {Hardal}}, \bibinfo {author} {\bibfnamefont {N.}~\bibnamefont
  {Aslan}}, \bibinfo {author} {\bibfnamefont {C.}~\bibnamefont {Wilson}}, \
  and\ \bibinfo {author} {\bibfnamefont {{\"O}.~E.}\ \bibnamefont
  {M{\"u}stecapl{\i}o{\u{g}}lu}},\ }\href {\doibase 10.1103/PhysRevE.96.062120}
  {\bibfield  {journal} {\bibinfo  {journal} {Physical Review E}\ }\textbf
  {\bibinfo {volume} {96}},\ \bibinfo {pages} {062120} (\bibinfo {year}
  {2017})}\BibitemShut {NoStop}%
\bibitem [{\citenamefont {Huang}\ \emph {et~al.}(2018)\citenamefont {Huang},
  \citenamefont {Sun}, \citenamefont {Guo},\ and\ \citenamefont
  {Yu}}]{Huang2018}%
  \BibitemOpen
  \bibfield  {author} {\bibinfo {author} {\bibfnamefont {X.}~\bibnamefont
  {Huang}}, \bibinfo {author} {\bibfnamefont {Q.}~\bibnamefont {Sun}}, \bibinfo
  {author} {\bibfnamefont {D.}~\bibnamefont {Guo}}, \ and\ \bibinfo {author}
  {\bibfnamefont {Q.}~\bibnamefont {Yu}},\ }\href {\doibase
  10.1016/j.physa.2017.09.104} {\bibfield  {journal} {\bibinfo  {journal}
  {Physica A}\ }\textbf {\bibinfo {volume} {491}},\ \bibinfo {pages} {604}
  (\bibinfo {year} {2018})}\BibitemShut {NoStop}%
\bibitem [{\citenamefont {Esposito}\ \emph {et~al.}(2010)\citenamefont
  {Esposito}, \citenamefont {Kawai}, \citenamefont {Lindenberg},\ and\
  \citenamefont {{Van den Broeck}}}]{Esposito2010}%
  \BibitemOpen
  \bibfield  {author} {\bibinfo {author} {\bibfnamefont {M.}~\bibnamefont
  {Esposito}}, \bibinfo {author} {\bibfnamefont {R.}~\bibnamefont {Kawai}},
  \bibinfo {author} {\bibfnamefont {K.}~\bibnamefont {Lindenberg}}, \ and\
  \bibinfo {author} {\bibfnamefont {C.}~\bibnamefont {{Van den Broeck}}},\
  }\href {\doibase 10.1103/PhysRevE.81.041106} {\bibfield  {journal} {\bibinfo
  {journal} {Phys. Rev. E}\ }\textbf {\bibinfo {volume} {81}},\ \bibinfo
  {pages} {041106} (\bibinfo {year} {2010})}\BibitemShut {NoStop}%
\bibitem [{\citenamefont {Abah}\ \emph {et~al.}(2012)\citenamefont {Abah},
  \citenamefont {Ro{\ss}nagel}, \citenamefont {Jacob}, \citenamefont {Deffner},
  \citenamefont {Schmidt-Kaler}, \citenamefont {Singer},\ and\ \citenamefont
  {Lutz}}]{Abah2012}%
  \BibitemOpen
  \bibfield  {author} {\bibinfo {author} {\bibfnamefont {O.}~\bibnamefont
  {Abah}}, \bibinfo {author} {\bibfnamefont {J.}~\bibnamefont {Ro{\ss}nagel}},
  \bibinfo {author} {\bibfnamefont {G.}~\bibnamefont {Jacob}}, \bibinfo
  {author} {\bibfnamefont {S.}~\bibnamefont {Deffner}}, \bibinfo {author}
  {\bibfnamefont {F.}~\bibnamefont {Schmidt-Kaler}}, \bibinfo {author}
  {\bibfnamefont {K.}~\bibnamefont {Singer}}, \ and\ \bibinfo {author}
  {\bibfnamefont {E.}~\bibnamefont {Lutz}},\ }\href {\doibase
  10.1103/PhysRevLett.109.203006} {\bibfield  {journal} {\bibinfo  {journal}
  {Phys. Rev. Lett.}\ }\textbf {\bibinfo {volume} {109}},\ \bibinfo {pages}
  {203006} (\bibinfo {year} {2012})}\BibitemShut {NoStop}%
\bibitem [{\citenamefont {Zagoskin}\ \emph {et~al.}(2012)\citenamefont
  {Zagoskin}, \citenamefont {Savel'ev}, \citenamefont {Nori},\ and\
  \citenamefont {Kusmartsev}}]{Zagoskin2012}%
  \BibitemOpen
  \bibfield  {author} {\bibinfo {author} {\bibfnamefont {A.~M.}\ \bibnamefont
  {Zagoskin}}, \bibinfo {author} {\bibfnamefont {S.}~\bibnamefont {Savel'ev}},
  \bibinfo {author} {\bibfnamefont {F.}~\bibnamefont {Nori}}, \ and\ \bibinfo
  {author} {\bibfnamefont {F.~V.}\ \bibnamefont {Kusmartsev}},\ }\href
  {\doibase 10.1103/PhysRevB.86.014501} {\bibfield  {journal} {\bibinfo
  {journal} {Phys. Rev. B}\ }\textbf {\bibinfo {volume} {86}},\ \bibinfo
  {pages} {014501} (\bibinfo {year} {2012})}\BibitemShut {NoStop}%
\bibitem [{\citenamefont {Abah}\ and\ \citenamefont {Lutz}(2014)}]{Abah2014}%
  \BibitemOpen
  \bibfield  {author} {\bibinfo {author} {\bibfnamefont {O.}~\bibnamefont
  {Abah}}\ and\ \bibinfo {author} {\bibfnamefont {E.}~\bibnamefont {Lutz}},\
  }\href {\doibase 10.1209/0295-5075/106/20001} {\bibfield  {journal} {\bibinfo
   {journal} {Europhysics Lett.}\ }\textbf {\bibinfo {volume} {106}},\ \bibinfo
  {pages} {20001} (\bibinfo {year} {2014})}\BibitemShut {NoStop}%
\bibitem [{\citenamefont {Ro{\ss}nagel}\ \emph {et~al.}(2014)\citenamefont
  {Ro{\ss}nagel}, \citenamefont {Abah}, \citenamefont {Schmidt-Kaler},
  \citenamefont {Singer},\ and\ \citenamefont {Lutz}}]{Rossnagel2014}%
  \BibitemOpen
  \bibfield  {author} {\bibinfo {author} {\bibfnamefont {J.}~\bibnamefont
  {Ro{\ss}nagel}}, \bibinfo {author} {\bibfnamefont {O.}~\bibnamefont {Abah}},
  \bibinfo {author} {\bibfnamefont {F.}~\bibnamefont {Schmidt-Kaler}}, \bibinfo
  {author} {\bibfnamefont {K.}~\bibnamefont {Singer}}, \ and\ \bibinfo {author}
  {\bibfnamefont {E.}~\bibnamefont {Lutz}},\ }\href {\doibase
  10.1103/PhysRevLett.112.030602} {\bibfield  {journal} {\bibinfo  {journal}
  {Phys. Rev. Lett.}\ }\textbf {\bibinfo {volume} {112}},\ \bibinfo {pages}
  {030602} (\bibinfo {year} {2014})}\BibitemShut {NoStop}%
\bibitem [{\citenamefont {del Campo}\ \emph {et~al.}(2014)\citenamefont {del
  Campo}, \citenamefont {Goold},\ and\ \citenamefont
  {Paternostro}}]{DelCampo2014}%
  \BibitemOpen
  \bibfield  {author} {\bibinfo {author} {\bibfnamefont {A.}~\bibnamefont {del
  Campo}}, \bibinfo {author} {\bibfnamefont {J.}~\bibnamefont {Goold}}, \ and\
  \bibinfo {author} {\bibfnamefont {M.}~\bibnamefont {Paternostro}},\ }\href
  {\doibase 10.1038/srep06208} {\bibfield  {journal} {\bibinfo  {journal} {Sci.
  Rep.}\ }\textbf {\bibinfo {volume} {4}},\ \bibinfo {pages} {6208} (\bibinfo
  {year} {2014})}\BibitemShut {NoStop}%
\bibitem [{\citenamefont {Mukherjee}\ \emph {et~al.}(2016)\citenamefont
  {Mukherjee}, \citenamefont {Niedenzu}, \citenamefont {Kofman},\ and\
  \citenamefont {Kurizki}}]{Mukherjee2016}%
  \BibitemOpen
  \bibfield  {author} {\bibinfo {author} {\bibfnamefont {V.}~\bibnamefont
  {Mukherjee}}, \bibinfo {author} {\bibfnamefont {W.}~\bibnamefont {Niedenzu}},
  \bibinfo {author} {\bibfnamefont {A.~G.}\ \bibnamefont {Kofman}}, \ and\
  \bibinfo {author} {\bibfnamefont {G.}~\bibnamefont {Kurizki}},\ }\href
  {\doibase 10.1103/PhysRevE.94.062109} {\bibfield  {journal} {\bibinfo
  {journal} {Phys. Rev. E}\ }\textbf {\bibinfo {volume} {94}},\ \bibinfo
  {pages} {062109} (\bibinfo {year} {2016})}\BibitemShut {NoStop}%
\bibitem [{\citenamefont {Stefanatos}\ and\ \citenamefont
  {Paspalakis}(2018)}]{Stefanatos2018}%
  \BibitemOpen
  \bibfield  {author} {\bibinfo {author} {\bibfnamefont {D.}~\bibnamefont
  {Stefanatos}}\ and\ \bibinfo {author} {\bibfnamefont {E.}~\bibnamefont
  {Paspalakis}},\ }\href {\doibase 10.1088/1367-2630/aac122} {\bibfield
  {journal} {\bibinfo  {journal} {New J. Phys.}\ }\textbf {\bibinfo {volume}
  {20}},\ \bibinfo {pages} {055009} (\bibinfo {year} {2018})}\BibitemShut
  {NoStop}%
\bibitem [{\citenamefont {Veps{\"{a}}l{\"{a}}inen}\ \emph
  {et~al.}(2017)\citenamefont {Veps{\"{a}}l{\"{a}}inen}, \citenamefont
  {Danilin},\ and\ \citenamefont {Paraoanu}}]{Vepsalainen2017}%
  \BibitemOpen
  \bibfield  {author} {\bibinfo {author} {\bibfnamefont {A.}~\bibnamefont
  {Veps{\"{a}}l{\"{a}}inen}}, \bibinfo {author} {\bibfnamefont
  {S.}~\bibnamefont {Danilin}}, \ and\ \bibinfo {author} {\bibfnamefont
  {S.}~\bibnamefont {Paraoanu}},\ }\href {https://arxiv.org/abs/1709.03731} {\
  (\bibinfo {year} {2017})},\ \Eprint {http://arxiv.org/abs/1709.03731}
  {arXiv:1709.03731} \BibitemShut {NoStop}%
\bibitem [{\citenamefont {Man}\ \emph {et~al.}(2016)\citenamefont {Man},
  \citenamefont {An},\ and\ \citenamefont {Xia}}]{Nguyen}%
  \BibitemOpen
  \bibfield  {author} {\bibinfo {author} {\bibfnamefont {Z.-X.}\ \bibnamefont
  {Man}}, \bibinfo {author} {\bibfnamefont {N.~B.}\ \bibnamefont {An}}, \ and\
  \bibinfo {author} {\bibfnamefont {Y.-J.}\ \bibnamefont {Xia}},\ }\href
  {\doibase 10.1103/PhysRevE.94.042135} {\bibfield  {journal} {\bibinfo
  {journal} {Phys. Rev. E}\ }\textbf {\bibinfo {volume} {94}},\ \bibinfo
  {pages} {042135} (\bibinfo {year} {2016})}\BibitemShut {NoStop}%
\bibitem [{\citenamefont {Stefanatos}(2017)}]{Stefanatos}%
  \BibitemOpen
  \bibfield  {author} {\bibinfo {author} {\bibfnamefont {D.}~\bibnamefont
  {Stefanatos}},\ }\href {\doibase 10.1109/TAC.2017.2684083} {\bibfield
  {journal} {\bibinfo  {journal} {IEEE Trans. Automat. Contr.}\ }\textbf
  {\bibinfo {volume} {62}},\ \bibinfo {pages} {4290} (\bibinfo {year}
  {2017})}\BibitemShut {NoStop}%
\bibitem [{\citenamefont {Man}\ and\ \citenamefont {Xia}(2017)}]{Xia}%
  \BibitemOpen
  \bibfield  {author} {\bibinfo {author} {\bibfnamefont {Z.-X.}\ \bibnamefont
  {Man}}\ and\ \bibinfo {author} {\bibfnamefont {Y.-J.}\ \bibnamefont {Xia}},\
  }\href {\doibase 10.1103/PhysRevE.96.012122} {\bibfield  {journal} {\bibinfo
  {journal} {Phys. Rev. E}\ }\textbf {\bibinfo {volume} {96}},\ \bibinfo
  {pages} {012122} (\bibinfo {year} {2017})}\BibitemShut {NoStop}%
\bibitem [{\citenamefont {Quan}\ \emph {et~al.}(2007)\citenamefont {Quan},
  \citenamefont {Liu}, \citenamefont {Sun},\ and\ \citenamefont
  {Nori}}]{Quan2007}%
  \BibitemOpen
  \bibfield  {author} {\bibinfo {author} {\bibfnamefont {H.~T.}\ \bibnamefont
  {Quan}}, \bibinfo {author} {\bibfnamefont {Y.-x.}\ \bibnamefont {Liu}},
  \bibinfo {author} {\bibfnamefont {C.~P.}\ \bibnamefont {Sun}}, \ and\
  \bibinfo {author} {\bibfnamefont {F.}~\bibnamefont {Nori}},\ }\href {\doibase
  10.1103/PhysRevE.76.031105} {\bibfield  {journal} {\bibinfo  {journal} {Phys.
  Rev. E}\ }\textbf {\bibinfo {volume} {76}},\ \bibinfo {pages} {031105}
  (\bibinfo {year} {2007})}\BibitemShut {NoStop}%
\bibitem [{\citenamefont {Lucia}\ and\ \citenamefont
  {A{\c{c}}ıkkalp}(2017)}]{Lucia}%
  \BibitemOpen
  \bibfield  {author} {\bibinfo {author} {\bibfnamefont {U.}~\bibnamefont
  {Lucia}}\ and\ \bibinfo {author} {\bibfnamefont {E.}~\bibnamefont
  {A{\c{c}}ıkkalp}},\ }\href {\doibase 10.1016/j.chemphys.2017.07.009}
  {\bibfield  {journal} {\bibinfo  {journal} {Chem. Phys.}\ }\textbf {\bibinfo
  {volume} {494}},\ \bibinfo {pages} {47} (\bibinfo {year} {2017})}\BibitemShut
  {NoStop}%
\bibitem [{\citenamefont {Fadaie}\ \emph {et~al.}(2018)\citenamefont {Fadaie},
  \citenamefont {Yunt},\ and\ \citenamefont
  {M{\"u}stecapl{\i}o{\u{g}}lu}}]{Fadaie2018}%
  \BibitemOpen
  \bibfield  {author} {\bibinfo {author} {\bibfnamefont {M.}~\bibnamefont
  {Fadaie}}, \bibinfo {author} {\bibfnamefont {E.}~\bibnamefont {Yunt}}, \ and\
  \bibinfo {author} {\bibfnamefont {{\"O}.~E.}\ \bibnamefont
  {M{\"u}stecapl{\i}o{\u{g}}lu}},\ }\href {\doibase 10.1103/PhysRevE.98.052124}
  {\bibfield  {journal} {\bibinfo  {journal} {Physical Review E}\ }\textbf
  {\bibinfo {volume} {98}},\ \bibinfo {pages} {052124} (\bibinfo {year}
  {2018})}\BibitemShut {NoStop}%
\bibitem [{\citenamefont {Rom{\'a}n-Ancheyta}\ \emph
  {et~al.}(2019)\citenamefont {Rom{\'a}n-Ancheyta}, \citenamefont
  {{\c{C}}akmak},\ and\ \citenamefont
  {M{\"u}stecapl{\i}o{\u{g}}lu}}]{Roman2019}%
  \BibitemOpen
  \bibfield  {author} {\bibinfo {author} {\bibfnamefont {R.}~\bibnamefont
  {Rom{\'a}n-Ancheyta}}, \bibinfo {author} {\bibfnamefont {B.}~\bibnamefont
  {{\c{C}}akmak}}, \ and\ \bibinfo {author} {\bibfnamefont {{\"O}.~E.}\
  \bibnamefont {M{\"u}stecapl{\i}o{\u{g}}lu}},\ }\href {\doibase
  10.1088/2058-9565/ab5e4f} {\bibfield  {journal} {\bibinfo  {journal} {Quantum
  Science and Technology}\ }\textbf {\bibinfo {volume} {5}},\ \bibinfo {pages}
  {015003} (\bibinfo {year} {2019})}\BibitemShut {NoStop}%
\bibitem [{\citenamefont {Silveri}\ \emph {et~al.}(2017)\citenamefont
  {Silveri}, \citenamefont {Tuorila}, \citenamefont {Thuneberg},\ and\
  \citenamefont {Paraoanu}}]{Silveri}%
  \BibitemOpen
  \bibfield  {author} {\bibinfo {author} {\bibfnamefont {M.~P.}\ \bibnamefont
  {Silveri}}, \bibinfo {author} {\bibfnamefont {J.~A.}\ \bibnamefont
  {Tuorila}}, \bibinfo {author} {\bibfnamefont {E.~V.}\ \bibnamefont
  {Thuneberg}}, \ and\ \bibinfo {author} {\bibfnamefont {G.~S.}\ \bibnamefont
  {Paraoanu}},\ }\href {\doibase 10.1088/1361-6633/aa5170} {\bibfield
  {journal} {\bibinfo  {journal} {Reports Prog. Phys.}\ }\textbf {\bibinfo
  {volume} {80}},\ \bibinfo {pages} {056002} (\bibinfo {year}
  {2017})}\BibitemShut {NoStop}%
\bibitem [{\citenamefont {del Campo}\ \emph {et~al.}(2018)\citenamefont {del
  Campo}, \citenamefont {Chenu}, \citenamefont {Deng},\ and\ \citenamefont
  {Wu}}]{delCampo2018}%
  \BibitemOpen
  \bibfield  {author} {\bibinfo {author} {\bibfnamefont {A.}~\bibnamefont {del
  Campo}}, \bibinfo {author} {\bibfnamefont {A.}~\bibnamefont {Chenu}},
  \bibinfo {author} {\bibfnamefont {S.}~\bibnamefont {Deng}}, \ and\ \bibinfo
  {author} {\bibfnamefont {H.}~\bibnamefont {Wu}},\ }in\ \href {\doibase
  10.1007/978-3-319-99046-0_5} {\emph {\bibinfo {booktitle} {Thermodynamics in
  the Quantum Regime. Fundamental Theories of Physics}}},\ Vol.\ \bibinfo
  {volume} {195},\ \bibinfo {editor} {edited by\ \bibinfo {editor}
  {\bibfnamefont {F.}~\bibnamefont {Binder}}, \bibinfo {editor} {\bibfnamefont
  {L.~A.}\ \bibnamefont {Correa}}, \bibinfo {editor} {\bibfnamefont
  {C.}~\bibnamefont {Gogolin}}, \bibinfo {editor} {\bibfnamefont
  {J.}~\bibnamefont {Anders}}, \ and\ \bibinfo {editor} {\bibfnamefont
  {G.}~\bibnamefont {Adesso}}}\ (\bibinfo  {publisher} {Springer},\ \bibinfo
  {address} {Cham},\ \bibinfo {year} {2018})\ Chap.~\bibinfo {chapter} {5}, p.\
  \bibinfo {pages} {127}\BibitemShut {NoStop}%
\bibitem [{\citenamefont {Kosloff}(2019)}]{Kosloff2019}%
  \BibitemOpen
  \bibfield  {author} {\bibinfo {author} {\bibfnamefont {R.}~\bibnamefont
  {Kosloff}},\ }\href {\doibase 10.1063/1.5096173} {\bibfield  {journal}
  {\bibinfo  {journal} {J. Chem. Phys.}\ }\textbf {\bibinfo {volume} {150}},\
  \bibinfo {pages} {204105} (\bibinfo {year} {2019})}\BibitemShut {NoStop}%
\bibitem [{\citenamefont {Kosloff}\ and\ \citenamefont {Rezek}(2017)}]{Rezek}%
  \BibitemOpen
  \bibfield  {author} {\bibinfo {author} {\bibfnamefont {R.}~\bibnamefont
  {Kosloff}}\ and\ \bibinfo {author} {\bibfnamefont {Y.}~\bibnamefont
  {Rezek}},\ }\href {\doibase 10.3390/e19040136} {\bibfield  {journal}
  {\bibinfo  {journal} {Entropy}\ }\textbf {\bibinfo {volume} {19}},\ \bibinfo
  {pages} {136} (\bibinfo {year} {2017})}\BibitemShut {NoStop}%
\bibitem [{\citenamefont {Peterson}\ \emph {et~al.}(2019)\citenamefont
  {Peterson}, \citenamefont {Batalh{\~a}o}, \citenamefont {Herrera},
  \citenamefont {Souza}, \citenamefont {Sarthour}, \citenamefont {Oliveira},\
  and\ \citenamefont {Serra}}]{Peterson2019}%
  \BibitemOpen
  \bibfield  {author} {\bibinfo {author} {\bibfnamefont {J.~P.}\ \bibnamefont
  {Peterson}}, \bibinfo {author} {\bibfnamefont {T.~B.}\ \bibnamefont
  {Batalh{\~a}o}}, \bibinfo {author} {\bibfnamefont {M.}~\bibnamefont
  {Herrera}}, \bibinfo {author} {\bibfnamefont {A.~M.}\ \bibnamefont {Souza}},
  \bibinfo {author} {\bibfnamefont {R.~S.}\ \bibnamefont {Sarthour}}, \bibinfo
  {author} {\bibfnamefont {I.~S.}\ \bibnamefont {Oliveira}}, \ and\ \bibinfo
  {author} {\bibfnamefont {R.~M.}\ \bibnamefont {Serra}},\ }\href {\doibase
  10.1103/PhysRevLett.123.240601} {\bibfield  {journal} {\bibinfo  {journal}
  {Physical Review Letters}\ }\textbf {\bibinfo {volume} {123}},\ \bibinfo
  {pages} {240601} (\bibinfo {year} {2019})}\BibitemShut {NoStop}%
\bibitem [{\citenamefont {Shi}\ \emph {et~al.}(2009)\citenamefont {Shi},
  \citenamefont {Wysocki},\ and\ \citenamefont
  {Belashchenko}}]{Shi2009_PhysRevB.79.104404}%
  \BibitemOpen
  \bibfield  {author} {\bibinfo {author} {\bibfnamefont {S.}~\bibnamefont
  {Shi}}, \bibinfo {author} {\bibfnamefont {A.~L.}\ \bibnamefont {Wysocki}}, \
  and\ \bibinfo {author} {\bibfnamefont {K.~D.}\ \bibnamefont {Belashchenko}},\
  }\href {\doibase 10.1103/PhysRevB.79.104404} {\bibfield  {journal} {\bibinfo
  {journal} {Phys. Rev. B}\ }\textbf {\bibinfo {volume} {79}},\ \bibinfo
  {pages} {104404} (\bibinfo {year} {2009})}\BibitemShut {NoStop}%
\bibitem [{\citenamefont {Mostovoy}\ \emph {et~al.}(2010)\citenamefont
  {Mostovoy}, \citenamefont {Scaramucci}, \citenamefont {Spaldin},\ and\
  \citenamefont {Delaney}}]{Mostovoy2010_PhysRevLett.105.087202}%
  \BibitemOpen
  \bibfield  {author} {\bibinfo {author} {\bibfnamefont {M.}~\bibnamefont
  {Mostovoy}}, \bibinfo {author} {\bibfnamefont {A.}~\bibnamefont
  {Scaramucci}}, \bibinfo {author} {\bibfnamefont {N.~A.}\ \bibnamefont
  {Spaldin}}, \ and\ \bibinfo {author} {\bibfnamefont {K.~T.}\ \bibnamefont
  {Delaney}},\ }\href {\doibase 10.1103/PhysRevLett.105.087202} {\bibfield
  {journal} {\bibinfo  {journal} {Phys. Rev. Lett.}\ }\textbf {\bibinfo
  {volume} {105}},\ \bibinfo {pages} {087202} (\bibinfo {year}
  {2010})}\BibitemShut {NoStop}%
\bibitem [{\citenamefont {Sawada}(1994)}]{Sawada1994}%
  \BibitemOpen
  \bibfield  {author} {\bibinfo {author} {\bibfnamefont {H.}~\bibnamefont
  {Sawada}},\ }\href {\doibase 10.1016/0025-5408(94)90019-1} {\bibfield
  {journal} {\bibinfo  {journal} {Mater. Res. Bull.}\ }\textbf {\bibinfo
  {volume} {29}},\ \bibinfo {pages} {239} (\bibinfo {year} {1994})}\BibitemShut
  {NoStop}%
\bibitem [{\citenamefont {Momma}\ and\ \citenamefont
  {Izumi}(2011)}]{Momma2011jac}%
  \BibitemOpen
  \bibfield  {author} {\bibinfo {author} {\bibfnamefont {K.}~\bibnamefont
  {Momma}}\ and\ \bibinfo {author} {\bibfnamefont {F.}~\bibnamefont {Izumi}},\
  }\href {\doibase 10.1107/S0021889811038970} {\bibfield  {journal} {\bibinfo
  {journal} {J. Appl. Crystallogr.}\ }\textbf {\bibinfo {volume} {44}},\
  \bibinfo {pages} {1272} (\bibinfo {year} {2011})}\BibitemShut {NoStop}%
\bibitem [{\citenamefont {Allahverdyan}\ and\ \citenamefont
  {Nieuwenhuizen}(2005)}]{Allahverdyan2005}%
  \BibitemOpen
  \bibfield  {author} {\bibinfo {author} {\bibfnamefont {A.~E.}\ \bibnamefont
  {Allahverdyan}}\ and\ \bibinfo {author} {\bibfnamefont {T.~M.}\ \bibnamefont
  {Nieuwenhuizen}},\ }\href {\doibase 10.1016/j.physe.2005.05.003} {\bibfield
  {journal} {\bibinfo  {journal} {Physica E}\ }\textbf {\bibinfo {volume}
  {29}},\ \bibinfo {pages} {74} (\bibinfo {year} {2005})}\BibitemShut {NoStop}%
\bibitem [{\citenamefont {Plastina}\ \emph {et~al.}(2014)\citenamefont
  {Plastina}, \citenamefont {Alecce}, \citenamefont {Apollaro}, \citenamefont
  {Falcone}, \citenamefont {Francica}, \citenamefont {Galve}, \citenamefont
  {Gullo},\ and\ \citenamefont {Zambrini}}]{Plastina2014}%
  \BibitemOpen
  \bibfield  {author} {\bibinfo {author} {\bibfnamefont {F.}~\bibnamefont
  {Plastina}}, \bibinfo {author} {\bibfnamefont {A.}~\bibnamefont {Alecce}},
  \bibinfo {author} {\bibfnamefont {T.~J.}\ \bibnamefont {Apollaro}}, \bibinfo
  {author} {\bibfnamefont {G.}~\bibnamefont {Falcone}}, \bibinfo {author}
  {\bibfnamefont {G.}~\bibnamefont {Francica}}, \bibinfo {author}
  {\bibfnamefont {F.}~\bibnamefont {Galve}}, \bibinfo {author} {\bibfnamefont
  {N.~L.}\ \bibnamefont {Gullo}}, \ and\ \bibinfo {author} {\bibfnamefont
  {R.}~\bibnamefont {Zambrini}},\ }\href {\doibase
  10.1103/PhysRevLett.113.260601} {\bibfield  {journal} {\bibinfo  {journal}
  {Phys. Rev. Lett.}\ }\textbf {\bibinfo {volume} {113}},\ \bibinfo {pages}
  {260601} (\bibinfo {year} {2014})}\BibitemShut {NoStop}%
\bibitem [{\citenamefont {Alecce}\ \emph {et~al.}(2015)\citenamefont {Alecce},
  \citenamefont {Galve}, \citenamefont {Gullo}, \citenamefont {Dell’Anna},
  \citenamefont {Plastina},\ and\ \citenamefont {Zambrini}}]{Alecce2015}%
  \BibitemOpen
  \bibfield  {author} {\bibinfo {author} {\bibfnamefont {A.}~\bibnamefont
  {Alecce}}, \bibinfo {author} {\bibfnamefont {F.}~\bibnamefont {Galve}},
  \bibinfo {author} {\bibfnamefont {N.~L.}\ \bibnamefont {Gullo}}, \bibinfo
  {author} {\bibfnamefont {L.}~\bibnamefont {Dell’Anna}}, \bibinfo {author}
  {\bibfnamefont {F.}~\bibnamefont {Plastina}}, \ and\ \bibinfo {author}
  {\bibfnamefont {R.}~\bibnamefont {Zambrini}},\ }\href {\doibase
  10.1088/1367-2630/17/7/075007} {\bibfield  {journal} {\bibinfo  {journal}
  {New J. Phys.}\ }\textbf {\bibinfo {volume} {17}},\ \bibinfo {pages} {075007}
  (\bibinfo {year} {2015})}\BibitemShut {NoStop}%
\bibitem [{\citenamefont {{\c{C}}akmak}\ \emph {et~al.}(2016)\citenamefont
  {{\c{C}}akmak}, \citenamefont {Altintas},\ and\ \citenamefont
  {M{\"u}stecapl{\i}o{\u{g}}lu}}]{Cakmak2016}%
  \BibitemOpen
  \bibfield  {author} {\bibinfo {author} {\bibfnamefont {S.}~\bibnamefont
  {{\c{C}}akmak}}, \bibinfo {author} {\bibfnamefont {F.}~\bibnamefont
  {Altintas}}, \ and\ \bibinfo {author} {\bibfnamefont {{\"O}.~E.}\
  \bibnamefont {M{\"u}stecapl{\i}o{\u{g}}lu}},\ }\href {\doibase
  10.1088/0031-8949/91/7/075101} {\bibfield  {journal} {\bibinfo  {journal}
  {Physica Scripta}\ }\textbf {\bibinfo {volume} {91}},\ \bibinfo {pages}
  {075101} (\bibinfo {year} {2016})}\BibitemShut {NoStop}%
\bibitem [{\citenamefont {Jarzynski}(2013)}]{Jarzynski2013}%
  \BibitemOpen
  \bibfield  {author} {\bibinfo {author} {\bibfnamefont {C.}~\bibnamefont
  {Jarzynski}},\ }\href {\doibase 10.1103/PhysRevA.88.040101} {\bibfield
  {journal} {\bibinfo  {journal} {Phys. Rev. A}\ }\textbf {\bibinfo {volume}
  {88}},\ \bibinfo {pages} {040101} (\bibinfo {year} {2013})}\BibitemShut
  {NoStop}%
\bibitem [{\citenamefont {del Campo}\ \emph {et~al.}(2012)\citenamefont {del
  Campo}, \citenamefont {Rams},\ and\ \citenamefont {Zurek}}]{DelCampo2012}%
  \BibitemOpen
  \bibfield  {author} {\bibinfo {author} {\bibfnamefont {A.}~\bibnamefont {del
  Campo}}, \bibinfo {author} {\bibfnamefont {M.~M.}\ \bibnamefont {Rams}}, \
  and\ \bibinfo {author} {\bibfnamefont {W.~H.}\ \bibnamefont {Zurek}},\ }\href
  {\doibase 10.1103/PhysRevLett.109.115703} {\bibfield  {journal} {\bibinfo
  {journal} {Phys. Rev. Lett.}\ }\textbf {\bibinfo {volume} {109}},\ \bibinfo
  {pages} {115703} (\bibinfo {year} {2012})}\BibitemShut {NoStop}%
\bibitem [{\citenamefont {del Campo}(2013)}]{DelCampo2013}%
  \BibitemOpen
  \bibfield  {author} {\bibinfo {author} {\bibfnamefont {A.}~\bibnamefont {del
  Campo}},\ }\href {\doibase 10.1103/PhysRevLett.111.100502} {\bibfield
  {journal} {\bibinfo  {journal} {Phys. Rev. Lett.}\ }\textbf {\bibinfo
  {volume} {111}},\ \bibinfo {pages} {100502} (\bibinfo {year}
  {2013})}\BibitemShut {NoStop}%
\bibitem [{\citenamefont {Demirplak}\ and\ \citenamefont
  {Rice}(2003)}]{Demirplak2003}%
  \BibitemOpen
  \bibfield  {author} {\bibinfo {author} {\bibfnamefont {M.}~\bibnamefont
  {Demirplak}}\ and\ \bibinfo {author} {\bibfnamefont {S.~A.}\ \bibnamefont
  {Rice}},\ }\href {\doibase 10.1021/jp030708a} {\bibfield  {journal} {\bibinfo
   {journal} {J. Phys. Chem. A}\ }\textbf {\bibinfo {volume} {107}},\ \bibinfo
  {pages} {9937} (\bibinfo {year} {2003})}\BibitemShut {NoStop}%
\bibitem [{\citenamefont {Berry}(2009)}]{Berry2009}%
  \BibitemOpen
  \bibfield  {author} {\bibinfo {author} {\bibfnamefont {M.~V.}\ \bibnamefont
  {Berry}},\ }\href {\doibase 10.1088/1751-8113/42/36/365303} {\bibfield
  {journal} {\bibinfo  {journal} {J. Phys. A Math. Theor.}\ }\textbf {\bibinfo
  {volume} {42}},\ \bibinfo {pages} {365303} (\bibinfo {year}
  {2009})}\BibitemShut {NoStop}%
\bibitem [{\citenamefont {Song}\ \emph {et~al.}(2016)\citenamefont {Song},
  \citenamefont {Zhang}, \citenamefont {Ai}, \citenamefont {Qiu},\ and\
  \citenamefont {Deng}}]{XueKeSong}%
  \BibitemOpen
  \bibfield  {author} {\bibinfo {author} {\bibfnamefont {X.-K.}\ \bibnamefont
  {Song}}, \bibinfo {author} {\bibfnamefont {H.}~\bibnamefont {Zhang}},
  \bibinfo {author} {\bibfnamefont {Q.}~\bibnamefont {Ai}}, \bibinfo {author}
  {\bibfnamefont {J.}~\bibnamefont {Qiu}}, \ and\ \bibinfo {author}
  {\bibfnamefont {F.-G.}\ \bibnamefont {Deng}},\ }\href {\doibase
  10.1088/1367-2630/18/2/023001} {\bibfield  {journal} {\bibinfo  {journal}
  {New J. Phys.}\ }\textbf {\bibinfo {volume} {18}},\ \bibinfo {pages} {023001}
  (\bibinfo {year} {2016})}\BibitemShut {NoStop}%
\bibitem [{\citenamefont {Abah}\ and\ \citenamefont {Lutz}(2017)}]{Abah2017}%
  \BibitemOpen
  \bibfield  {author} {\bibinfo {author} {\bibfnamefont {O.}~\bibnamefont
  {Abah}}\ and\ \bibinfo {author} {\bibfnamefont {E.}~\bibnamefont {Lutz}},\
  }\href {\doibase 10.1209/0295-5075/118/40005} {\bibfield  {journal} {\bibinfo
   {journal} {EPL (Europhysics Letters)}\ }\textbf {\bibinfo {volume} {118}},\
  \bibinfo {pages} {40005} (\bibinfo {year} {2017})}\BibitemShut {NoStop}%
\bibitem [{\citenamefont {Abah}\ and\ \citenamefont {Lutz}(2018)}]{Abah2018}%
  \BibitemOpen
  \bibfield  {author} {\bibinfo {author} {\bibfnamefont {O.}~\bibnamefont
  {Abah}}\ and\ \bibinfo {author} {\bibfnamefont {E.}~\bibnamefont {Lutz}},\
  }\href {\doibase 10.1103/PhysRevE.98.032121} {\bibfield  {journal} {\bibinfo
  {journal} {Physical Review E}\ }\textbf {\bibinfo {volume} {98}},\ \bibinfo
  {pages} {032121} (\bibinfo {year} {2018})}\BibitemShut {NoStop}%
\bibitem [{\citenamefont {Abah}\ and\ \citenamefont
  {Paternostro}(2019)}]{Abah2019}%
  \BibitemOpen
  \bibfield  {author} {\bibinfo {author} {\bibfnamefont {O.}~\bibnamefont
  {Abah}}\ and\ \bibinfo {author} {\bibfnamefont {M.}~\bibnamefont
  {Paternostro}},\ }\href {\doibase 10.1103/PhysRevE.99.022110} {\bibfield
  {journal} {\bibinfo  {journal} {Physical Review E}\ }\textbf {\bibinfo
  {volume} {99}},\ \bibinfo {pages} {022110} (\bibinfo {year}
  {2019})}\BibitemShut {NoStop}%
\bibitem [{\citenamefont {{\c{C}}akmak}\ and\ \citenamefont
  {M{\"u}stecapl{\i}o{\u{g}}lu}()}]{Cakmak2019}%
  \BibitemOpen
  \bibfield  {author} {\bibinfo {author} {\bibfnamefont {B.}~\bibnamefont
  {{\c{C}}akmak}}\ and\ \bibinfo {author} {\bibfnamefont {{\"O}.~E.}\
  \bibnamefont {M{\"u}stecapl{\i}o{\u{g}}lu}},\ }\href {\doibase
  10.1103/PhysRevE.99.032108} {\bibfield  {journal} {\bibinfo  {journal}
  {Physical Review E}\ }10.1103/PhysRevE.99.032108}\BibitemShut {NoStop}%
\bibitem [{\citenamefont {Breuer}\ \emph {et~al.}(2002)\citenamefont {Breuer},
  \citenamefont {Petruccione} \emph {et~al.}}]{Breuer2002}%
  \BibitemOpen
  \bibfield  {author} {\bibinfo {author} {\bibfnamefont {H.-P.}\ \bibnamefont
  {Breuer}}, \bibinfo {author} {\bibfnamefont {F.}~\bibnamefont {Petruccione}},
   \emph {et~al.},\ }\href {\doibase 10.1093/acprof:oso/9780199213900.001.0001}
  {\emph {\bibinfo {title} {The theory of open quantum systems}}}\ (\bibinfo
  {publisher} {Oxford University Press on Demand},\ \bibinfo {year}
  {2002})\BibitemShut {NoStop}%
\end{thebibliography}%

\end{document}